\documentclass[10pt]{article}
\usepackage{dcolumn}
\usepackage{latexsym}
\usepackage{graphicx}
\usepackage{pstricks}
\usepackage{epsfig}
\usepackage{color}

\input epsf
\usepackage{amsmath,amssymb}
\newcommand{\be}{\begin{equation}}
\newcommand{\ee}{\end{equation}}
\newcommand{\bea}{\begin{eqnarray}}
\newcommand{\eea}{\end{eqnarray}}

\newcommand{\lbl}[1]{\label{eq:#1}}
\newcommand{ \rf}[1]{(\ref{eq:#1})}

\newcommand{\lapprox}{%
\mathrel{%
\setbox0=\hbox{$<$}
\raise0.6ex\copy0\kern-\wd0
\lower0.65ex\hbox{$\sim$}
}}
\newcommand{\gapprox}{%
\mathrel{%
\setbox0=\hbox{$>$}
\raise0.6ex\copy0\kern-\wd0
\lower0.65ex\hbox{$\sim$}
}}

\def\theequation{\arabic{section}.\arabic{equation}}

\begin{document}

\renewcommand{\thefootnote}{\fnsymbol{footnote}}

\begin{center}
{\Large\bf A dispersive study of final-state interactions in $K\to\pi\pi\pi$ amplitudes}

\vspace{0.75cm}

{V\'eronique Bernard$^1$}\footnote{veronique.bernard@ijclab.in2p3.fr}, S\'ebastien Descotes-Genon$^1$\footnote{sebastien.descotes-genon@ijclab.in2p3.fr}, 
{Marc Knecht$^2$}\footnote{knecht@cpt.univ-mrs.fr},
Bachir Moussallam$^1$\footnote{bachir.moussallam@ijclab.in2p3.fr}

\indent

{$^1${\small\it{Universit\'e Paris-Saclay, CNRS/IN2P3, IJCLab, 91405 Orsay, France}} }

\vspace{0.2cm}

{$^2${\small\it{Centre de Physique Th\'{e}orique, CNRS/Aix-Marseille Univ./Univ. de Toulon (UMR7332)\\
CNRS-Luminy Case 907, 13288 Marseille Cedex 9, France}} }

\begin{abstract}
A system of dispersive representations of the Omn\`es-Khuri-Treiman-Sawyer-Wali type  for the final-state interactions
in the amplitudes of the $K\to\pi\pi\pi$ weak transitions is constructed, under the assumptions that
CP and isospin symmetries are conserved. Both the $\Delta I = 1/2$ and $\Delta I = 3/2$ transition channels 
are considered. The set of single-variable functions involved in these representations is identified,
and the polynomial ambiguities in their definitions are discussed. Numerical solutions for the
system of coupled integral relations satisfied by these functions are provided, and the determination 
of the subtraction constants in terms of the experimental information on the amplitudes in the decay 
region is addressed.

\end{abstract}

\end{center}

\renewcommand{\thefootnote}{\arabic{footnote}}

\section{Introduction}\label{sec:intro}
\setcounter{equation}{0}

The dispersive treatment of final-state interactions in three-body decays of mesons was originally 
introduced in refs. \cite{Khuri:1960zz} and \cite{Sawyer:1960hrc} in order to describe the observed 
deviation from a constant of the three-pion decay amplitudes of the charged kaons. 
This formalism was subsequently further studied, improved and developed by various authors,
see refs. \cite{Bronzan:1963mby,Kacser:1963zz,Bronzan:1964zz,Aitchison:1664,Aitchison:1966lpz,Pasquier:1968zz,Pasquier:1969dt,Neveu:1970tn} 
for a representative but incomplete list. Additional references and an informative account can be found in \cite{Aitchison:2015jxa}.
Although the early literature was mainly motivated by the 
description of $\pi\pi$ rescattering in the $K\to\pi\pi\pi$ amplitudes \cite{Khuri:1960zz,Sawyer:1960hrc,Kacser:1963zz,Neveu:1970tn}, the more recent works
were rather devoted to the three-pion decay of the eta meson  
\cite{Anisovich:1993kn,Kambor:1995yc,Anisovich:1996tx,Guo:2015zqa,Colangelo:2018jxw,Albaladejo:2017hhj,Gasser:2018qtg}. 
But dispersive approaches of this Omn\`es-Khuri-Treiman-Sawyer-Wali\footnote{From now on, we will follow the common practice
in the literature and refer to Omn\`es-Khuri-Treiman or even simply to Khuri-Treiman equations.} type were 
also devised in a variety of other contexts, see for instance the articles 
\cite{Descotes-Genon:2014tla,Niecknig:2012sj,Kubis:2014gka,Isken:2017dkw,Albaladejo:2018gif,JPAC:2020umo,Dax:2020dzg,Stamen:2022eda}
for a sample of examples and for further references. Interestingly enough, the $K\to\pi\pi\pi$ transitions were 
not part of the more recent surge of interest for this kind of dispersive approach.

The main motivation for the present work, however, is not to simply fill this gap. It rather takes its origin in the 
observation that the amplitudes for the $K\to\pi\pi\pi$ transitions also enter the theoretical studies 
of the amplitudes of a certain number of rare decay modes of the kaons, like $K_L\to\pi^0\gamma\gamma$, 
$K_L\to\pi^0\gamma \ell^+\ell^-$, $K^\pm\to\pi^\pm\gamma\gamma$, $K\to\pi\ell^+\ell^-$,
or $K^\pm\to\pi^\pm\gamma e^+ e^-$,  when considered from the point of view of the low-energy expansion. For these amplitudes, the lowest-order contribution 
starts at the one-loop level, with a pion loop rescattering into the leptonic/photonic component of the final state and originating from a tree-level 
$K\pi\pi\pi$ vertex. The latter is described
by the two low-energy constants that also give the tree-level contribution to the $K\to\pi\pi$ amplitudes \cite{Kambor:1991ah,Bijnens:2002vr}. 
However, most of the time these one-loop amplitudes do not provide a sufficiently accurate description of experimental data, and higher-order 
effects need to be included. In order to bypass a full-fledged two-loop calculation, the corrections that are applied to these amplitudes
often amount, among other things, to the replacement of the two couplings entering the description of the $K\pi\pi\pi$ vertex
at lowest order by appropriate combinations of parameters extracted from 
the phenomenological study of the experimental Dalitz plots of the $K\to\pi\pi\pi$ decays \cite{Devlin:1978ye,Kambor:1991ah,Bijnens:2002vr,DAmbrosio:2022jmd}.
For the amplitudes listed above, a procedure of this type was implemented in the articles \cite{Cappiello:1992kk,Kambor:1993tv,Cohen:1993ta}, 
\cite{Donoghue:1997rr,Donoghue:1998ur},
\cite{DAmbrosio:1996cak}, \cite{DAmbrosio:1998gur}, and \cite{Gabbiani:1998tj}, respectively.\footnote{Ref. \cite{Kambor:1993tv} stands 
somewhat apart in this list, since the authors actually consider a treatment \textit{\`a la} Khuri-Treiman but only for the subset consisting of the 
$K_L\to\pi^0\pi^0\pi^0$ and $K_L\to \pi^+ \pi^-\pi^0$ amplitudes and with a certain number of additional approximations. In particular,
contributions from $\Delta I = 3/2$ transitions and from P waves are not included.\label{footnote}}
With the recent increase of precision in the experimental 
measurements of some of these processes, and with further improvements on the experimental side to be expected in the future
\cite{Goudzovski:2022scl,Aebischer:2022vky,NA62KLEVER:2022nea,HIKE:2022qra}, these simple unitarization procedures,
as they are often presented, might no longer be sufficient for an appropriate description of the data. Having at disposal a more
solid description of final-state interactions in the $K\pi\to\pi\pi$ amplitudes constitutes a first step toward improving the situation
on the theoretical side. 
The purpose of the present article is to construct and to provide such a tool.
The subsequent applications to specific radiative kaon-decay amplitudes will, however, not be considered here and will be left for future studies.

On the practical side, our study will be performed under similar conditions as in refs. \cite{Khuri:1960zz,Sawyer:1960hrc}, in the sense 
that CP-violating and isospin-breaking effects will be ignored.\footnote{Since we stay in the isospin limit  we will thus not be able
to describe effect related to the difference between the charged and neutral pion masses, like the cusp in the decay distributions of 
the decay modes $K^\pm\to\pi^\pm\pi^0\pi^0$ and $K_L\to\pi^0\pi^0\pi^0$ observed experimentally by the NA48/2 \cite{NA482:2005wht}
and KTeV \cite{KTeV:2008gel} collaborations, respectively.} But in contrast to these references
we will include both S- and P-wave contributions to the final-state rescattering of pion pairs, as done routinely in more recent studies, 
assuming that the contributions from higher partial waves
to the absorptive parts can be neglected, at least in the range of energies where the resulting
expressions can be applied. From this point of view, the general framework is quite similar to the one
currently considered in the study of final-state interactions in the decay of the eta meson
into three pions \cite{Anisovich:1993kn,Kambor:1995yc,Anisovich:1996tx,Guo:2015zqa,Colangelo:2018jxw,Albaladejo:2017hhj,Gasser:2018qtg}. 
The main difference lies in the fact that the number of amplitudes to start with is somewhat larger
in the case of the kaons, which can be charged or neutral.
This is even more so the case to the extent that we will consider both the $\Delta I = 1/2$ and 
$\Delta I = 3/2$ components of the amplitudes. Fortunately, these can (and will of course) be treated separately.

The remainder of this article is organized in the following manner. In section \ref{sec:isospin_crossing} 
the set of initial 13 independent amplitudes is expressed in terms of nine isospin amplitudes. Crossing
properties further reduce this set to only four independent amplitudes. In section \ref{sec:single_var}
the isospin amplitudes are decomposed into single-variable functions. The polynomial ambiguities 
of this decomposition are discussed and, with a choice of appropriate asymptotic conditions,
dispersion relations obeyed by the single-variable functions are obtained. The corresponding absorptive 
parts as provided by elastic unitarity are the subject of section \ref{sec:pw_unitarity}. The final set 
of coupled Omn\`es-Khuri-Treiman dispersion relations for the single-variable functions are established 
in section \ref{sec:integral_eqs}, where their iterative numerical solution is also provided. Section \ref{sec:subtraction_csts}
discusses the determination of the subtraction constants and considers two applications of the formalism:
the description of the amplitudes as second-order polynomials in the Dalitz-plot variables, and 
the evaluation of the strong phases of the amplitudes for the decay modes of the charged kaon
over the decay region. A summary and a few conclusions form the content of section \ref{sec:conclusions}.
In order not to interrupt the narrative of the main text too frequently with issues and asides of a more technical 
kind, details related to several specific aspects that are only briefly discussed or mentioned in the 
bulk of the article have been gathered in a series of appendices.

\indent

\section{Isospin and crossing properties of the $K\pi\to\pi\pi$ amplitudes}\label{sec:isospin_crossing}
\setcounter{equation}{0}

We start from the amplitudes for the processes $K^+(p_K)\pi(p_3)\to\pi(p_1)\pi(p_2)$ and $K^0(p_K)\pi(p_3)\to\pi(p_1)\pi(p_2)$ with all 
possible assignments of pion charges compatible with the conservation of the electric charge, and with
all possible pion permutations. This gives altogether 13 amplitudes ${\cal A}_i (s,t,u)$, $i=1,\ldots,13$,
with the labels chosen as indicated in Table \ref{table:amplitudes}. Since violation of CP symmetry
is not important for the present study, we assume CP to be conserved. 
We thus need not consider explicitly
the $K\pi\to\pi\pi$ amplitudes involving $K^-$ or ${\bar K}^0$, since they can be obtained as CP conjugate of those
involving $K^+$ or $K^0$, respectively. The Mandelstam variables are defined as usual,
\be
s = (p_K + p_3)^2 , \quad t = (p_K - p_1)^2 , \quad u = (p_K - p_2)^2
.
\ee
The amplitudes for the decay processes $K\to\pi\pi\pi$ are obtained from the previous ones
through crossing, and are usually expressed in terms of the variables $s_i = (p_K - p_i)^2$,
$i=1,2,3$. In either case, one has $s+t+u = 3 s_0$ or $s_1 + s_2 + s_3 = 3 s_0$, with
\be
s_0 = M_\pi^2 + M_K^2/3 .
\ee
We also assume isospin to be conserved, so that charged and neutral kaons (pions) share the same mass
$M_K$ ($M_\pi$). For numerical applications, we take for the value of $M_\pi$ the mass of the 
charged pion, and for $M_K$ the average of the masses of the neutral and charged kaons. With the values 
provided by the Review of Particle Properties \cite{ParticleDataGroup:2022pth} this means
\be
M_\pi = 0.13957039~{\rm GeV} , \quad M_K = 0.495644~{\rm GeV}.
\ee
The phases inside a given isospin multiplet, i.e. either $(K^0 , K^+)$ or $(\pi^- , \pi^0 , \pi^+)$,
are chosen according to the Condon and Shortley convention \cite{CondonShortley1953}. Phases between these two isospin multiplets
are set according to the prescription of de Swart \cite{deSwart:1963pdg}. These choices explain the crossing phases displayed in
the two last columns of Table \ref{table:amplitudes}. Each initial $K\pi$ state is therefore characterized 
by its total isospin projection ${\cal I}_z$. Likewise, the final $\pi\pi$ states have total isospin projections $I_z$.
These values are given in the third and fourth columns of Table \ref{table:amplitudes}, respectively.
As to the total isospin, it can take the values ${\cal I}=\frac{1}{2}, \frac{3}{2}$ for the $K\pi$ states 
or $I=0,1,2$ for the two-pion states.

The crossing properties and isospin symmetry will reduce the number of independent amplitudes.
Crossing will only operate among subsets of amplitudes separated by the horizontal lines in
Table \ref{table:amplitudes}. Constraints involving amplitudes that belong to different subsets 
will only occur through the implementation of isospin symmetry. In order to achieve this, we need 
to specify the symmetry properties of the operators that are responsible for the $K\pi\to\pi\pi$ transitions.
As is well known, at first order in the Fermi constant, and in the absence of electroweak corrections
(required by our assumption that isospin is conserved) these transitions are described by matrix elements
\be
\langle \pi\pi \vert {\rm  T}^{\Delta S = 1} \vert K\pi \rangle
,
\ee
where ${\rm T}^{\Delta S = 1}$ is given by the integral of the effective Lagrangian for weak non-leptonic $\Delta S = 1$ transitions
\cite{Gaillard:1974nj,Altarelli:1974exa,Shifman:1975tn,Witten:1976kx,Wise:1979at,Gilman:1979bc}.
What matters here is that ${\rm T}^{\Delta S = 1}$ proceeds by either $\Delta I = \frac{1}{2}$ or $\Delta I = \frac{3}{2}$
isospin transitions,
\be
{\rm T}^{\Delta S = 1} = {\rm T}^{\frac{1}{2}}_{\frac{1}{2}} + {\rm T}^{\frac{3}{2}}_{\frac{1}{2}}
,
\ee
where the superscript gives the total isospin $\Delta I$ and the subscript the corresponding isospin projection,
which is necessarily $\frac{1}{2}$. In order to simplify the notation, we will omit this subscript in the sequel.
Upon expanding the $\vert K \pi\rangle$ and $\vert\pi\pi\rangle$ states
over isospin states $\vert {\cal I} , {\cal I}_z \rangle$ and $\vert I , I_z \rangle$, respectively, we can implement the Wigner-Eckart
theorem and express each transition amplitude in terms of a small number of reduced matrix elements, multiplied by the
appropriate Clebsch-Gordan coefficients,
\be
\langle I , I_z \vert {\rm T}^{\frac{1}{2}} \vert {\cal I} , {\cal I}_z \rangle =
\Big( {\cal I} , {\cal I}_z ; \frac{1}{2} , \frac{1}{2} \Big\vert I , I_z \Big)
\langle I \vert\!\vert {\rm T}^{\frac{1}{2}} \vert\!\vert {\cal I} \rangle
,~~
\langle I , I_z \vert {\rm T}^{\frac{3}{2}} \vert {\cal I} , {\cal I}_z \rangle =
\Big( {\cal I} , {\cal I}_z ; \frac{3}{2} , \frac{1}{2} \Big\vert I , I_z \Big)
\langle I \vert\!\vert {\rm T}^{\frac{3}{2}} \vert\!\vert {\cal I} \rangle
.
\ee
With the phase convention adopted here, the Clebsch-Gordan coefficients can be read off,
for instance, from the tables provided in ref. \cite{ParticleDataGroup:2022pth}.
\begin{table}[t]
\begin{center}
\begin{tabular}{c|c|c|c|c|c|c}
 $i$ & reaction & ${\cal I}_z$ & $I_z$ & $t \leftrightarrow u$ & $s \leftrightarrow t$ & $u \leftrightarrow s$
\\
\hline\hline\noalign{\smallskip}
1 & $K^+\pi^+\to\pi^+\pi^+$  & $+\frac{3}{2}$ & $\!\!{+2}$ & 1  &  3  &  2
\\[0.15cm]
2 & $K^+\pi^-\to\pi^+\pi^-$  & $-\frac{1}{2}$ & $~0$ & 3  &  2  &  1
\\[0.15cm]
3 & $K^+\pi^-\to\pi^-\pi^+$  & $-\frac{1}{2}$ & $~0$ & 2  &  1  &  3
\\[0.15cm]
\hline\noalign{\smallskip}
4 & $K^+\pi^-\to \pi^0\pi^0$ & $-\frac{1}{2}$ & $~0$ & 4  & $\!\!\!\!{-6}$ & $\!\!\!\!{-5}$
\\[0.15cm]
5 & $K^+\pi^0\to \pi^0\pi^+$ & $+\frac{1}{2}$ & $\!\!{+1}$ & 6  & 5 & $\!\!\!\!{-4}$
\\[0.15cm]
6 & $K^+\pi^0\to \pi^+\pi^0$ & $+\frac{1}{2}$ & $\!\!{+1}$ & 5  & $\!\!\!\!{-4}$ & 6
\\[0.15cm]
\hline\noalign{\smallskip}
7 & $K^0\pi^0\to \pi^+\pi^-$ & $-\frac{1}{2}$ & $~0$ & 8  & $\!\!\!\!\!{-12}$ & $\!\!\!\!{-9}$
\\[0.15cm]
8 & $K^0\pi^0\to \pi^-\pi^+$ & $-\frac{1}{2}$ & $~0$ & 7  & $\!\!\!\!\!{-10}$ & $\!\!\!\!\!{-11}$
\\[0.15cm]
9 & $K^0\pi^+\to \pi^+\pi^0$ & $+\frac{1}{2}$ & $\!\!{+1}$ & $\!\!{10}$  & $\!{11}$ & $\!\!\!\!{-7}$
\\[0.15cm]
$\!\!{10}$ & $K^0\pi^+\to \pi^0\pi^+$ & $+\frac{1}{2}$ & $\!\!{+1}$ & 9  & $\!\!\!\!{-8}$ & $\!\!{12}$
\\[0.15cm]
$\!\!{11}$ & $K^0\pi^-\to \pi^-\pi^0$ & $-\frac{3}{2}$ & $\!\!{-1}$ & $\!\!{12}$  & 9 & $\!\!\!\!{-8}$
\\[0.15cm]
$\!\!{12}$ & $K^0\pi^-\to \pi^0\pi^-$ & $-\frac{3}{2}$ & $\!\!{-1}$ & $\!\!{11}$  & $\!\!\!\!{-7}$ & $\!\!{10}$
\\[0.15cm]
\hline\noalign{\smallskip}
$\!\!{13}$ & $K^0\pi^0\to \pi^0\pi^0$ & $-\frac{1}{2}$ & $~0$ & $\!\!{13}$  & $\!\!{13}$ & $\!\!{13}$
\\[0.15cm]
\hline\hline
\end{tabular}
\caption{The 13 amplitudes ${\cal A}_i(s,t,u)$ with the corresponding isospin projections ${\cal I}_z$ and $I_z$ of the $K\pi$
and $\pi\pi$ pairs, respectively. The first column gives the label $i$ assigned to the amplitude that describes the process 
displayed in the second column. The crossing properties of the amplitudes are also indicated in the last three columns. These 
entries should be read, for instance, as follows: ${\cal A}_1 (t,s,u) = + {\cal A}_3 (s,t,u)$ or
${\cal A}_4 (u,t,s) = - {\cal A}_5 (t,s,u)$. The horizontal lines delimit subsets of amplitudes that are closed under crossing.}
\label{table:amplitudes}
\end{center}
\end{table}
There are four independent reduced amplitudes of the type $\langle I \vert\!\vert {\rm T}^{\frac{1}{2}} \vert\!\vert {\cal I} \rangle (s,t,u)$,
and five of the type $\langle I \vert\!\vert {\rm T}^{\frac{3}{2}} \vert\!\vert {\cal I} \rangle (s,t,u)$. For reasons of convenience,
we will work with the isospin amplitudes ${\cal M}_1^{I ; {\cal I}} (s,t,u)$, defined as
\be
{\vec{\cal M}}_1 (s,t,u) \equiv
\left(\!\!\!
\begin{tabular}{c}
 ${\cal M}_1^{2;\frac{3}{2}} (s,t,u)$
\\
 ${\cal M}_1^{1;\frac{3}{2}} (s,t,u)$
\\
 ${\cal M}_1^{1;\frac{1}{2}} (s,t,u)$
\\
 ${\cal M}_1^{0;\frac{1}{2}} (s,t,u)$
\end{tabular}
\!\!\!\right)
=
\left(\!\!
\begin{tabular}{cccc}
$1$ & 0 & 0 & 0
\\ 
0 & $-\sqrt{3}$ & 0 & 0 
\\ 
0 & 0 & $\sqrt{\frac{3}{2}}$ & 0 
\\ 
0 & 0 & 0 & $-1$
\end{tabular}
\!\!\right)
\left(\!\!\!
\begin{tabular}{c}
$\langle 2 \vert\!\vert {\rm T}^{\frac{1}{2}} \vert\!\vert \frac{3}{2} \rangle (s,t,u)$
\\
$\langle 1 \vert\!\vert {\rm T}^{\frac{1}{2}} \vert\!\vert \frac{3}{2} \rangle (s,t,u)$
\\
$\langle 1 \vert\!\vert {\rm T}^{\frac{1}{2}} \vert\!\vert \frac{1}{2} \rangle (s,t,u)$
\\
$\langle 0 \vert\!\vert {\rm T}^{\frac{1}{2}} \vert\!\vert \frac{1}{2} \rangle (s,t,u)$
\end{tabular}
\!\!\!\right)
,
\lbl{isospin_amplitudes_1}
\ee
for the $\Delta I = \frac{1}{2}$ transitions, and, for the $\Delta I = \frac{3}{2}$ transitions, 
with the isospin amplitudes ${\cal M}_3^{I ; {\cal I}} (s,t,u)$, where
\be
{\vec{\cal M}}_3 (s,t,u) \equiv
\left(\!\!\!
\begin{tabular}{c}
 ${\cal M}_3^{2;\frac{3}{2}} (s,t,u)$
\\
 ${\cal M}_3^{2;\frac{1}{2}} (s,t,u)$
\\
 ${\cal M}_3^{1;\frac{3}{2}} (s,t,u)$
\\
 ${\cal M}_3^{1;\frac{1}{2}} (s,t,u)$
\\
 ${\cal M}_3^{0;\frac{3}{2}} (s,t,u)$
\end{tabular}
\!\!\!\right)
=
\left(\!\!
\begin{tabular}{ccccc}
$\frac{1}{\sqrt{2}}$ & 0 & 0 & 0 & 0
\\ 
0 & $\frac{1}{\sqrt{2}}$ & 0 & 0 & 0
\\ 
0 & 0 & $-\sqrt{\frac{3}{10}}$ & 0 & 0 
\\  
0 & 0 & 0 & $-\sqrt{\frac{3}{2}}$ & 0 
\\ 
0 & 0 & 0 & 0 & $\frac{1}{2}$ 
\end{tabular}
\!\!\right)
\left(\!\!\!
\begin{tabular}{c}
$\langle 2 \vert\!\vert {\rm T}^{\frac{3}{2}} \vert\!\vert \frac{3}{2} \rangle (s,t,u)$
\\
$\langle 2 \vert\!\vert {\rm T}^{\frac{3}{2}} \vert\!\vert \frac{1}{2} \rangle (s,t,u)$
\\
$\langle 1 \vert\!\vert {\rm T}^{\frac{3}{2}} \vert\!\vert \frac{3}{2} \rangle (s,t,u)$
\\
$\langle 1 \vert\!\vert {\rm T}^{\frac{3}{2}} \vert\!\vert \frac{1}{2} \rangle (s,t,u)$
\\
$\langle 0 \vert\!\vert {\rm T}^{\frac{3}{2}} \vert\!\vert \frac{3}{2} \rangle (s,t,u)$
\end{tabular}
\!\!\!\right)
.
\lbl{isospin_amplitudes_3}
\ee
In matrix notation, the relations between the 13 amplitudes ${\cal A}_i$ of Table \ref{table:amplitudes}
and these isospin amplitudes write as
\be
{\vec{\cal A}} (s,t,u) = {\mathbf{R}}_1 {\vec{\cal M}}_1 (s,t,u)
+ {\mathbf{R}}_3 {\vec{\cal M}}_3 (s,t,u)   ,
\lbl{amplitudes_iso}
\ee
with the $4\times 13$ and $5\times 13$ matrices ${\mathbf{R}}_1$ and ${\mathbf{R}}_3$ given by
\be
{\mathbf{R}}_1 =
\left(\!\!
\begin{tabular}{cccc}
1  &  0  &  0  &  0
\\
$\frac{1}{6}$ & $\frac{1}{6}$ & $\frac{1}{3}$ & $\frac{1}{3}$ 
\\[0.1cm]
$\frac{1}{6}$ & $-\frac{1}{6}~~$ & $-\frac{1}{3}~~$ & $\frac{1}{3}$  
\\[0.1cm]
$\frac{1}{3}$ & 0 & 0 & $-\frac{1}{3}~~$ 
\\[0.1cm]
$\frac{1}{2}$ & $-\frac{1}{6}~~$ & $-\frac{1}{3}~~$ & 0 
\\[0.1cm]
$\frac{1}{2}$ & $\frac{1}{6}$ & $\frac{1}{3}$ & 0  
\\[0.15cm]
$\frac{1}{3\sqrt{2}}$ & $\frac{1}{3\sqrt{2}}$ & $-\frac{1}{3\sqrt{2}}~~$ & $-\frac{1}{3\sqrt{2}}~~$
\\[0.15cm]
$\frac{1}{3\sqrt{2}}$ & $-\frac{1}{3\sqrt{2}}~~$ & $\frac{1}{3\sqrt{2}}$ & $-\frac{1}{3\sqrt{2}}~~$
\\[0.15cm]
$\frac{1}{2\sqrt{2}}$ & $\frac{1}{6\sqrt{2}}$ & $-\frac{\sqrt{2}}{3}~~$ & 0
\\[0.15cm]
$\frac{1}{2\sqrt{2}}$ & $-\frac{1}{6\sqrt{2}}~~$ & $\frac{\sqrt{2}}{3}$ & 0
\\[0.15cm]
$\frac{1}{2\sqrt{2}}$ & $-\frac{1}{2\sqrt{2}}~~$ & 0 & 0
\\[0.15cm]
$\frac{1}{2\sqrt{2}}$ & $\frac{1}{2\sqrt{2}}$ & 0 & 0
\\[0.15cm]
$\frac{\sqrt{2}}{3}$ & 0 & 0 & $\frac{1}{3\sqrt{2}}$ 
\end{tabular}
\!\!\!\right)
,~
{\mathbf{R}}_3 = 
\left(\!\!
\begin{tabular}{ccccc}
1  &  0  &  0  &  0  &  0
\\
$-\frac{1}{6}~~$ & $\frac{1}{3}$ & $\frac{1}{6}$ & $\frac{1}{3}$ & $\frac{1}{3}$ 
\\[0.1cm]
$-\frac{1}{6}~~$ & $\frac{1}{3}$ & $-\frac{1}{6}~~$ & $-\frac{1}{3}~~$ & $\frac{1}{3}$  
\\[0.1cm]
$-\frac{1}{3}~~$ & $\frac{2}{3}$ & 0 & 0 & $-\frac{1}{3}~~$ 
\\[0.1cm]
0  & $\frac{1}{2}$ & $-\frac{2}{3}~~$ & $\frac{1}{6}$ & 0 
\\[0.1cm]
0  & $\frac{1}{2}$ & $\frac{2}{3}$ & $-\frac{1}{6}~~$ & 0  
\\[0.15cm]
$-\frac{1}{3\sqrt{2}}~~$ & $-\frac{1}{3\sqrt{2}}~~$ & $\frac{1}{3\sqrt{2}}$ & $-\frac{1}{3\sqrt{2}}~~$ & $\frac{\sqrt{2}}{3}$
\\[0.15cm]
$-\frac{1}{3\sqrt{2}}~~$ & $-\frac{1}{3\sqrt{2}}~~$ & $-\frac{1}{3\sqrt{2}}~~$ & $\frac{1}{3\sqrt{2}}$ & $\frac{\sqrt{2}}{3}$
\\[0.1cm]
0  &  $-\frac{1}{\sqrt{2}}~~$ & $\frac{\sqrt{2}}{3}$ & $\frac{1}{3\sqrt{2}}$ & 0
\\[0.1cm]
0  &  $-\frac{1}{2\sqrt{2}}~~$ & $-\frac{\sqrt{2}}{3}~~$ & $-\frac{1}{3\sqrt{2}}~~$ & 0
\\[0.1cm]
$-\frac{1}{\sqrt{2}}~~$ &  0  & $\frac{1}{\sqrt{2}}$ & 0 & 0
\\[0.15cm]
$-\frac{1}{\sqrt{2}}~~$ &  0  & $-\frac{1}{\sqrt{2}}~~$ & 0 & 0
\\[0.15cm]
$-\frac{\sqrt{2}}{3}~~$ &  $-\frac{\sqrt{2}}{3}~~$ & 0 & 0 & $-\frac{\sqrt{2}}{3}~~$  
\end{tabular}
\!\!\!\right)
.
\ee
The crossing relations among the isospin amplitudes follow from those among 
the amplitudes ${\cal A}_i (s,t,u)$, and express themselves in the form
\be
{\vec{\cal M}}_{1,3} (s,u,t) = C_{tu}^{(1,3)} {\vec{\cal M}}_{1,3} (s,t,u)
, \quad
{\vec{\cal M}}_{1,3} (t,s,u) = C_{st}^{(1,3)} {\vec{\cal M}}_{1,3} (s,t,u)
, \quad
{\vec{\cal M}}_{1,3} (u,t,s) = C_{us}^{(1,3)} {\vec{\cal M}}_{1,3} (s,t,u)
.
\lbl{crossing_isospin-amplitudes}
\ee
All six crossing matrices $C_{tu}^{(1,3)}$, $C_{st}^{(1,3)}$, $C_{us}^{(1,3)}$, must square to the unit matrix and satisfy, in addition,
\be
C_{tu}^{(1,3)} C_{st}^{(1,3)} C_{tu}^{(1,3)} = C_{us}^{(1,3)} = C_{st}^{(1,3)} C_{tu}^{(1,3)} C_{st}^{(1,3)}
.
\ee
One finds that the matrices $C_{tu}^{(1,3)}$ are diagonal,\footnote{Exchanging $t$ and $u$ obviously will not modify $I$,
but flips the sign of $I_z$ when $I$ is odd.}
\be
C_{tu}^{(1)} = {\rm diag}(+1 , -1 , -1 , +1)
, ~~
C_{tu}^{(3)} = {\rm diag}(+1 , +1 , -1 , -1 , +1)
,
\ee
whereas the remaining crossing matrices are given by
\be
C_{st}^{(1)} = \left(
\begin{tabular}{cccc}
$+ \frac{1}{6}$ & $- \frac{1}{6}$ & $- \frac{1}{3}$ & $+ \frac{1}{3}$
\\[0.1cm]
$- \frac{5}{6}$ & $- \frac{1}{2}$ & $+ 1$ & $+ \frac{1}{3}$
\\[0.1cm]
$- \frac{5}{6}$ & $+ \frac{1}{2}$ & $~~0$ & $+ \frac{1}{3}$
\\[0.1cm]
$+ \frac{5}{3}$ & $+ \frac{1}{3}$ & $+ \frac{2}{3}$ & $+ \frac{1}{3}$
\end{tabular}
\right)
\!, ~~
C_{us}^{(1)} = \left(
\begin{tabular}{cccc}
$+ \frac{1}{6}$ & $+ \frac{1}{6}$ & $+ \frac{1}{3}$ & $+ \frac{1}{3}$
\\[0.1cm]
$+ \frac{5}{6}$ & $- \frac{1}{2}$ & $+ 1$ & $- \frac{1}{3}$
\\[0.1cm]
$+ \frac{5}{6}$ & $+ \frac{1}{2}$ & $~~0$ & $- \frac{1}{3}$
\\[0.1cm]
$+ \frac{5}{3}$ & $- \frac{1}{3}$ & $- \frac{2}{3}$ & $+ \frac{1}{3}$
\end{tabular}
\right)
\!,
\lbl{crossing_1}
\ee
and
\be
C_{st}^{(3)} = \left(
\begin{tabular}{ccccc}
$- \frac{1}{6}$ & $+ \frac{1}{3}$ & $- \frac{1}{6}$ & $- \frac{1}{3}$ & $+ \frac{1}{3}$
\\[0.1cm]
$+ \frac{1}{3}$ & $- \frac{1}{6}$ & $- \frac{2}{3}$ & $+ \frac{1}{6}$ & $+ \frac{1}{3}$
\\[0.1cm]
$- \frac{1}{6}$ & $- \frac{2}{3}$ & $+ \frac{1}{2}$ & $~~0$ & $+ \frac{1}{3}$
\\[0.1cm]
$- \frac{5}{3}$ & $+ \frac{5}{6}$ & $~~0$ & $+ \frac{1}{2}$ & $+ \frac{1}{3}$
\\[0.1cm]
$+ \frac{5}{6}$ & $+ \frac{5}{6}$ & $+ \frac{5}{6}$ & $+ \frac{1}{6}$ & $+ \frac{1}{3}$
\end{tabular}
\right)
\!, ~~
C_{us}^{(3)} = \left(
\begin{tabular}{ccccc}
$- \frac{1}{6}$ & $+ \frac{1}{3}$ & $+ \frac{1}{6}$ & $+ \frac{1}{3}$ & $+ \frac{1}{3}$
\\[0.1cm]
$+ \frac{1}{3}$ & $- \frac{1}{6}$ & $+ \frac{2}{3}$ & $- \frac{1}{6}$ & $+ \frac{1}{3}$
\\[0.1cm]
$+ \frac{1}{6}$ & $+ \frac{2}{3}$ & $+ \frac{1}{2}$ & $~~0$ & $- \frac{1}{3}$
\\[0.1cm]
$+ \frac{5}{3}$ & $- \frac{5}{6}$ & $~~0$ & $+ \frac{1}{2}$ & $- \frac{1}{3}$
\\[0.1cm]
$+ \frac{5}{6}$ & $+ \frac{5}{6}$ & $- \frac{5}{6}$ & $- \frac{1}{6}$ & $+ \frac{1}{3}$
\end{tabular}
\right)
\!.
\lbl{crossing_3}
\ee
\\
It is a straightforward exercise to check that these matrices indeed satisfy all the properties stated above.
We will indicate a simple way to work out these crossing matrices at the end of this section, once we have
introduced a representation of the isospin functions in terms of a more reduced set of independent amplitudes.

While the isospin amplitudes take into account the restrictions coming from isospin 
symmetry, they do not represent the most compact description of the amplitudes
${\cal A}_i (s,t,u)$, and one may expect that the number of independent functions 
that are required to this effect can be further reduced.
This expectation is, for instance, borne out by the similar and well-known situation 
in the case of elastic pion-pion scattering in the absence of isospin-breaking effects.
There are only three isospin amplitudes in that case, but once crossing is enforced,
a single function $A(s \vert t,u)$, with $A(s \vert t,u) = A(s \vert u,t)$, is 
enough to describe the amplitudes of all the scattering channels \cite{Chew:1960iv,Petersen:1977cs}.
In the present case, one can indeed construct a representation of the isospin amplitudes 
in terms of only four independent functions, two in each of the two sectors, 
${\cal M}_1$ and ${\cal N}_1$ in the $\Delta I = 1/2$ sector, ${\cal M}_3$ and ${\cal N}_3$ in the $\Delta I = 3/2$ sector. 
The details of this construction are given in appendix \ref{app:crossing}, 
whereas appendix \ref{app:spurions} provides, for the interested reader, an alternative 
derivation based on the so-called spurion formalism. Here we simply present the result 
of this analysis, which allows the isospin amplitudes to be expressed as
\bea
{\cal M}^{2;\frac{3}{2}}_1  (s,t,u) &=& {\cal M}_1 (t \vert s,u) 
+ {\cal M}_1 (u \vert s,t)
,
\nonumber\\ 
{\cal M}^{1;\frac{3}{2}}_1  (s,t,u) &=& {\cal M}_1 (t \vert s,u) 
- {\cal M}_1 (u \vert s,t) - 2 {\cal N}_1 (s, t, u) 
,
\nonumber\\ 
{\cal M}^{1;\frac{1}{2}}_1  (s,t,u) &=& {\cal M}_1 (t \vert s,u) 
- {\cal M}_1 (u \vert s,t) +  {\cal N}_1 (s, t, u)
,
\nonumber\\
{\cal M}^{0;\frac{1}{2}}_1  (s,t,u) &=& 3 {\cal M}_1 (s \vert t,u) 
+ {\cal M}_1 (t \vert s,u) + {\cal M}_1 (u \vert s,t) 
,
\lbl{inv_amps_1}
\eea
and
\bea
{\cal M}^{2;\frac{3}{2}}_3  (s,t,u) &=& 
{\cal M}_3 (t \vert s,u) + {\cal M}_3 (u \vert s,t)
+ {\cal N}_3 (t,s,u) + {\cal N}_3 (u,s,t) 
,
\nonumber\\
{\cal M}^{2;\frac{1}{2}}_3  (s,t,u) &=& 
{\cal M}_3 (t \vert s,u) + {\cal M}_3 (u \vert s,t)
,
\nonumber\\  
{\cal M}^{1;\frac{3}{2}}_3  (s,t,u) &=& 
{\cal M}_3 (t \vert s,u) - {\cal M}_3 (u \vert s,t) 
-  {\cal N}_3 (s,t,u)
,
\nonumber\\ 
{\cal M}^{1;\frac{1}{2}}_3  (s,t,u) &=& 
{\cal M}_3 (t \vert s,u) - {\cal M}_3 (u \vert s,t)
- 4  {\cal N}_3 (s,t,u)
,
\nonumber\\ 
{\cal M}^{0;\frac{3}{2}}_3  (s,t,u) &=& 3 \, {\cal M}_3 (s \vert t,u) 
+ {\cal M}_3 (t \vert s,u) + {\cal M}_3 (u \vert s,t) 
-  {\cal N}_3 (t,s,u) -  {\cal N}_3 (u,s,t)
,
\lbl{inv_amps_3}
\eea
where the following restrictions apply:
\begin{enumerate}
 \item[-] both functions ${\cal M}_1 (s \vert t,u) $ and ${\cal M}_3 (s \vert t,u) $
 are symmetric under permutation of the variables $t$ and $u$;
 \item[-] the function ${\cal N}_1 (s,t,u)$ is antisymmetric under permutation of any two of its variables;
 \item[-] the function ${\cal N}_3 (s,t,u)$ is antisymmetric under permutation of the variables
 $t$ and $u$, and is in addition subject to the condition
 \be
 {\cal N}_3 (s,t,u) + {\cal N}_3 (t,u,s) + {\cal N}_3 (u,s,t) = 0
 .
 \ee
\end{enumerate}
Notice that some of these symmetry properties reflect the fact that the amplitudes with $I=0$ or $I=2$ ($I=1$) will
only produce partial-wave projections with even (odd) values of the total angular momentum, as
required by Bose symmetry. 

Coming now back to the crossing matrices introduced in eq. \rf{crossing_isospin-amplitudes}, one notices,
for instance, that eq. \rf{inv_amps_1} tells us that ${\cal M}^{2;\frac{3}{2}}_1  (t,s,u)$
is equal to ${\cal M}_1 (s \vert t,u) + {\cal M}_1 (u \vert s,t)$. The
two amplitudes in this sum can in turn  be expressed in terms of the isospin amplitudes 
${\cal M}_1^{I ; {\cal I}} (s , t , u)$ upon inverting Eq. \rf{inv_amps_1}, giving
\be
{\cal M}^{2;\frac{3}{2}}_1  (t,s,u) = \frac{1}{6} {\cal M}^{2;\frac{3}{2}}_1  (s,t,u) 
- \frac{1}{6} {\cal M}^{1;\frac{3}{2}}_1  (s,t,u) - \frac{1}{3} {\cal M}^{1;\frac{1}{2}}_1  (s,t,u) + \frac{1}{3} {\cal M}^{0;\frac{1}{2}}_1  (s,t,u) .
\ee
Working out the other cases along similar lines, one easily reproduces the crossing matrices 
for the isospin amplitudes given in eqs. \rf{crossing_1} and \rf{crossing_3} above. This 
actually proves that the decompositions \rf{inv_amps_1} and \rf{inv_amps_3} are indeed 
compatible with the crossing properties of the amplitudes ${\vec{\cal A}} (s,t,u)$.

\indent

\section{Representation of the amplitudes in terms of sin\-gle-variable functions}\label{sec:single_var}
\setcounter{equation}{0}

\subsection{Decomposition of the amplitudes into single-variable functions}

On general grounds, we expect that the amplitudes ${\cal A}_i (s,t,u)$, or the 
isospin amplitudes ${\cal M}_{1,3}^{I ; {\cal I}} (s,t,u)$ discussed in the previous 
section, satisfy fixed-$t$ dispersion relations.
In order to set up a system of dispersion relations of the Khuri-Treiman type for, say,
the isospin amplitudes, we need to express them 
in terms of a set of functions of a single variable. Such a representation will hold in 
the energy region where the absorptive parts of the partial waves with total angular momentum 
equal to or higher than two can be neglected in the dispersive integrals. Requiring that 
the resulting amplitudes also satisfy the appropriate crossing relations then leads to the 
expressions (numerical factors have been chosen for convenience)
\be
{\vec{\cal M}}_1 (s,t,u) =
\left(
\begin{tabular}{c}
$ 2 M_{1}^{2;\frac{3}{2}} (s) $
\\
$ 2 (t-u) M_{1}^{1;\frac{3}{2}} (s) $
\\
$ 2 (t-u) M_{1}^{1;\frac{1}{2}} (s) $
\\
$ 3 M_{1}^{0;\frac{1}{2}} (s) $
\end{tabular}
\right)
+ C_{st}^{(1)}
\left(
\begin{tabular}{c}
$ 2 M_{1}^{2;\frac{3}{2}} (t) $
\\
$ 2 (s-u) M_{1}^{1;\frac{3}{2}} (t) $
\\
$ 2 (s-u) M_{1}^{1;\frac{1}{2}} (t) $
\\
$ 3 M_{1}^{0;\frac{1}{2}} (t) $
\end{tabular}
\right)
+ C_{tu}^{(1)} C_{st}^{(1)}
\left(
\begin{tabular}{c}
$ 2 M_{1}^{2;\frac{3}{2}} (u) $
\\
$ 2 (s-t) M_{1}^{1;\frac{3}{2}} (u) $
\\
$ 2 (s-t) M_{1}^{1;\frac{1}{2}} (u) $
\\
$ 3 M_{1}^{0;\frac{1}{2}} (u) $
\end{tabular}
\right)
\lbl{one_var_1}
\ee
and
\be
{\vec{\cal M}}_3 (s,t,u) =
\left(
\begin{tabular}{c}
$ 2 M_{3}^{2;\frac{3}{2}} (s) $
\\
$ 2 M_{3}^{2;\frac{1}{2}} (s) $
\\
$ 2 (t-u) M_{3}^{1;\frac{3}{2}} (s) $
\\
$ 2 (t-u) M_{3}^{1;\frac{1}{2}} (s) $
\\
$ 3 M_{3}^{0;\frac{3}{2}} (s) $
\end{tabular}
\right)
+ C_{st}^{(3)}
\left(
\begin{tabular}{c}
$ 2 M_{3}^{2;\frac{3}{2}} (t) $
\\
$ 2 M_{3}^{2;\frac{1}{2}} (t) $
\\
$ 2 (s-u) M_{3}^{1;\frac{3}{2}} (t) $
\\
$ 2 (s-u) M_{3}^{2;\frac{1}{2}} (t) $
\\
$ 3 M_{3}^{0;\frac{3}{2}} (t) $
\end{tabular}
\right)
+ C_{tu}^{(3)} C_{st}^{(3)}
\left(
\begin{tabular}{c}
$ 2 M_{3}^{2;\frac{3}{2}} (u) $
\\
$ 2 M_{3}^{2;\frac{1}{2}} (u) $
\\
$ 2 (s-t) M_{3}^{1;\frac{3}{2}} (u) $
\\
$ 2 (s-t) M_{3}^{1;\frac{1}{2}} (u) $
\\
$ 3 M_{3}^{0;\frac{3}{2}} (u) $
\end{tabular}
\right)
\lbl{one_var_3}
\ee
The derivation of these representations, which has by now become standard and is amply documented in the existing literature,%
\footnote{
Although the construction was first shown to hold in the case of the $\pi\pi$ scattering amplitude to
two loops in the low-energy expansion \cite{Stern:1993rg}, it was subsequently applied to the low-energy 
expansion of other meson-meson scattering amplitudes \cite{Zdrahal:2008bd}, and even beyond this framework, under
the assumption that the absorptive parts of the D and higher partial waves can be neglected, see the references 
quoted in the first paragraph of the introduction. A discussion more specifically devoted to the $K\to\pi\pi\pi$ amplitudes
in the presence of isospin breaking has been given in refs. \cite{Kampf:2008ts,Zdrahal:2009ns,Kampf:2019bkf}
}
also shows that the functions of a single variable involved are analytic 
except for a right-hand cut starting at the two-pion threshold, with the discontinuity
given by unitarity. Before constructing the corresponding dispersion relations, we need to give consideration 
to two issues. The first one concerns the non-unicity of the decomposition into single-variable 
functions. The second one concerns the asymptotic conditions that the amplitudes are required to satisfy.
The remaining part of the present section will successively address these two aspects. We also briefly 
discuss the connection of the single-variable representations discussed here with other decompositions
used in the literature.

\subsection{Polynomial ambiguities}

The decompositions of the amplitudes ${\cal A}_i (s,t,u)$ in terms of the single-variable functions $M_{1,3}^{I ; {\cal I}}$,
are unique only modulo some polynomials of finite degrees in $s$. In other words, alternative decompositions, in 
terms of functions ${\widetilde M}_{1,3}^{I;{\cal I}}(s)$, are possible, where the differences 
$\delta M_{1,3}^{I ; {\cal I}} (s) \equiv {\widetilde M}_{1,3}^{I;{\cal I}} (s) -  M_{1,3}^{I;{\cal I}} (s)$
are polynomials. In order to determine the most general form of the polynomials $\delta M_{1,3}^{I;{\cal I}} (s)$
one may easily adapt the method used in Ref. \cite{Colangelo:2018jxw} to the case at hand.
It is convenient, and equivalent, to discuss this issue at the level of the invariant functions ${\cal M}_{1,3} (s \vert t,u)$
and ${\cal N}_{1,3} (s,t,u)$. Inverting the relations \rf{inv_amps_1} and \rf{inv_amps_3} and replacing the 
isospin amplitudes by their expressions \rf{one_var_1} and \rf{one_var_3}, one obtains\footnote{
It is straightforward to verify that these representations fulfill the requirements listed after eq. \rf{inv_amps_3}.}
\bea
{\cal M}_1 (s \vert t , u) &=& M_{1}^{0;\frac{1}{2}}(s) - \frac{2}{3} M_{1}^{2;\frac{3}{2}} (s)
+ M_{1}^{2;\frac{3}{2}} (t) + M_{1}^{2;\frac{3}{2}} (u)
\nonumber\\
&&\!\!\!\!\!
+ \, \frac{1}{3} (s-u) \left[ M_{1}^{1;\frac{3}{2}} (t) + 2 M_{1}^{1;\frac{1}{2}} (t) \right]
+ \frac{1}{3} (s-t) \left[ M_{1}^{1;\frac{3}{2}} (u) + 2 M_{1}^{1;\frac{1}{2}} (u) \right]
,
\nonumber\\[0.1cm]
{\cal N}_1 (s ,t ,u) &=& \frac{2}{3} (t-u) \left[ M_{1}^{1;\frac{1}{2}} (s) - M_{1}^{1;\frac{3}{2}} (s) \right]
+ \frac{2}{3} (u-s) \left[ M_{1}^{1;\frac{1}{2}} (t) - M_{1}^{1;\frac{3}{2}} (t) \right] 
\nonumber\\
&&\!\!\!\!\!
+ \, \frac{2}{3} (s-t) \left[ M_{1}^{1;\frac{1}{2}} (u)  -M_{1}^{1;\frac{3}{2}} (u) \right]
,
\nonumber\\
\lbl{iso_single}
\\
{\cal M}_3 (s \vert t , u) &=& M_{3}^{0;\frac{3}{2}} (s) + \frac{2}{3} \left[ M_{3}^{2;\frac{3}{2}} (s) - 2 M_{3}^{2;\frac{1}{2}} (s) \right]
+ M_{3}^{2;\frac{1}{2}} (t) + M_{3}^{2;\frac{1}{2}} (u)
\nonumber\\
&&\!\!\!\!\!
+ \, \frac{1}{3} (s-u) \left[ 4 M_{3}^{1;\frac{3}{2}} (t) -  M_{3}^{1;\frac{1}{2}} (t) \right]
+ \frac{1}{3} (s-t) \left[ 4 M_{3}^{1;\frac{3}{2}} (u) - M_{3}^{1;\frac{1}{2}} (u) \right]
,
\nonumber\\[0.1cm]
{\cal N}_3 (s ,t ,u) &=& \frac{2}{3} (t-u) \left[ M_{3}^{1;\frac{3}{2}} (s) - M_{3}^{1;\frac{1}{2}} (s) \right]
+ M_{3}^{2;\frac{3}{2}} (t) - M_{3}^{2;\frac{1}{2}} (t) - M_{3}^{2;\frac{3}{2}} (u) + M_{3}^{2;\frac{1}{2}} (u)
\nonumber\\
&&\!\!\!\!\!
+ \, \frac{1}{3} (s-u) \left[ M_{3}^{1;\frac{3}{2}} (t) - M_{3}^{1;\frac{1}{2}} (t) \right] 
- \frac{1}{3} (s-t) \left[ M_{3}^{1;\frac{3}{2}} (u) - M_{3}^{1;\frac{1}{2}} (u) \right] 
.
\nonumber
\eea 
One then finds (for the derivation, we refer the interested reader to Appendix \ref{app:polynomial}) 
that these four functions are not modified under the following polynomial transformations of the single-variable functions
\bea
\delta M_1^{2 ; \frac{3}{2}} (s) &=& \frac{b_1}{3} s^3 - (c_1 + 3 s_0 b_1) s^2 + g_1 s + h_1
,
\nonumber\\
\delta M_1^{1 ; \frac{3}{2}} (s) &=& 2 a_1 s^3 + (b_1 - 6 s_0 a_1) s^2 + ( c_1 + 2 d_1) s + ( e_1 + 2 f_1)
,
\nonumber\\
\delta M_1^{1 ; \frac{1}{2}} (s) &=& - a_1 s^3 + (b_1 + 3 s_0 a_1) s^2 + ( c_1 - d_1) s + ( e_1 - f_1)
.
\nonumber
\\
\delta M_1^{0 ; \frac{1}{2}} (s) &=& -\frac{4}{9} b_1 s^3 + \frac{4}{3} (c_1 + 3 b_1 s_0) s^2
- \frac{4}{3} g_1 s + 3 (s-s_0) (g_1 - e_1 - 3 c_1 s_0 - 6 b_1 s_0^2) 
- \frac{4}{3} h_1 
,
\lbl{transf_1}
\eea
\bea
\delta M_3^{2 ; \frac{3}{2}} (s) \!\!\!&=&\!\!\! \frac{1}{9} (a_3 + 2 c_3) s^3 - \frac{1}{3} (b_3 + 2 d_3 + 3 a_3 s_0 + 6 c_3 s_0) s^2 + g_3 s + h_3
,
\nonumber\\[0.1cm]
\delta M_3^{2 ; \frac{1}{2}} (s) \!\!\!&=&\!\!\! \frac{1}{9} (4 a_3 - c_3) s^3 - \frac{1}{3} (4b_3 - d_3 + 12 a_3 s_0 - 3 c_3 s_0) s^2
+ \, (e_3 - f_3 + g_3 + 3 b_3 s_0 - 3 d_3 s_0 + 6 a_3 s_0^2 - 6 c_3 s_0^2) s + k_3, ~~~~~~
\nonumber\\[0.1cm]
\delta M_3^{1 ; \frac{3}{2}} (s) \!\!&=&\!\! a_3 s^2 + b_3 s + e_3 
,
\nonumber\\[0.1cm]
\delta M_3^{1 ; \frac{1}{2}} (s) \!\!&=&\!\! c_3 s^2 + d_3 s + f_3
,
\nonumber\\[0.1cm]
\delta M_3^{0 ; \frac{3}{2}} (s) \!\!\!&=&\!\!\! - \, \frac{2}{27} (5 a_3 + c_3) s^3 + \frac{2}{9} (5 b_3 + d_3 + 15 a_3 s_0 + 3 c_3 s_0 ) s^2
- \left( \frac{5}{3} e_3 + \frac{4}{3} f_3 - \frac{5}{3} g_3 + 5 b_3 s_0 + 4 d_3 s_0 + 10 a_3 s_0^2 + 8 c_3 s_0^2 \right) \!s 
\nonumber\\
&&\!\! 
+ \, 6 (a_3 + 2 c_3) s_0^3 + 3 (b_3 + 2 d_3 ) s_0^2 + (e_3 + 2 f_3 - 3 g_3 ) s_0 - \frac{2}{3} (h_3 + k_3)
.
\lbl{transf_3}
\eea
These polynomial ambiguities involve a much larger set of free parameters 
than in the case of the $\eta\to\pi\pi\pi$ amplitudes: instead of the five parameters
\cite{Colangelo:2018jxw} one finds in the case of these  $\Delta I = 1$ transitions, there are now eight such
parameters ($a_1$, $b_1,\ldots$ $h_1$) in the case of the $\Delta I = 1/2$
transitions, and nine ($a_3$, $b_3,\ldots$ $h_3$, $k_3$) in the case of 
the $\Delta I = 3/2$ transitions. In the next subsection, we will use these transformations
in order to put constraints on the single-variable functions and on their dispersive 
representations, once we have also specified the asymptotic conditions they are required to satisfy.

\subsection{Asymptotic behaviour and dispersion relations}

In the complex-$s$ plane, the single-variable functions $M_{1,3}^{I ; {\cal I}} (s)$ are analytic
except for a cut that runs along the positive real-$s$ axis, for ${\rm Re}\,s \ge 4 M_\pi^2$. They
thus obey dispersion relations of the generic form
\be
M(s) = P(s) + \frac{s^n}{\pi} \int_{4 M_\pi^2}^{+\infty} \frac{dx}{x^n} \, \frac{{\rm Abs} \, M(x)}{x-s}
\lbl{disp-rel_M}
,
\ee
where the order $n-1$ of the subtraction polynomial $P(s)$ depends on the function
under consideration and requires some knowledge of the asymptotic behaviour of the 
amplitudes. We thus discuss the two issues together. Two possibilities have been considered
in the recent literature. The first one assumes that all amplitudes grow at most linearly in $s$ and 
$\tau\equiv t-u$ in the limit where these two variables become simultaneously large. As discussed 
in ref. \cite{Anisovich:1996tx}, this is the situation expected from Regge phenomenology and from 
the assumption that the asymptotic behaviour of the $K\pi\to\pi\pi$ amplitudes is the same as 
for the amplitudes of elastic $\pi\pi$ scattering. This linear asymptotic behaviour was
adopted in the treatment of the $\eta\to\pi\pi\pi$ amplitude in refs. 
\cite{Anisovich:1996tx,Descotes-Genon:2014tla,Albaladejo:2017hhj}. A different, less restrictive, 
asymptotic condition, where the amplitudes are assumed to grow at most quadratically with $s$ and
$\tau$, was adopted in Ref. \cite{Colangelo:2018jxw} for the $\eta\to\pi\pi\pi$ amplitude.
Here we will consider the first option and assume a linear asymptotic behaviour for all 
amplitudes. This leads to a set of integral equations with a minimal number of
subtraction constants and that is well suited for an iterative numerical solution. 
For completeness, we briefly discuss in Appendix \ref{app:quadratic} how the dispersive 
representations would change upon considering instead a quadratic asymptotic behaviour,
but without entering their full numerical treatment.

Since we assume a linear condition for the asymptotic growth of the amplitudes,
\be
\lim_{\lambda\to + \infty} {\cal M}_{1,3} (\lambda s , \lambda\tau) = {\cal O} (\lambda)
, \ \ \lim_{\lambda\to + \infty} {\cal N}_{1,3} (\lambda s , \lambda\tau) = {\cal O} (\lambda)
,
\lbl{asympt_1}
\ee
the asymptotic behaviours of the single-variable functions are restricted to
\bea
M_1^{2 ; \frac{3}{2}} (s) \!&\asymp &\! \frac{B_1}{3} s^3 + D_1 s^2 + E_1 s + \cdots  ,
\nonumber\\
M_1^{1 ; \frac{3}{2}} (s) \!&\asymp &\! 2 A_1 s^3 + (B_1 - 6 s_0 A_1) s^2 + C_1 s + \cdots   ,
\nonumber\\
M_1^{1 ; \frac{1}{2}} (s) \!&\asymp &\! - A_1 s^3 + (B_1 + 3 s_0 A_1) s^2 - \frac{1}{2} (C_1 + 3 D_1 + 9 s_0 B_1) s + \cdots   ,
\nonumber\\
M_1^{0 ; \frac{1}{2}} (s) \!&\asymp &\! - \frac{4B_1}{9} s^3 - \frac{4D_1}{3} s^2 + H_1 s + \cdots  
,
\lbl{asympt_M1_lin}
\eea
for the $\Delta I = 1/2$ amplitudes, and, for the $\Delta I = 3/2$ amplitudes, to
\bea 
M_3^{2 ; \frac{3}{2}} (s) \!&\asymp &\! \frac{1}{9} (A_3 + 2 C_3) s^3 - \frac{1}{3} (B_3 + 2 D_3 + 3 A_3 s_0 + 6 C_3 s_0 ) s^2 + \cdots  ,
\nonumber\\
M_3^{2 ; \frac{1}{2}} (s) \!&\asymp &\! \frac{1}{9} (4A_3 - C_3) s^3 - \frac{1}{3} ( 4 B_3 - D_3 + 12 A_3 s_0 - 3 C_3 s_0) s^2 + \cdots  ,
\nonumber\\
M_3^{1 ; \frac{3}{2}} (s) \!&\asymp &\! A_3 s^2 + B_3 s + \cdots   , 
\nonumber\\
M_3^{1 ; \frac{1}{2}} (s) \!&\asymp &\! C_3 s^2 + D_3 s + \cdots   ,
\nonumber\\
M_3^{0 ; \frac{3}{2}} (s) \!&\asymp &\! - \frac{2}{27} (5 A_3 + C_3) s^3  + \frac{2}{9} (5 B_3 + D_3 + 15 A_3 s_0 + 3 C_3 s_0) s^2 + \cdots  
,
\lbl{asympt_M3_lin}
\eea
for some unknown constants $A_1$, $B_1$, $\ldots$. The ellipses stand for terms that grow less rapidly
than the terms displayed explicitly.
We see that generically the leading asymptotic behaviour of the individual isospin functions is 
cubic in $s$, while it is only quadratic for the $\Delta I = 3/2$ transitions to a pair of final pions
in an isospin $I=1$ state. Altogether, these asymptotic conditions would require a rather large number 
of subtraction constants in order to obtain convergent dispersion relations for the isospin functions.
We can substantially improve this situation by making use of the freedom to suitably redefine these functions by
polynomials, in a manner that was worked out in eqs. \rf{transf_1} and \rf{transf_3}, and thus tame the 
asymptotic behaviour of the single-variable functions. Let us first discuss the amplitudes describing the 
$\Delta I = 1/2$ sector. Making the polynomial shifts with the choices
\be
a_1 = - A_1 ,~ b_1 = -B_1,~ c_1 = D_1 + 3 s_0 B_1 ,~d_1 = - \frac{1}{2} (C_1 + D_1 + 3 s_0 B_1) ,
\ee
one may achieve that the various single-variable functions actually behave asymptotically as
\be
M_1^{0 ; \frac{1}{2}} (s) \asymp {\cal O}(s) , \ M_1^{1 ; \frac{3}{2}} (s) \asymp {\rm constant} ,
\ M_1^{1 ; \frac{1}{2}} (s) \asymp {\rm constant} , \ \ M_1^{2 ; \frac{3}{2}} (s) \asymp {\cal O}(s) .
\lbl{M1_asympt_lin}
\ee
The remaining four parameters of the transformations \rf{transf_1} can then be used in order to impose
conditions on their behaviours at the origin, i.e.
\be
M_1^{1 ; \frac{3}{2}} (0) = M_1^{1 ; \frac{1}{2}} (0) = 0 ,
\ M_1^{2 ; \frac{3}{2} \prime} (0) = M_1^{2 ; \frac{3}{2}} (0) = 0   ,
\lbl{M1_origin_lin}
\ee
without altering the asymptotic behaviour \rf{M1_asympt_lin}.
The dispersion relations that follow are
\bea
M_1^{2 ; \frac{3}{2}} (s) &=&  \frac{s^{2}}{\pi}
\int_{4 M_\pi^2}^\infty \frac{dx}{x^2} \, \frac{{\rm Abs}\,M_1^{2 ; \frac{3}{2}} (x)}{x - s}
, 
\nonumber\\
M_1^{1 ; \frac{3}{2}} (s) &=& \frac{s}{\pi}
\int_{4 M_\pi^2}^\infty \frac{dx}{x} \, \frac{{\rm Abs}\,M_1^{1 ; \frac{3}{2}} (x)}{x - s}
, 
\nonumber\\
M_1^{1 ; \frac{1}{2}} (s) &=& \frac{s}{\pi}
\int_{4 M_\pi^2}^\infty \frac{dx}{x} \, \frac{{\rm Abs}\,M_1^{1 ; \frac{1}{2}} (x)}{x - s}
, 
\nonumber\\
M_1^{0 ; \frac{1}{2}} (s) &=& {\tilde\alpha}_{1} + {\tilde\beta}_{1} s +  
\frac{s^{2}}{\pi} \int_{4 M_\pi^2}^\infty \frac{dx}{x^2} \, \frac{{\rm Abs}\,M_1^{0 ; \frac{1}{2}} (x)}{x - s}
.
\lbl{M1_disp-rel_lin}
\eea
With only two subtraction constants ${\tilde\alpha}_{1}$ and ${\tilde\beta}_{1}$,
this choice of the parameters of the polynomial shifts thus leads to a situation similar 
to the one found for the $\eta\to\pi\pi\pi$ amplitude when its
asymptotic growth is assumed to be linear. We thus expect that the conditions that has been
imposed on the single-variable functions will likewise lead to a stable and rapidly converging
numerical treatment of the dispersion relations. Without wishing to enter a complete and exhaustive 
discussion, let us simply mention here that other choices could in principle be considered. 
These alternative possibilities would however 
require that not all of the resulting subtraction constants remain free, but that some of them are
fixed by dispersion sum rules. As a matter of principle, there is nothing to complain about such a 
situation, but it would make the iterative numerical treatment of the final dispersion relations
more cumbersome and more delicate, since these subtraction constants would need to be adjusted
anew at each iteration.

Turning next to the amplitudes describing the $\Delta I = 3/2$ sector, by making the polynomial shifts with the choices
\be
a_3 = - A_3 ,~ b_3 = - B_3 ,~ c_3 = - C_3 ,~ d_3 = - D_3,
\ee
one may achieve that the various single-variable functions behave asymptotically as
\be
M_3^{1 ; \frac{3}{2}} (s) \asymp {\rm constant},
\ \ M_3^{1 ; \frac{1}{2}} (s) \asymp {\rm constant} ,
\ \ M_3^{2 ; \frac{3}{2}} (s) \asymp {\cal O}(s) ,
\ \ M_3^{2 ; \frac{1}{2}} (s) \asymp {\cal O}(s) ,
\ \ M_3^{0 ; \frac{3}{2}} (s) \asymp {\cal O}(s)
.
\lbl{M3_asympt_lin}
\ee
Of the five remaining parameters  of the transformations \rf{transf_3}, 
$g_3$, $h_3$, and $k_3$ can be used in order to impose
\be
M_3^{2 ; \frac{3}{2}} (0) = M_3^{2 ; \frac{3}{2}\prime} (0) = 0 ,
\ \ M_3^{2 ; \frac{1}{2}} (0) = 0   .
\lbl{M3_origin_lin}
\ee
As for the remaining two parameters $e_3$ and $f_3$, one may observe that they occur as
\be
\delta M_3^{1 ; \frac{3}{2}} (0) = e_3 , \qquad \delta M_3^{1 ; \frac{1}{2}} (0) = f_3 , \qquad 
\delta M_3^{2 ; \frac{1}{2}\prime} (0) = e_3 - f_3 + X_3 ,
\ee
where $X_3$ is a quantity already fixed by the conditions set on the other functions. It is thus possible, 
in general, to choose $M_3^{2 ; \frac{1}{2}\prime} (0) = 0$ and to make a specific linear combination of 
$ M_3^{1 ; \frac{3}{2}} (0)$ and of $ M_3^{1 ; \frac{1}{2}} (0)$ vanish, with the exception of 
the combination $ M_3^{1 ; \frac{3}{2}} (0) -  M_3^{1 ; \frac{1}{2}} (0)$. As will become
clear in section \ref{sec:integral_eqs}, it is convenient to take this combination as $ 5 M_3^{1 ; \frac{3}{2}} (0) +  M_3^{1 ; \frac{1}{2}} (0)$.
This leads to the following dispersion relations for the functions 
describing the amplitudes of the $\Delta I = 3/2$ transitions:
\bea
M_3^{2 ; \frac{3}{2}} (s) &=& 
\frac{s^{2}}{\pi}
\int_{4 M_\pi^2}^\infty \frac{dx}{x^2} \, \frac{{\rm Abs}\,M_3^{2 ; \frac{3}{2}} (x)}{x - s}
, 
\nonumber\\
M_3^{2 ; \frac{1}{2}} (s) &=& \frac{s^{2}}{\pi}
\int_{4 M_\pi^2}^\infty \frac{dx}{x^2} \, \frac{{\rm Abs}\,M_3^{2 ; \frac{3}{2}} (x)}{x - s}
, 
\nonumber\\
M_3^{1 ; \frac{3}{2}} (s) &=& {\tilde\gamma}_{3} + \frac{s}{\pi}
\int_{4 M_\pi^2}^\infty \frac{dx}{x} \, \frac{{\rm Abs}\,M_3^{1 ; \frac{3}{2}} (x)}{x - s}
, 
\nonumber\\
M_3^{1 ; \frac{1}{2}} (s) &=& - 5 {\tilde\gamma}_{3} + \frac{s}{\pi}
\int_{4 M_\pi^2}^\infty \frac{dx}{x} \, \frac{{\rm Abs}\,M_3^{1 ; \frac{1}{2}} (x)}{x - s}
, 
\nonumber\\
M_3^{0 ; \frac{3}{2}} (s) &=& {\tilde\alpha}_{3} + {\tilde\beta}_{3} s 
+ \frac{s^{2}}{\pi}
\int_{4 M_\pi^2}^\infty \frac{dx}{x^2} \, \frac{{\rm Abs}\,M_3^{0 ; \frac{3}{2}} (x)}{x - s}
.
\eea
These dispersion relations involve only three subtraction constants: ${\tilde\alpha}_{3}$,  ${\tilde\beta}_{3}$, and ${\tilde\gamma}_{3}$.

\subsection{A few comments on the existing literature}

The isospin decomposition of the $K\to\pi\pi\pi$ amplitudes has been discussed several times 
in the literature, see for instance \cite{Dalitz:1956da,Weinberg:1960zza,Barton:1963mg,Zemach:1963bc,DAmbrosio:1994vba,Bijnens:2002vr}.
These previous studies were focusing on the three-pion decays of the kaons, so that the decomposition
was made according to the total isospin of the three pions. Since the Khuri-Treiman equations are 
conveniently derived in terms of the scattering amplitudes $K\pi\to\pi\pi$, the decomposition
in terms of the isospin carried by the two pions in the final state is more relevant. But the conclusion 
that only four independent invariant amplitudes are required to describe all the possible 
channels once crossing properties are enforced holds, of course, in both cases.

The decomposition of the amplitudes into single-variable functions and the associated
ambiguities were also discussed in ref. \cite{Bijnens:2002vr}. 
The authors of ref. \cite{Bijnens:2002vr} start from a redundant set of thirteen single-variable functions
among which they establish four linear relations. This eventually leads also to only nine independent single-variable 
functions. Our set of single-variable functions has the advantage that the 
contributions from $\Delta I = 1/2$ and $\Delta I = 3/2$ transitions are well 
identified and kept separated, whereas the choice made by the authors of ref. \cite{Bijnens:2002vr} 
leads to functions that receive contributions from both sectors. As far as the polynomial 
ambiguities in the definition of the single-variable functions is concerned, the authors 
of ref. \cite{Bijnens:2002vr} find that they can be described by 23 parameters. But this
holds for the redundant set of functions. The four relations among them also induce 
relations among these parameters, so that only 17 parameters are eventually 
found to be independent, in agreement with our findings. The interested reader will find the \textit{in-extenso} relations between 
the single-variable functions introduced in the present work and those of ref. \cite{Bijnens:2002vr} 
in appendix \ref{app:comp_BDP}.

\indent

\section{Partial-waves and absorptive parts from elastic unitarity}\label{sec:pw_unitarity}
\setcounter{equation}{0}

The dispersion relations that have been established for the single-variable functions 
will become useful only provided some knowledge of the absorptive parts is available. We next determine the
form of these absorptive parts in the regime where elastic unitarity holds.
For this, we first introduce the partial-wave projections of the isospin amplitudes ${\cal M}^{I;{\cal I}}_{1,3}$.
They are defined as
\be
16 \pi f^{I;{\cal I}}_{1,3 ; l} (s) = \frac{1}{2} \int _{-1}^{+1} dz P_l (z) {\cal M}^{I;{\cal I}}_{1,3} (s , t(s , z) , u(s , z) )
,
\lbl{PW-def}
\ee
where the variables $t$ and $u$ are related to $s$ and to $z\equiv \cos \theta$, $\theta$ denoting the 
scattering angle in the centre-of-mass frame, by
\be  
t , u = \frac{1}{2} \left( 3 M_\pi^2 + M_K^2 - s \pm \kappa (s) z \right)
\lbl{t_u_in_terms_of_s_z}
\ee
with 
\be
\kappa^2 (s) = \left( 1 - \frac{4 M_\pi^2}{s} \right) \left[ s - (M_K - M_\pi)^2 \right]  \left[ s - (M_K + M_\pi)^2 \right]
.
\ee
Performing the projection on the S and P waves of the isospin amplitudes expressed
in terms of the single-variable functions in Eqs. \rf{one_var_1} and \rf{one_var_3}, one finds
\bea
16 \pi f_{1,3;0}^{2;\frac{3}{2}} (s) &=&
2 \big[ M_{1,3}^{2;\frac{3}{2}} (s) + {\hat M}_{1,3}^{2;\frac{3}{2}} (s) \big],
\hspace{2.0cm}
16 \pi f_{3;0}^{2;\frac{1}{2}} (s) ~=~
2 \big[ M_{3}^{2;\frac{1}{2}} (s) + {\hat M}_{3}^{2;\frac{1}{2}} (s) \big],
\nonumber\\
16 \pi f_{1,3;1}^{1;\frac{3}{2}} (s) &=&
\frac{2}{3} \kappa (s) \big[ M_{1,3}^{1;\frac{3}{2}} (s) + {\hat M}_{1,3}^{1;\frac{3}{2}} (s) \big],
\nonumber\\
16 \pi f_{1,3;1}^{1;\frac{1}{2}} (s) &=&
\frac{2}{3} \kappa (s) \big[ M_{1,3}^{1;\frac{1}{2}} (s) + {\hat M}_{1,3}^{1;\frac{1}{2}} (s) \big],
\nonumber\\
16 \pi f_{1;0}^{0;\frac{1}{2}} (s) &=&
3 \big[ M_{1}^{0;\frac{1}{2}} (s) + {\hat M}_{1}^{0;\frac{1}{2}} (s) \big],
\hspace{2.cm}
16 \pi f_{3;0}^{0;\frac{3}{2}} (s) ~=~
3 \big[ M_{3}^{0;\frac{3}{2}} (s) + {\hat M}_{3}^{0;\frac{3}{2}} (s) \big].
\eea
where
\bea
{\hat M}_{1}^{2;\frac{3}{2}}  &=& \langle M_{1}^{0;\frac{1}{2}} \rangle  + \frac{1}{3} \langle M_{1}^{2;\frac{3}{2}} \rangle
- \frac{1}{2} (s - s_0) \langle M_{1}^{1;\frac{3}{2}} + 2 M_{1}^{1;\frac{1}{2}} \rangle 
- \frac{1}{6} \kappa \langle z ( M_{1}^{1;\frac{3}{2}} + 2 M_{1}^{1;\frac{1}{2}} ) \rangle
,
\nonumber\\
{\hat M}_{1}^{1;\frac{3}{2}}  &=& \frac{1}{\kappa} \left[ 3 \langle z M_{1}^{0;\frac{1}{2}} \rangle - 5 \langle z M_{1}^{2;\frac{3}{2}} \rangle
- \frac{9}{2} (s - s_0) \langle z ( M_{1}^{1;\frac{3}{2}} - 2 M_{1}^{1;\frac{1}{2}} ) \rangle 
- \frac{3}{2} \kappa \langle z^2 ( M_{1}^{1;\frac{3}{2}} - 2 M_{1}^{1;\frac{1}{2}} ) \rangle
\right]
,
\nonumber\\
{\hat M}_{1}^{1;\frac{1}{2}}  &=& \frac{1}{\kappa} \left[ 3 \langle z M_{1}^{0;\frac{1}{2}} \rangle - 5 \langle z M_{1}^{2;\frac{3}{2}} \rangle
+ \frac{9}{2} (s - s_0) \langle z M_{1}^{1;\frac{3}{2}} \rangle 
+ \frac{3}{2} \kappa \langle z^2 M_{1}^{1;\frac{3}{2}} \rangle
\right]
,
\nonumber\\
{\hat M}_{1}^{0;\frac{1}{2}}  &=& \frac{2}{3} \langle M_{1}^{;\frac{3}{2}} \rangle  + \frac{20}{9} \langle M_{1}^{2;\frac{3}{2}} \rangle
+ \frac{2}{3} (s - s_0) \langle M_{1}^{1;\frac{3}{2}} + 2 M_{1}^{1;\frac{1}{2}} \rangle 
+ \frac{2}{9} \kappa \langle z ( M_{1}^{1;\frac{3}{2}} + 2 M_{1}^{1;\frac{1}{2}} ) \rangle
,
\lbl{hatM_1}
\eea
and
\bea
{\hat M}_{3}^{2;\frac{3}{2}}  &=& \langle M_{3}^{0;\frac{3}{2}} \rangle  - \frac{1}{3} \langle M_{3}^{2;\frac{3}{2}} \rangle 
+ \frac{2}{3} \langle M_{3}^{2;\frac{1}{2}} \rangle
- \frac{1}{2} (s - s_0) \langle M_{3}^{1;\frac{3}{2}} + 2 M_{3}^{1;\frac{1}{2}} \rangle 
- \frac{1}{6} \kappa \langle z ( M_{3}^{1;\frac{3}{2}} + 2 M_{3}^{1;\frac{1}{2}} ) \rangle
,
\nonumber\\
{\hat M}_{3}^{2;\frac{1}{2}}  &=& \langle M_{3}^{0;\frac{3}{2}} \rangle  + \frac{2}{3} \langle M_{3}^{2;\frac{3}{2}} \rangle 
- \frac{1}{3} \langle M_{3}^{2;\frac{1}{2}} \rangle
- \frac{1}{2} (s - s_0) \langle 4 M_{3}^{1;\frac{3}{2}} - M_{3}^{1;\frac{1}{2}} \rangle 
- \frac{1}{6} \kappa \langle z ( 4 M_{3}^{1;\frac{3}{2}} - M_{3}^{1;\frac{1}{2}} ) \rangle
,
\nonumber\\
{\hat M}_{3}^{1;\frac{3}{2}}  &=& \frac{1}{\kappa} \left[ 3 \langle z M_{3}^{0;\frac{3}{2}} \rangle - \langle z M_{3}^{2;\frac{3}{2}} \rangle 
- 4 \langle z M_{3}^{2;\frac{1}{2}} \rangle
+ \frac{9}{2} (s - s_0) \langle z M_{3}^{1;\frac{3}{2}} \rangle 
+ \frac{3}{2} \kappa \langle z^2 M_{3}^{1;\frac{3}{2}} \rangle
\right]
,
\nonumber\\
{\hat M}_{3}^{1;\frac{1}{2}}  &=& \frac{1}{\kappa} \left[ 3 \langle z M_{3}^{0;\frac{3}{2}} \rangle - 10 \langle z M_{3}^{2;\frac{3}{2}} \rangle 
+ 5 \langle z M_{3}^{2;\frac{1}{2}} \rangle
+ \frac{9}{2} (s - s_0) \langle z M_{3}^{1;\frac{1}{2}} \rangle 
+ \frac{3}{2} \kappa \langle z^2 M_{3}^{1;\frac{1}{2}} \rangle
\right]
,
\nonumber\\
{\hat M}_{3}^{0;\frac{3}{2}}  &=& \frac{2}{3} \langle M_{3}^{0;\frac{3}{2}} \rangle  + \frac{10}{9} \langle M_{3}^{2;\frac{3}{2}} \rangle  
+ \frac{10}{9} \langle M_{3}^{2;\frac{1}{2}} \rangle
+ \frac{1}{3} (s - s_0) \langle 5 M_{3}^{1;\frac{3}{2}} + M_{3}^{1;\frac{1}{2}} \rangle 
+ \frac{1}{9} \kappa \langle z ( 5 M_{3}^{1;\frac{3}{2}} + M_{3}^{1;\frac{1}{2}} ) \rangle . ~~~~~~~~~~
\lbl{hatM_3}
\eea
In these expressions, we have used the short-hand notation
\be
\langle z^n {\cal W} \rangle (s) \equiv \frac{1}{2} \int_{-1}^{+1} dz
z^n {\cal W} \left( t(s,z) \right)
\ee
with $t(s,z)$ given in eq. \rf{t_u_in_terms_of_s_z}.
Since the functions $ {\hat M}_{1,3}^{I,{\cal I}} (s)$ have no discontinuities on the unitarity cut, 
the absorptive parts of the functions $M_{1,3}^{I,{\cal I}} (s)$ thus coincide with those 
of the corresponding partial-wave projections along their right-hand cut, which are given by unitarity.
In order to determine the discontinuities of the partial waves $ f_{1,3;l}^{I,{\cal I}} (s)$,
let us recall that the amplitudes ${\cal M}^{I;{\cal I}}_{1,3}$ correspond to the matrix elements, 
\be
{\cal M}^{I;{\cal I}}_{1,3} (s,t,u) = 
{ }_{\rm out}\!\langle (\pi\pi)^I \vert {\rm T}^{\frac{1}{2} , \frac{3}{2}} \vert (K\pi)^{\cal I} \rangle_{\rm in}
.
\ee
The two-meson states occurring in this expression belong to the spectrum of three-flavour QCD,
and the in and out states refer to the corresponding QCD S matrix.
Since we ignore violations of CP, 
the operators ${\rm T}^{\frac{1}{2}}$ and ${\rm T}^{\frac{3}{2}}$ 
commute with the operator of time reversal. 
It follows that
\be
{\cal M}^{I;{\cal I}}_{1,3} { }^* \equiv
{ }_{\rm in}\!\langle (K\pi)^{\cal I} \vert \big( {\rm T}^{\frac{1}{2} , \frac{3}{2}} \big)^\dagger \vert  (\pi\pi)^I \rangle_{\rm out}
=
{ }_{\rm in}\!\langle (\pi\pi)^I \vert {\rm T}^{\frac{1}{2} , \frac{3}{2}} \vert (K\pi)^{\cal I} \rangle_{\rm out}
.
\ee
The insertion of complete sets of states of three-flavour QCD into this relation then gives
\bea
{\cal M}^{I;{\cal I}}_{1,3} { }^*
&=&
\!\!
\sum_{\alpha , \beta} {\!\!\!\!\!\!\!\!\int} ~
{ }_{\rm in}\!\langle (\pi\pi)^I \vert \alpha \rangle_{\rm out} \,
{ }_{\rm out}\!\langle \alpha \vert {\rm T}^{\frac{1}{2} , \frac{3}{2}} \vert \beta \rangle_{\rm in}
\, { }_{\rm in}\!\langle \beta \vert (K\pi)^{\cal I} \rangle_{\rm out}
\lbl{unitarity_amps}
\\
&=&
\!\!
\sum_{\alpha , \beta} {\!\!\!\!\!\!\!\!\int} ~
[ { }_{\rm in}\!\langle (\pi\pi)^I \vert \alpha \rangle_{\rm in} -
{ }_{\rm in}\!\langle (\pi\pi)^I \vert i \big( {\rm T}^{\frac{1}{2} , \frac{3}{2}} \big)^\dagger \vert \alpha \rangle_{\rm in} ] \,
{ }_{\rm out}\!\langle \alpha \vert {\rm T}^{\frac{1}{2} , \frac{3}{2}} \vert \beta \rangle_{\rm in}
\, { }_{\rm in}\!\langle \beta \vert (K\pi)^{\cal I} \rangle_{\rm out}
. ~~~
\nonumber
\eea
We will exploit this relation in the approximation where we limit the sum over $\alpha$ to a two-pion
state and the sum over $\beta$ to a $K \pi$ state, but neglecting rescattering in the initial state. As a further
step one might contemplate the possibility to also include initial-state rescattering, either by simply adding the 
corresponding $K\pi$ scattering phases to the $\pi\pi$ phases in eq. \rf{PW_unitarity} and replacing the asymtotic 
conditions in eq. \rf{asympt_phases} by appropriate conditions for the sum, or by considering a more elaborate 
multi-channel analysis, as done in ref. \cite{Albaladejo:2017hhj} in a different context, adding also inelastic channels
like $K{\bar K}$.

In the present study we will address the simplest situation, considering only elastic rescattering of the final-state
two-pion system.
With this restriction, the relation \rf{unitarity_amps} reduces to the following condition on the partial waves:
\be
f^{I;{\cal I}}_{1,3 ; l} (s) { }^* = f^{I;{\cal I}}_{1,3 ; l} (s)  \left[ 1 - 2 i \sigma_\pi (s) f_l^I (s) { }^* \right] 
, 
\ee
where
\be
\sigma_\pi (s) \equiv \sqrt{1 -  \frac{4M_\pi^2}{s}}
\ee
comes from the two-pion phase space in Eq. \rf{unitarity_amps}, whereas $f_l^I (s)$ are the partial-wave projections, 
defined analogously to Eq. \rf{PW-def}, of the $\pi\pi$ scattering amplitudes.
In the elastic domain, these partial waves are expressed in terms of the $\pi\pi$ strong scattering phases $\delta_l^I (s)$,
\be
f_l^I (s) = \frac{1}{\sigma_\pi (s)} \, e^{i \delta_l^I (s)} \sin \delta_l^I (s)
,
\ee
so that we obtain the unitarity relation for the partial waves $f^{I;{\cal I}}_{1,3 ; l} (s)$
in a form that only involves these phases,
\be
f^{I;{\cal I}}_{1,3 ; l} (s) { }^* = e^{-2i \delta_l^I (s)} \, f^{I;{\cal I}}_{1,3 ; l} (s)   .
\ee
Let us stress that this relation, which can be rewritten in a form that gives the discontinuity 
of $f^{I;{\cal I}}_{1,3 ; l} (s)$ along the right-hand cut $s \ge 4 M_\pi^2$,
\be
f^{I;{\cal I}}_{1,3 ; l} (s-i\epsilon) = e^{-2i \delta_l^I (s)} \, f^{I;{\cal I}}_{1,3 ; l} (s+i\epsilon)
,
\lbl{PW_unitarity}
\ee
is only valid in the energy region where the processes $\pi\pi \to \pi\pi$ is elastic
and rescattering in the initial state can be omitted. Below we will also need to specify 
the asymptotic behaviour of the $\pi\pi$ phases. We assume the following properties when 
$s \to \infty$:\footnote{In practice we have assumed that these asymptotic values are 
reached at a finite energy, $s_{\rm cut} = 10~{\rm GeV}^2$.}
\be
\lim_{s\to +\infty} \left\{ 
\begin{tabular}{l}
$\delta_0^0 (s)$
\\
$\delta_1^1 (s)$ 
\\
$\delta_0^2 (s)$ 
\end{tabular}
\right.
= \, \left\{ 
\begin{tabular}{l}
$ \pi $
\\
$ \pi $
\\
$ 0 $
\end{tabular}
\right.
.
\lbl{asympt_phases}
\ee
Notice that since we will deal only with S and P partial-wave projections, we will henceforth drop the subscript $l$ 
in the partial waves and in the phase shifts, writing simply $f^{I;{\cal I}}_{1,3 ; l} \equiv f^{I;{\cal I}}_{1,3}$ 
and $\delta_l^I \equiv \delta_{I}$, with the understanding that the angular momentum is necessarily $l=0$ when $I=0,2$ 
and $l=1$ when $I=1$.
In the region of energy where elastic unitarity holds, we may now obtain the absorptive parts
of the single-variable functions from the above:
\be
{\rm Abs} \, M_{1,3}^{I ; {\cal I}} (s) = 16 \pi \, e^{- i \delta_I (s)} \sin \delta_I (s) 
\left[ M_{1,3}^{I ; {\cal I}} (s) + {\hat M}_{1,3}^{I ; {\cal I}} (s) \right] \theta(s-4M_\pi^2)
.
\lbl{abs_M}
\ee
\begin{figure}[ht]
\centering
\includegraphics[width=0.45\linewidth]{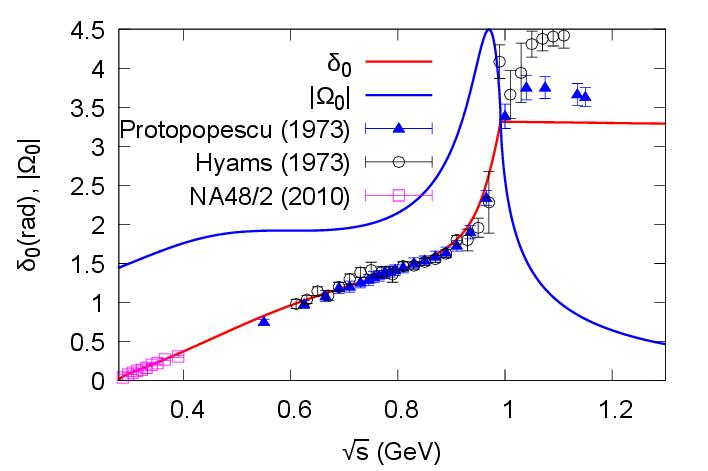}\includegraphics[width=0.45\linewidth]{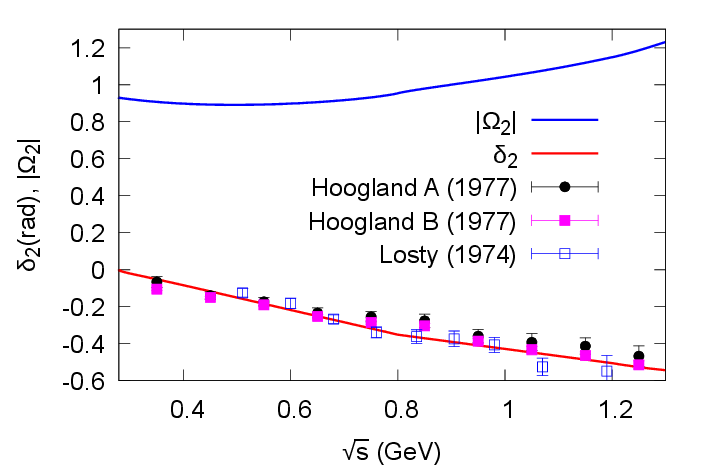}
\includegraphics[width=0.45\linewidth]{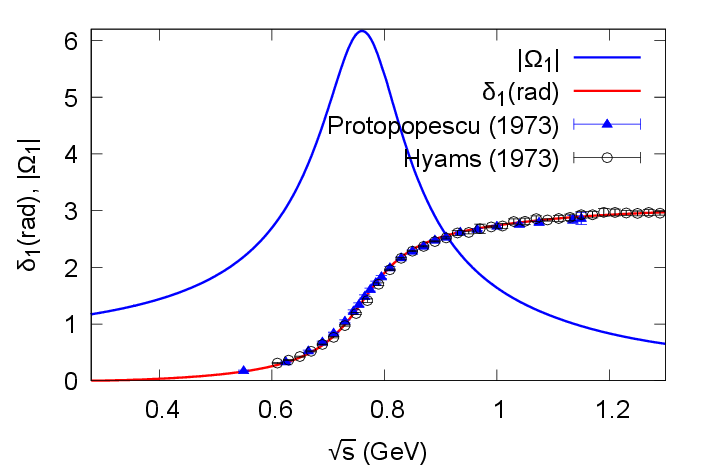}\
\caption{The S- and P-wave $\pi\pi$ scattering phase-shifts and
the  corresponding Omn\`es functions. The experimental data shown are from
  refs. \cite{NA482:2010dug,Hyams:1973zf,Losty:1973et,Hoogland:1977kt}, see the discussion in the text.} 
\label{fig:pipiphases}
\end{figure}
In the energy range where elastic unitarity holds, the absorptive parts are therefore 
known in terms of the angular averages \rf{hatM_1} and \rf{hatM_3} supplemented by some 
input for the elastic $\pi\pi$ phase shifts. 
Measurements of the pion-pion phase-shifts have been performed
in a large number of production experiments, $\pi N\to \pi\pi N,
\pi\pi\Delta$ in the 1970's, see~\cite{Martin:1976mb} for a
review. Alternative information on the phase-shifts can be obtained
from $\pi\pi$ form factors in semi-leptonic or electromagnetic
processes. For instance, information on the phase $\delta_0 (s)$ at low
energy can be extracted from $K_{l4}$ decays, the most precise results
coming from the NA48/2 experiment~\cite{NA482:2010dug}. The difference
of the two $I=0$ and $I=2$ S-wave phases at $s=M_K^2$ can be derived
from the widths of the three $K\to \pi\pi$ decay modes, up to some isospin
breaking corrections. An updated evaluation of these has been
performed in ref.~\cite{Cirigliano:2009rr}, leading to the following
result
\be
\Delta\equiv\delta_0(M_K^2)-\delta_2(M_K^2)=(52.5\pm0.8\pm2.8)^\circ 
\ee
which is compatible, within one sigma, with the result obtained from the
production experiments also constrained with the dispersive Roy equations:
$\Delta= (47.7\pm1.5)^\circ$ (ref.~\cite{Colangelo:2001df}),
$\Delta=(50.5\pm1.2)^\circ$ (ref.~\cite{Kaminski:2006qe}). Finally, data on
the pion vector form factor, as measured in $e^+e^-$ scattering or $\tau$
decays can be used to obtain an alternative determination of the P-wave
phase $\delta_1$ below 1 GeV, see for instance ref.~\cite{Ananthanarayan:2000ht}.
The phases used in
the present work, and the corresponding Omn\`es functions, are illustrated on
Fig. \ref{fig:pipiphases}. The fitting curves (shown as red lines in the figure)
correspond to Roy equations solutions from ref.~\cite{Ananthanarayan:2000ht}
in the region $\sqrt{s} \le 0.8$ GeV and the three phases are assumed to be constant,
equal to their asymptotic values \rf{asympt_phases}, when $\sqrt{s} \ge 3.16$ GeV. 
In the intermediate
energy region, in the case of $\delta_1$ the curve corresponds to a fit of
the data in the range $[0.8,1.45]$ GeV, using a K-matrix description as in
ref.~\cite{Hyams:1973zf} and a simple interpolation \cite{Moussallam:1999aq} to the asymptotic value
above $1.45$ GeV.  In the case of $\delta_0$, the data is fitted in the range $[0.8,1]$
GeV using a functional form which is also constrained by the Roy equations,
see ref.~\cite{Moussallam:2011zg}, and interpolated to $\pi$ beyond this energy range \cite{Moussallam:1999aq}. 
Finally, for $\delta_2$ the experimental data is fitted in the range $[0.8,1.2]$ GeV
with a simple polynomial function and interpolated to zero above $1.2$ GeV.

\section{Omn\`es-Khuri-Treiman integral equations}\label{sec:integral_eqs}
\setcounter{equation}{0}

In the approximation where the final state interactions are restricted to elastic
rescattering of the pion pairs, the system of dispersion relations \rf{disp-rel_M} 
seems now to be complete, with the absorptive parts given by eq. \rf{abs_M}.
There are, however, a couple of well-known issues that still need to be dealt with
before proceeding with the numerical solution of this system. On the one side, it has 
been observed long ago \cite{Bronzan:1964zz,Neveu:1970tn} that without the second 
contribution, given by ${\hat M}_{1,3}^{I ; {\cal I}} (s)$, in the absorptive parts, 
the dispersion relations would take the form of an Omn\`es problem. On the other 
side, in the form provided by eqs. \rf{disp-rel_M} and \rf{abs_M}, the dispersion 
relations do in general not lead to a unique set of solutions \cite{Anisovich:1996tx}. 
Both issues can be dealt with, and in particular unique solutions are expected, 
if the dispersion relations are written instead for the functions $M_{1,3}^{I ; {\cal I}} (s) / \Omega_I (s)$,
where $\Omega_I (s)$ denote the Omn\`es factors 
\be
\Omega_I (s) = {\rm exp} \left[ \frac{s}{\pi} \int_{4 M_\pi^2}^{+\infty} \frac{dx}{x} \, \frac{\delta_I (x)}{x-s} \right]
,
\lbl{omnes}
\ee 
with their asymptotic behaviours for $s\to\infty$ given, according to Eq. \rf{asympt_phases}, by
\be
\Omega_0 (s) \sim \frac{1}{s}, \ \ \Omega_1 (s) \sim \frac{1}{s} , \ \ \Omega_2 (s) \sim {\rm constant}
.
\lbl{omnes_asympt}
\ee
This gives the set of Omn\`es-Khuri-Treiman integral equations
\be
M_{1,3}^{I ; {\cal I}} (s) = \Omega_I (s) \left\{ P_{1,3}^{I ; {\cal I}} (s) + \frac{s^{n_{1,3}^{I ; {\cal I}}}}{\pi}
\int_{4 M_\pi^2}^\infty \frac{dx}{x^{n_{1,3}^{I ; {\cal I}}}} \, 
\frac{\sin \delta_I (x) {\hat M}_{1,3}^{I ; {\cal I}} (x)}{\vert \Omega_I (x) \vert (x - s)}
\right\}
, \quad I=0,1,2, \ {\cal I} = \frac{1}{2},\,\frac{3}{2} .
\lbl{disp_rel_omnes}
\ee
Here, 
$P_{1,3}^{I ; {\cal I}} (s)$ are subtraction polynomials whose orders 
depend on the asymptotic conditions assumed for the amplitudes and on the additional constraints
that were put on their behaviour at $s=0$. Before giving the explicit form of the 
subtraction polynomials that follow from these restrictions, we observe that the 
expression of the absorptive part of any function ${\hat M}_{1,3}^{I ; {\cal I}} (s)$
in eqs. \rf{hatM_1} or \rf{hatM_3} involves almost all the other functions, $M_{1}^{I ; {\cal I}} (s)$ 
or $M_{3}^{I ; {\cal I}} (s)$, respectively. By forming appropriate linear combinations
of the isospin functions, one may decompose the relations \rf{hatM_1} or \rf{hatM_3} into 
subsets that do not mix with the other subsets when taking the partial-wave projections. 
Let us therefore define\footnote{These definitions mix initial states with different values 
of the isospin of the $K\pi$ system, which is allowed to the extend that we disregard the 
influence of $K\pi$ rescattering here, see the discussion following eq. \rf{unitarity_amps}.}
\be
\left(
\begin{tabular}{c}
$ M_2 (s) $
\\
$  M_1 (s) $
\\
$ {\tilde M}_1 (s) $
\\
$ M_0 (s) $
\end{tabular}
\right)
=
\left(
\begin{tabular}{cccc}
$1$ & 0 & 0 & 0
\\[0.1cm]
0 & $\frac{1}{3}$ & $\frac{2}{3}$ & 0
\\[0.1cm]
0 & $\frac{2}{3}$ & $-\frac{2}{3}~~$ & 0
\\[0.1cm]
0 & 0 & 0 & 1
\end{tabular}
\right)
\left(
\begin{tabular}{c}
$ M_{1}^{2;\frac{3}{2}} (s) $
\\
$ M_{1}^{1;\frac{3}{2}} (s) $
\\
$ M_{1}^{1;\frac{1}{2}} (s) $
\\
$ M_{1}^{0;\frac{1}{2}} (s) $
\end{tabular}
\right) 
\lbl{new_basis_1}
\ee
and
\be
\left(
\begin{tabular}{c}
$ {N}_2 (s) $
\\
$ {\tilde {N}}_2 (s) $
\\
$  {N}_1 (s) $
\\
$ {\tilde{N}}_1 (s) $
\\
$ {N}_0 (s) $
\end{tabular}
\right)
=
\left(
\begin{tabular}{ccccc}
$\frac{1}{2}$ & $\!\!\frac{1}{2}$ & $\!\!\!\!\!0$ & $\!\!\!0$ & $\!\!\!0$
\\[0.1cm]
$1$ & $\!\!\! -1~~$ & $\!\!\!\!\!0$ & $\!\!\!0$ & $\!\!\!0$
\\[0.1cm]
0 & $\!\!\!0$ & $\!\!\!\!\!\frac{5}{6}$ & $\!\!\!\frac{1}{6}$ & $\!\!\!0$
\\[0.1cm]
0 & $\!\!\!0$ & $\!\!\!\!\!- \frac{1}{3}~~$ & $\!\!\!\frac{1}{3}$ & $\!\!\!0$
\\[0.1cm]
0 & $\!\!\!0$ & $\!\!\!\!\!0$ & $\!\!\!0$ & $\!\!\!1$
\end{tabular}
\right) \!
\left(
\begin{tabular}{c}
$ M_{3}^{2;\frac{3}{2}} (s) $
\\
$ M_{3}^{2;\frac{1}{2}} (s) $
\\
$ M_{3}^{1;\frac{3}{2}} (s) $
\\
$ M_{3}^{1;\frac{1}{2}} (s) $
\\
$ M_{3}^{0;\frac{3}{2}} (s) $
\end{tabular}
\right)     .
\lbl{new_basis_3}
\ee
The corresponding absorptive parts are then given by
\bea\lbl{Mhat}
{\hat M}_2  &=& \langle M_0 \rangle  + \frac{1}{3} \langle M_2 \rangle
- \frac{3}{2} (s - s_0) \langle M_{1} \rangle 
- \frac{1}{2} \kappa \langle z M_{1} \rangle
,
\nonumber\\
{\hat M}_1  &=& \frac{1}{\kappa} \left[ 3 \langle z M_{0} \rangle  - 5 \langle z M_{2} \rangle
+ \frac{9}{2} (s - s_0) \langle z M_1 \rangle
+ \frac{3}{2} \kappa \langle z^2 M_{1} \rangle
\right]
,
\nonumber\\
{\hat M}_0  &=& \frac{2}{3} \langle M_0 \rangle + \frac{20}{9} \langle M_2 \rangle
+ 2 (s - s_0) \langle M_{1} \rangle + \frac{2}{3} \kappa \langle z M_{1} \rangle
,
\nonumber\\
\\
{\hat{\tilde M}}_1  &=& \frac{1}{\kappa} \left[ - 9 (s - s_0) \langle z {\tilde M}_1 \rangle - 3 \kappa \langle z^2 {\tilde M}_1 \rangle \right]
\nonumber
\eea
and
\bea\lbl{Nhat}
{\hat {N}}_2 &=& \langle {N}_0 \rangle + \frac{1}{3} \langle {N}_2 \rangle - \frac{3}{2} (s - s_0) \langle {N}_1 \rangle 
- \frac{1}{2} \kappa \langle z {N}_1 \rangle
,
\nonumber\\
{\hat {N}}_1 &=& \frac{1}{\kappa} \left[ 3 \langle z {N}_0 \rangle - 5 \langle z {N}_2 \rangle + \frac{9}{2} (s - s_0) \langle z {N}_1 \rangle 
+ \frac{3}{2} \kappa \langle z^2 {N}_1 \rangle \right]
,
\nonumber\\
{\hat {N}}_0 &=& \frac{2}{3} \langle {N}_0 \rangle  + \frac{20}{9} \langle {N}_2 \rangle
+ 2 (s - s_0) \langle {N}_1 \rangle 
+ \frac{2}{3} \kappa \langle z {N}_1 \rangle ,
\nonumber\\
\\
{\hat{\tilde {N}}}_2 &=& - \langle {\tilde {N}}_2 \rangle - \frac{9}{2} (s - s_0) \langle {\tilde {N}}_1 \rangle 
- \frac{3}{2} \kappa \langle z {\tilde {N}}_2 \rangle ,
\nonumber\\
{\hat{\tilde {N}}}_1 &=& \frac{1}{\kappa} \left[ - 3 \langle z {\tilde {N}}_2 \rangle + \frac{9}{2} (s - s_0) \langle z {\tilde {N}}_1 \rangle 
+ \frac{3}{2} \kappa \langle z^2 {\tilde {N}}_1 \rangle \right]   .
\nonumber
\eea
For the functions related to the $\Delta I = 1/2$ transitions, we observe that there are 
two disconnected subsets, the second one consisting of the single function ${\tilde M}_1 (s)$,
whose absorptive part depends on none of the other functions, while this function in turn does not
appear in the absorptive parts of the other three functions $M_{0,1,2} (s)$. Likewise, the
functions related to the $\Delta I = 3/2$ transitions separate into two disconnected sets,
consisting of the three functions $N_{0,1,2} (s)$ and the two functions ${\tilde N}_{1,2} (s)$.
It is also of interest to notice that the numerical coefficients in front of the functions ${M}_{0,1,2}$
entering the definitions of the functions ${\hat M}_{0,1,2}$ are the same as those that define the functions
${\hat N}_{0,1,2}$ in terms of the functions $N_{0,1,2}$. In addition,
the function $N_1 (s)$ is proportional to the combination
$5 M_3^{1;\frac{3}{2}} (s) + M_3^{1;\frac{1}{2}} (s)$ that we had singled out at the 
end of Section \ref{sec:single_var} for precisely this purpose.
These features will be useful in determining the number of independent solutions of the 
system of dispersion relations. In terms of these new functions, and taking 
all the contraints discussed so far into account, the dispersion relations 
finally take the form
\bea\lbl{disp_M}
M_2 (s) &=& \Omega_2 (s)  s^2 I_2 (s)  \hspace {5.13cm}  
I_2 (s) = \frac{1}{\pi} \int_{4 M_\pi^2}^\infty \frac{dx}{x^2} \, \frac{\sin \delta_2 (x) {\hat M}_2 (x)}{\vert \Omega_2 (x) \vert (x - s)}
, 
\nonumber\\
M_1 (s) &=& \Omega_1 (s) \left\{ \mu_3  s + s I_1 (s) \right\}    \hspace {3.875cm}  
I_1 (s) = \frac{1}{\pi} \int_{4 M_\pi^2}^\infty \frac{dx}{x^2} \, \frac{\sin \delta_1 (x) {\hat M}_1 (x)}{\vert \Omega_1 (x) \vert (x - s)}
, 
\nonumber\\
{\tilde M}_1 (s) &=&\Omega_1 (s) \left\{ {\tilde\mu}_1 s + s {\tilde I}_1 (s) \right\}  \hspace {3.76cm}  
{\tilde I}_1 (s) = \frac{1}{\pi} \int_{4 M_\pi^2}^\infty \frac{dx}{x^2} \, \frac{\sin \delta_1 (x) {\hat{\tilde M}}_1 (x)}{\vert \Omega_1 (x) \vert (x - s)}
, 
\nonumber\\
M_0 (s) &=& \Omega_0 (s) \left\{ \mu_{0} + \mu_1 s + \mu_2 s^2 + s^2 I_0 (s) \right\}    \hspace {1.75cm}  
I_0 (s) = \frac{1}{\pi} \int_{4 M_\pi^2}^\infty \frac{dx}{x^2} \, \frac{\sin \delta_0 (x) {\hat M}_0 (x)}{\vert \Omega_0 (x) \vert (x - s)}
\eea
and
\bea\lbl{disp_N}
N_2 (s) &=& \Omega_2 (s) s^2 J_2 (s)  \hspace{5.0cm} 
J_2 (s) = \frac{1}{\pi}
\int_{4 M_\pi^2}^\infty \frac{dx}{x^2} \, \frac{\sin \delta_2 (x) {\hat N}_2 (x)}{\vert \Omega_2 (x) \vert (x - s)}
, 
\nonumber\\
{\tilde N}_2 (s) &=& \Omega_2 (s) s^2 {\tilde J}_2 (s)  \hspace{5.0cm} 
{\tilde J}_2 (s) =  \frac{1}{\pi} \int_{4 M_\pi^2}^\infty \frac{dx}{x^2} \, \frac{\sin \delta_2 (x) {\hat{\tilde N}}_2 (x)}{\vert \Omega_2 (x) \vert (x - s)}
, 
\nonumber\\
N_1 (s) &=& \Omega_1 (s) \left\{ \nu_{3} s + s J_1 (s) \right\}  \hspace{3.75cm}
J_1 (s) =  \frac{1}{\pi} \int_{4 M_\pi^2}^\infty \frac{dx}{x} \, \frac{\sin \delta_1 (x) {\hat N}_1 (x)}{\vert \Omega_1 (x) \vert (x - s)}
, 
\nonumber\\
{\tilde N}_1 (s) &=&\Omega_1 (s) \left\{ {\tilde\nu}_0 + {\tilde\nu}_{1} s + s {\tilde J}_1 (s) \right\}    \hspace{2.85cm}
{\tilde J}_1 (s) = \frac{1}{\pi}
\int_{4 M_\pi^2}^\infty \frac{dx}{x} \, \frac{\sin \delta_1 (x) {\hat{\tilde N}}_1 (x)}{\vert \Omega_1 (x) \vert (x - s)}
, 
\nonumber\\
N_0 (s) &=& \Omega_0 (s) \left\{ \nu_{0} + {\nu}_{1} s + {\nu}_{2} s^2 + s^2 J_0 (s) \right\}   \hspace{1.65cm}
J_0 (s) = \frac{1}{\pi}
\int_{4 M_\pi^2}^\infty \frac{dx}{x^2} \, \frac{\sin \delta_0 (x) {\hat N}_0 (x)}{\vert \Omega_0 (x) \vert (x - s)}
.
\eea
When restricted to either one of the two subsets $M_{0,1,2} (s)$ and $N_{0,1,2} (s)$,
the equations \rf{Mhat} and \rf{disp_M} or \rf{Nhat} and \rf{disp_N} are exactly the same
as those found in the literature for the $\eta\to\pi\pi\pi$ amplitudes, except that
the function $\kappa(s)$ and the constant $s_0$ are now defined in terms of the kaon mass $M_K$ and not of the 
eta-meson mass. In the case of the $\eta\to\pi\pi\pi$ transitions, the total isospin of the 
three pions is necessarily equal to unity (if only first-order isospin-breaking effects
are retained). It is thus obvious that the two sets of functions $M_{0,1,2} (s)$ and $N_{0,1,2} (s)$ 
describe the kaon decays into three pions in a state of total isospin one.
The functions ${\tilde M}_1$, ${\tilde N}_1$, and ${\tilde N}_2$ have no 
counterpart in the case of the eta meson, and reflect the transition to a state of 
total isospin zero in the case of ${\tilde M}_1$, or of total isospin two in the case of
${\tilde N}_1$ and ${\tilde N}_2$.

\begin{figure}[ht]
\centering
\includegraphics[width=0.333\linewidth]{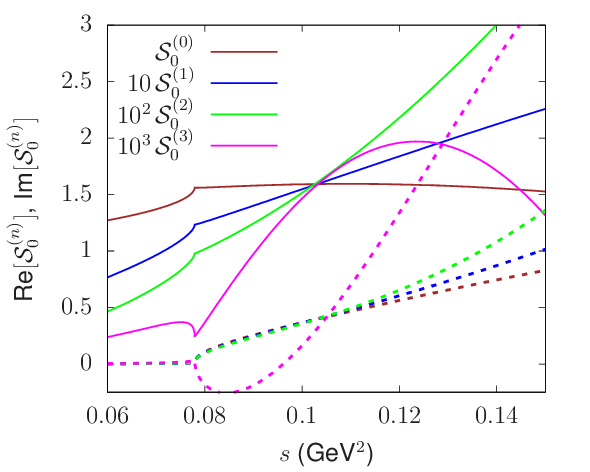}\includegraphics[width=0.333\linewidth]{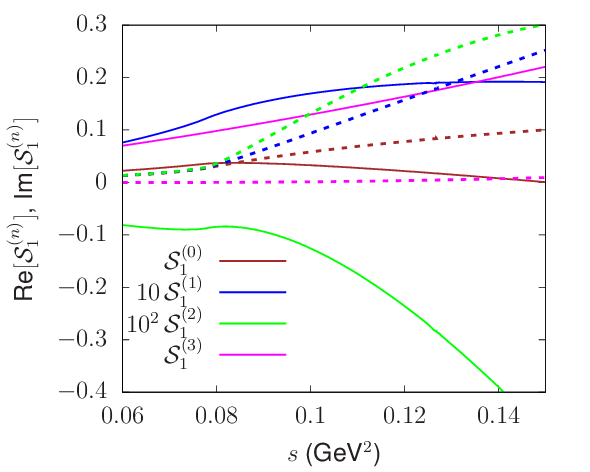}\includegraphics[width=0.333\linewidth]{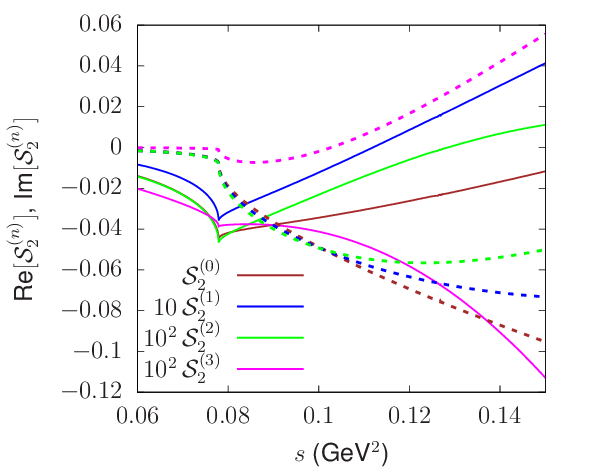}\\
\includegraphics[width=0.333\linewidth]{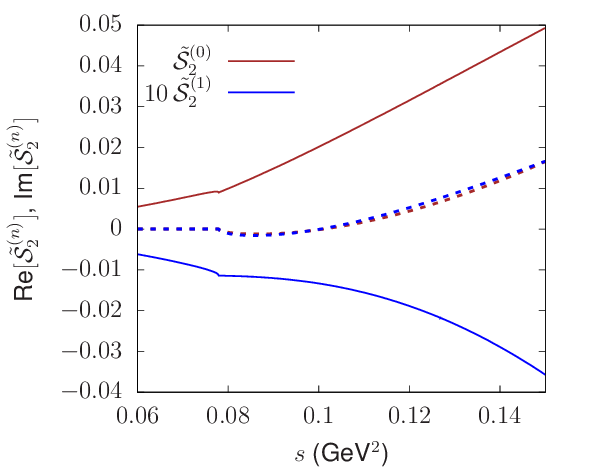}\includegraphics[width=0.333\linewidth]{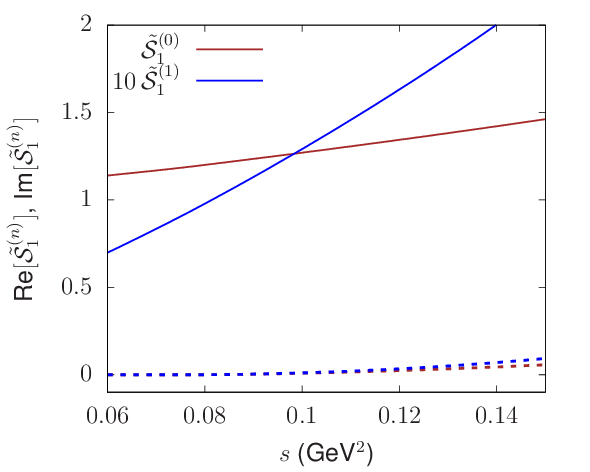}\includegraphics[width=0.333\linewidth]{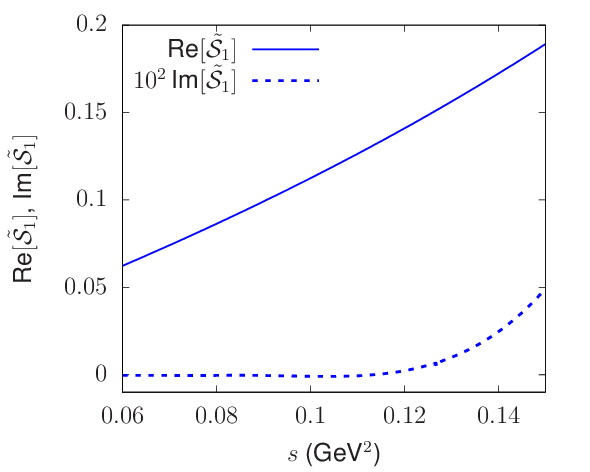}
\caption{The upper panel displays the plots of the fundamental solutions ${\cal S}^{(n)}_I (s)$ associated with the single-variable functions $M_I (s)$
and $N_I (s)$.  The left, central, and right windows correspond to $I=0,1,2$, respectively.
The lower panel shows the plots of the fundamental solutions ${\tilde{\cal S}}_2^{(0)}(s)$ (in red) and ${\tilde{\cal S}}_2^{(1)}(s)$ (in blue), associated with 
the function $\tilde{N}_2(s)$ (left window), ${\tilde{\cal S}}_1^{(0)}(s)$ (in red) and ${\tilde{\cal S}}_1^{(1)}(s)$ (in blue), associated with the function $\tilde{N}_1(s)$ 
(central window), and ${\tilde{\cal S}}_1 (s)$ corresponding to ${\tilde M}_1 (s)$
(right window).
The solid lines represent the real parts of the functions and the dashed lines the corresponding imaginary parts.  
The cusps originating from two-pion thresholds at $s=4 M_\pi^2 = 0.078$ GeV$^2$ are clearly visible in most of the plots.
The decay region corresponds to the interval $4 M_\pi^2 \le s \le (M_K-M_\pi)^2 = 0.127\,{\rm GeV}^2$.
Notice that the units on the vertical axes depend on the function under consideration and are expressed in powers of GeV
according to Table \ref{dimensions}.
}
\label{fig:fund_sols}
\end{figure}

Solutions to the set of integral equations \rf{disp_M} and \rf{disp_N} are
generated numerically, given as input a set of $\pi\pi$ scattering phases.
The latter have been discussed at the end of section \ref{sec:pw_unitarity}.
The solutions are obtained through an iterative process, putting all the hatted 
functions ${\hat M}_I$, ${\hat N}_I$, ... equal to zero as the starting point of the iteration. 
This numerical treatment is well documented in the literature, see for instance the references
\cite{Kambor:1995yc,Anisovich:1996tx,Descotes-Genon:2014tla,Colangelo:2018jxw,Gasser:2018qtg}
for more complete discussions. Let us simply recall here that since the equations are linear,
each single-variable function will take the form of a linear combination of the
subtraction constants multiplied by certain coefficient functions. The latter
can be obtained by successively taking one subtraction constant equal
to unity while putting all the remaining ones to zero. Therefore, they do not
depend on the subtraction constants anymore. Taking also into account the observations
made after eq. \rf{Nhat}, the solution of the system of integral equations
\rf{disp_M} and \rf{disp_N} thus writes as
\bea\lbl{gen_sol}
\left(\!\!
\begin{tabular}{c}
$M_I (s)$ \\
$N_I (s)$
\end{tabular}
\!\!\right) &=&
\left(\!\!
\begin{tabular}{c}
$\mu_0$ \\
$\nu_0$
\end{tabular}
\!\!\right)  {\cal S}_I^{(0)} (s) +
\left(\!\!
\begin{tabular}{c}
$\mu_1$ \\
$\nu_1$
\end{tabular}
\!\!\right)  {\cal S}_I^{(1)} (s) +
\left(\!\!
\begin{tabular}{c}
$\mu_2$ \\
$\nu_2$
\end{tabular}
\!\!\right)  {\cal S}_I^{(2)} (s) +
\left(\!\!
\begin{tabular}{c}
$\mu_3$ \\
$\nu_3$
\end{tabular}
\!\!\right)  {\cal S}_I^{(3)} (s) ,
\quad I=0,1,2
\nonumber\\[-0.1cm]
\\[-0.1cm]
{\tilde M}_1 (s) &=& {\tilde\mu}_1 {\tilde{\cal S}}_1 (s) , ~~~~~
{\tilde N}_I (s) = {\tilde\nu}_0 {\tilde{\cal S}}_I^{(0)} + {\tilde\nu}_1 {\tilde{\cal S}}_I^{(1)}  ,
\quad I=1,2  .
\nonumber
\eea
This general solution depends only on 17 independent functions, called fundamental solutions, that remain to be determined numerically,
through the iterative process described above. Tables of these numerical solutions, together with the set of
phase shifts used as input, are provided as ancillary files. Plots of the fundamental solutions are shown
in Figs. \ref{fig:fund_sols}.  

\begin{table}[ht]
\begin{center}
\begin{tabular}{|c|c|c|}
\hline\hline
& 
\\[-0.35cm]
$N$   &  subtraction csts.   &  fundamental functions 
\\[0.05cm]
\hline
& 
\\[-0.35cm]
$-4$  &   $\mu_2 , \mu_3 , \nu_2 , \nu_3, {\tilde\mu}_1 , {\tilde\nu}_1$ &  $-$  
\\[0.15cm]
$-2$  &   $\mu_1 , \nu_1 , {\tilde\nu}_0$  & ${\cal S}_1^{(0)}$  
\\[0.15cm]
$~~0$  &   $\mu_0 , \nu_0$  &  ${\cal S}_0^{(0)}, {\cal S}_2^{(0)}, {\cal S}_1^{(1)}, {\tilde{\cal S}}_1^{(0)}$  
\\                                    
      &                       &             
\\[0.15cm]
$+2$  &       $-$  &  ${\cal S}_0^{(1)}, {\cal S}_2^{(1)}, {\cal S}_1^{(2)}, {\cal S}_1^{(3)}, {\tilde{\cal S}}_1^{(1)}, %
                       {\tilde{\cal S}}_2^{(0)}, {\tilde{\cal S}}_1$  
\\[0.15cm]
$+4$  &       $-$  &  ${\cal S}_0^{(2)}, {\cal S}_2^{(2)}, {\cal S}_0^{(3)}, {\cal S}_2^{(3)}, {\tilde{\cal S}}_2^{(1)}$  
\\[0.15cm]
\hline\hline
\end{tabular}
\end{center}
\caption{Dimensions in GeV$^N$ of the various quantities involved in eq. \rf{gen_sol}. \label{dimensions}}
\end{table}

\noindent
The corresponding expressions for the $K\to\pi\pi\pi$ decay amplitudes read
\bea\lbl{decay_amps}
{\cal A}^{K^\pm\to\pi^\pm\pi^\pm\pi^\mp} (s_1,s_2,s_3) \!\!\!&=&\!\!\!  
- 2 M_2 (s_3) - \frac{1}{3} \left[ M_2 (s_1) + M_2 (s_2)  \right] - M_0 (s_1) - M_0 (s_2)
+ (s_3-s_1) M_1 (s_2) + (s_3-s_2) M_1 (s_1) 
\nonumber\\
&&\!\!\!
- \, 2 N_2 (s_3) - {\tilde N}_2 (s_3) 
- \frac{1}{6} \left[ 2 N_2 (s_1) - 3 {\tilde N}_2 (s_1) \right] 
- N_0 (s_1)
- \frac{1}{6} \left[ 2 N_2 (s_2) - 3 {\tilde N}_2 (s_2) \right]  
- N_0 (s_2)
\nonumber\\
&&\!\!\!
+ \, \frac{1}{2} (s_3-s_2) \left[  2 N_1 (s_1) + 3 {\tilde N}_1 (s_1) \right]
+ \frac{1}{2} (s_3-s_1) \left[  2 N_1 (s_2) + 3 {\tilde N}_1 (s_2) \right]    ,
\nonumber\\
{\cal A}^{K^\pm\to\pi^0\pi^0\pi^\pm} (s_1,s_2,s_3) \!\!\!&=&\!\!\!
M_0 (s_3) - \frac{2}{3} M_2 (s_3) + M_2 (s_1) + M_2 (s_2) + (s_3-s_1) M_1 (s_2) + (s_3-s_2) M_1 (s_1)
\nonumber\\
&&\!\!\!
+ \, N_0 (s_3) - \frac{1}{3} \left[ 2 N_2 (s_3) - 3 {\tilde N}_2 (s_3) \right] 
+ \frac{1}{2} \left[ 2 N_2 (s_1) - {\tilde N}_2 (s_1) \right]  
+ \frac{1}{2} \left[ 2 N_2 (s_2) - {\tilde N}_2 (s_2) \right] 
\nonumber\\
&&\!\!\!
+ \, \frac{1}{2} (s_3-s_2) \left[ 2 N_1 (s_1) - 3 {\tilde N}_1 (s_1) \right]
+ \frac{1}{2} (s_3-s_1) \left[ 2 N_1 (s_2) - 3 {\tilde N}_1 (s_2) \right]   ,
\nonumber\\
{\cal A}^{K_L\to\pi^+\pi^-\pi^0}  (s_1,s_2,s_3) \!\!\!&=&\!\!\!
- M_0(s_3) + \frac{2}{3} M_2 (s_3) - M_2 (s_1) - M_2 (s_2) - (s_3-s_1) M_1 (s_2) -  (s_3-s_2) M_1 (s_1) 
\nonumber\\
&&\!\!\!
+ \, 2 N_0 (s_3) - \frac{4}{3} N_2 (s_3)
+ 2 N_2 (s_1) + 2 N_2 (s_2) + 2(s_3-s_1) N_1 (s_2) +  2(s_3-s_2) N_1 (s_1),
\nonumber\\
{\cal A}^{K_S\to\pi^+\pi^-\pi^0}  (s_1,s_2,s_3) \!\!\!&=&\!\!\! (s_1-s_2) {\tilde M}_1 (s_3) + (s_2-s_3) {\tilde M}_1 (s_1) + (s_3-s_1) {\tilde M}_1 (s_2) 
\nonumber\\
&&\!\!\!
+ \, {\tilde N}_2 (s_1) - {\tilde N}_2 (s_2) + (s_3-s_1) {\tilde N}_1 (s_2) - (s_3-s_2) {\tilde N}_1 (s_1) 
- 2 (s_1-s_2) {\tilde N}_1 (s_3)    ,
\nonumber\\
{\cal A}^{K_L\to\pi^0\pi^0\pi^0}  (s_1,s_2,s_3) \!\!\!&=&\!\!\!
\left[  M_0 (s_1) + M_0 (s_2) + M_0 (s_3) \right]
+ \frac{4}{3} \left[ M_2 (s_1) + M_2 (s_2) + M_2 (s_3)  \right]
\nonumber\\
&&\!\!\!
- \, 2 \left[ N_0 (s_1)  + N_0 (s_2) + N_0 (s_3) \right]
- \frac{8}{3} \left[ N_2 (s_1) + N_2 (s_2) + N_2 (s_3)  \right]   .
\eea
The $K\pi\to\pi\pi$ amplitudes listed in Table \ref{table:amplitudes} can then be obtained from the above amplitudes
through crossing. In particular, with the phase convention we have been using, one has
${\cal A}_1 (s,t,u) = - {\cal A}^{K^\pm\to\pi^\pm\pi^\pm\pi^\mp} (t,u,s)$,
${\cal A}_4 (s,t,u) = - {\cal A}^{K^\pm\to\pi^0\pi^0\pi^\pm} (t,u,s)$,
$\sqrt{2} \, {\cal A}_{13} (s,t,u)$ $ = {\cal A}^{K_L\to\pi^0\pi^0\pi^0} (s,t,u)$ and
\be
\sqrt{2} \, {\cal A}_7 (s,t,u) = {\cal A}^{K_L\to\pi^+\pi^-\pi^0}  (t,u,s) + {\cal A}^{K_S\to\pi^+\pi^-\pi^0}  (t,u,s)   .
\ee
Finally, one also needs to fix the numerical values of the eleven
subtraction constants in eq. \rf{gen_sol}. This last issue is addressed in the next section.

\section{Fixing the subtraction constants and a few applications}\label{sec:subtraction_csts}
\setcounter{equation}{0}

Since the system of Omn\`es-Khuri-Treiman equations does not provide information on the subtraction constants themselves,
some external information is required for their determination. Moreover, the subtraction constants are in 
general complex numbers. We will follow a procedure that allows for a determination of all 
subtraction constants but one, ${\tilde\mu}_1$, based on the two following observations:

\begin{itemize}
\item 
In the region close to $s=0$, where the low-energy expansion is reliable, the imaginary parts of the 
single-variable functions are only generated at the two-loop level, and are thus expected 
to remain small near $s=0$. Neglecting these imaginary parts in the first coefficients of the 
Taylor expansion at $s=0$ of the single-variable functions will provide a relation between the real and 
imaginary parts of the subtraction constants.
\item 
From an experimental point of view, information on the Dalitz-plot structure and partial widths 
of all the $K\to\pi\pi\pi$ decay modes is available. 
\end{itemize}
Taken together, these two points allow us to obtain numerical values for ten out of the 
eleven complex subtraction constants. The remainder of this section is devoted to the 
implementation of this procedure and to providing the necessary details. The particular 
case of ${\tilde\mu}_1$, which escapes this analysis, will be briefly addressed separately.
We will then provide two simple illustrations of our results, the first one regarding the description 
of the $K\to\pi\pi\pi$ amplitudes as complex second-order polynomials in the variables $s_i-s_0$, 
the second one concerning the determination of the strong phases of these amplitudes.

\subsection{Relating the real and imaginary parts of the subtraction constants}

For small values of $s$, the low-energy expansion is expected to provide an accurate
description of the single-variable functions, in the form of an expansion in 
powers of $s$. This description thus amounts to retaining only the first terms of the Taylor 
expansions of the single-variable functions at $s=0$. In particular, the Taylor
coefficients of the constant and linear terms (quadratic terms), being generated
at lowest (next-to-lowest) order in the low-energy expansion, will only receive 
imaginary parts from higher orders, which will thus be quite small compared to their
real parts. It is therefore a good approximation
to assume that the imaginary parts of these lowest Taylor coefficients vanish.
At this point, we need to recall that the 
single-variable functions are defined only modulo polynomial ambiguities, as discussed in
section \ref{sec:single_var}. These ambiguities will also affect the coefficients of their 
Taylor expansions, and one has to work in terms of Taylor invariants \cite{Colangelo:2018jxw},
i.e. combinations of the Taylor coefficients that remain unchanged under these polynomial 
shifts in the single-variable functions. Expressed in terms of the basis of functions
introduced in eqs. \rf{new_basis_1} and \rf{new_basis_3}, but with the same parameters 
as in eqs \rf{transf_1} and \rf{transf_3}, these ambiguities read
\bea 
\delta M_2 &=& \frac{b_1}{3} s^3 - (c_1 + 3 s_0 b_1) s^2 + g_1 s + h_1   ,
\nonumber\\ 
\delta M_1 &=& b_1 s^2 + c_1 s + e_1   ,
\nonumber\\ 
\delta {\tilde M}_1 &=& 2 a_1 s^3 - 6 s_0 a_1 s^2 + 2 d_1 s + 2 f_1   ,
\nonumber\\ 
\delta M_0 &=& -\frac{4}{9} b_1 s^3 + \frac{4}{3} (c_1 + 3 b_1 s_0) s^2
- \frac{4}{3} g_1 s + 3 (s-s_0) (g_1 - e_1 - 3 c_1 s_0 - 6 b_1 s_0^2) 
- \frac{4}{3} h_1    ,
\lbl{poly_amb_1}
\eea
and 
\bea  
\delta N_2 \!\!\!&=&\!\!\! \frac{1}{18} (5 a_3 + c_3) s^3 - \frac{1}{6} (5b_3 + d_3 + 15 a_3 s_0 + 3 c_3 s_0) s^2 + 
\frac{1}{2}(e_3 - f_3 + 2g_3 + 3 b_3 s_0 - 3 d_3 s_0 + 6 a_3 s_0^2 - 6 c_3 s_0^2) s + \frac{h_3+k_3}{2} , 
\nonumber\\ 
\delta {\tilde N}_2 \!\!\!&=&\!\!\! - \frac{1}{3} (a_3 - c_3) s^3 + (b_3 - d_3 + 3 a_3 s_0 - 3 c_3 s_0) s^2 - 
(e_3 - f_3 + 3 b_3 s_0 - 3 d_3 s_0 + 6 a_3 s_0^2 - 6 c_3 s_0^2) s + h_3 - k_3   ,
\nonumber\\ 
\delta N_1 \!\!\!&=&\!\!\! \frac{1}{6} ( 5 a_3 + c_3) s^2 + \frac{1}{6} ( 5 b_3 + d_3) s + \frac{1}{6} ( 5 e_3 + f_3)  ,
\nonumber\\ 
\delta {\tilde N}_1 \!\!\!&=&\!\!\! - \frac{1}{3} ( a_3 - c_3) s^2 - \frac{1}{3} ( b_3 - d_3) s - \frac{1}{3} ( e_3 - f_3)  ,
\nonumber\\ 
\delta N_0 \!\!\!&=&\!\!\! - \, \frac{2}{27} (5 a_3 + c_3) s^3 + \frac{2}{9} (5 b_3 + d_3 + 15 a_3 s_0 + 3 c_3 s_0 ) s^2
-  \left( \frac{5}{3} e_3 + \frac{4}{3} f_3 - \frac{5}{3} g_3 + 5 b_3 s_0 + 4 d_3 s_0 + 10 a_3 s_0^2 + 8 c_3 s_0^2 \right) \!s   
\nonumber\\
&&\!\! 
+ \, 6 (a_3 + 2 c_3) s_0^3 + 3 (b_3 + 2 d_3 ) s_0^2 + (e_3 + 2 f_3 - 3 g_3 ) s_0 - \frac{2}{3} (h_3 + k_3)   .
\lbl{poly_amb_3}
\eea
This means that the Taylor coefficients up to the third (second) order for the functions $M_2$, ${\tilde M}_1$, $M_0$,
$N_2$, ${\tilde N}_2$, $N_0$ ($M_1$, $N_1$, ${\tilde N}_1$) will be ambiguous.\footnote{The fact that some 
of the Taylor coefficients are actually already fixed by the conditions \rf{M1_origin_lin} and \rf{M3_origin_lin} does not impinge 
on this discussion, which is quite general at this stage.}
A couple of useful remarks\footnote{One may also notice in passing that only the five parameters $b_1$, $c_1$, $e_1$, $g_1$, $h_1$ are required 
to describe the transformations of the functions $M_{0,1,2}$ in eq. \rf{poly_amb_1}. Likewise, only five independent combinations of the set of parameters
$a_3$...$k_3$ appear in the transformations \rf{poly_amb_3} of the functions $N_{0,1,2}$. This number is in agreement 
with what is found in order to describe the polynomial ambiguities of the three single-variable functions involved in 
the $\eta\to\pi\pi\pi$ decay \cite{Colangelo:2018jxw}.} 
can be made concerning these expressions:
\begin{itemize}
 \item The parameters $a_1$, $d_1$ and $f_1$ occur only in $\delta {\tilde M}_1$, therefore there
 will be no Taylor invariants constructed with the low-order Taylor coefficients of the function ${\tilde M}_1 (s)$. 
 On the other hand, as can be seen in eq. \rf{decay_amps}, the function ${\tilde M}_1 (s)$ contributes only to the 
 decay amplitude $K_S\to\pi^+\pi^-\pi^0$, and in the combination
 $$
 (s_1-s_2) {\tilde M}_1 (s_3) + (s_2-s_3) {\tilde M}_1 (s_1) + (s_3-s_1) {\tilde M}_1 (s_2) ,
 $$
 to which the terms constant and linear in $s$ do not contribute anyway. Moreover, the cubic and quadratic terms
 enter $\delta {\tilde M}_1 (s)$ in the combination $2 a_1 (s-s_0)^3$ (up to linear and constant terms),
 which also does not contribute to the amplitude for $K_S\to\pi^+\pi^-\pi^0$.
 \item The three combinations of the parameters $a_3$...$f_3$ occurring in $\delta {\tilde N}_1$
 are also present in $\delta {\tilde N}_2$, but the combination $h_3-k_3$ is only to be found
 in $\delta {\tilde N}_2$. Therefore, it will not be possible to construct a Taylor invariant 
 involving ${\tilde N}_2 (0)$, the lowest Taylor coefficient of ${\tilde N}_2(s)$ at $s=0$.
 Again, this constant term drops out of the amplitudes where ${\tilde N}_2(s)$ appears, see eq. \rf{decay_amps}.
\end{itemize}
The Taylor expansions of the single-variable functions themselves are in turn constructed from 
the corresponding expansions of the 17 independent fundamental solutions of eq. \rf{gen_sol}, 
and which we write in the following way
\bea
{\cal S}_I^{(n)} (s) = 
A_I^{(n)} + s B_I^{(n)} + s^2 C_I^{(n)} + s^3 D_I^{(n)} + \cdots    ,
~~~~~
{\tilde{\cal S}}_I^{(n)} (s) = 
{\tilde A}_I^{(n)} + s {\tilde B}_I^{(n)} + s^2 {\tilde C}_I^{(n)} + s^3 {\tilde D}_I^{(n)} + \cdots    ,
\lbl{Taylor_sol_1}
\eea
and 
\be
{\tilde{\cal S}}_1 (s) = 
{\tilde A}_1 + s {\tilde B}_1 + s^2 {\tilde C}_1 + s^3 {\tilde D}_1 + \cdots   .
\lbl{Taylor_sol_2}
\ee
The ambiguities in the Taylor coefficients of the single-variable functions
are readily extracted from the expressions \rf{poly_amb_1} and \rf{poly_amb_3}. 
Since there are\footnote{The two numbers exhibited by each sum correspond to the amplitudes relevant
for the $\Delta I=1/2$ and $\Delta I=3/2$ transitions, respectively.} $15+18$ ambiguous Taylor coefficients and $8+9$ parameters
to describe the polynomial ambiguities, one expects $7+9$ Taylor invariants formed with the former.
Of these, $4+6$ can be chosen such as to receive only contributions from the constant, linear and quadratic 
terms of the single-variable functions. They read
\bea\lbl{Taylor_coeffs_at_0}
&&
\sum_{n=0}^3 \left(\!\!
\begin{tabular}{c}
 $\mu_n$ \\
 $\nu_n$
\end{tabular}
\!\!\right)
\left[ A_0^{(n)} + s_0 B_0^{(n)} + \frac{4}{3} ( A_2^{(n)} + s_0 B_2^{(n)} )\right]   ,
\nonumber\\ 
&&
\sum_{n=0}^3 \left(\!\!
\begin{tabular}{c}
 $\mu_n$ \\
 $\nu_n$
\end{tabular}
\!\!\right) 
\left[ 3 A_1^{(n)} + B_0^{(n)} - \frac{5}{3} B_2^{(n)} + 9 s_0 ( B_1^{(n)} + 2 s_0 C_1^{(n)} ) \right]   ,
\nonumber\\ 
&&
\sum_{n=0}^3 \left(\!\!
\begin{tabular}{c}
 $\mu_n$ \\
 $\nu_n$
\end{tabular}
\!\!\right) 
\left[ 3 C_0^{(n)} + 4 C_2^{(n)} \right]   ,
\nonumber\\ 
&&
\sum_{n=0}^3 \left(\!\!
\begin{tabular}{c}
 $\mu_n$ \\
 $\nu_n$
\end{tabular}
\!\!\right)  
\left[ C_2^{(n)} + B_1^{(n)} + 3 s_0 C_1^{(n)} \right]    ,
\nonumber\\
\\
&&
\sum_{n=0,1} {\tilde\nu}_n [ 3 {\tilde A}^{(n)}_1 - {\tilde B}^{(n)}_2 + 9 s_0 ( {\tilde B}_1^{(n)} + 2 {\tilde C}_1^{(n)} ) ]   ,
\nonumber\\
&&
\sum_{n=0,1} {\tilde\nu}_n [ 3 {\tilde B}^{(n)}_1 + {\tilde C}^{(n)}_2 + 9 s_0 {\tilde C}_1^{(n)} ]    .
\nonumber
\eea
Requiring, as argued at the beginning of this subsection, 
that the imaginary parts of these ten Taylor invariants vanish leads to a linear relation 
between the imaginary parts of the ten subtraction constants $\mu_n$, $\nu_n$ and ${\tilde\nu}_n$ 
and their real parts. The numerical determination of the fundamental solutions obtained previously
provides the values of the Taylor coefficients \rf{Taylor_sol_1} and \rf{Taylor_sol_2} involved in the 
relations \rf{Taylor_coeffs_at_0}. Note that ${\tilde\mu}_1$ does not show up in the latter, as could be 
anticipated from the discussion after eq. \rf{poly_amb_3}.

\subsection{Determination of the real parts of the subtraction constants from data}\label{subsec:fit}

In order to determine the real parts of the subtraction coefficients, we now turn to the experimental information
that is available concerning the $K\to\pi\pi\pi$ amplitudes.
In the isospin-symmetric limit, these amplitudes are smooth functions
inside the decay region, since the cusps that are produced by $\pi\pi$ rescattering sit
on its boundary. It takes isospin breaking and the small difference between the
charged and the neutral pion masses to see some of these cusps moving inside the decay region,
where they have been observed experimentally, both in the $K^\pm\to\pi^0\pi^0\pi^\pm$ 
and $K_L\to\pi^0\pi^0\pi^0$ modes, by the NA48/2 \cite{NA482:2005wht} and the KTeV  
\cite{KTeV:2008gel} collaborations, respectively. In the absence of isospin breaking
the measured decay distributions for the three-pion transitions of $K_L$ and $K^\pm$ 
(for $K_S$ see below) are currently described in terms of linear and quadratic slopes in the variables
\be
Y \equiv \frac{s_3 - s_0}{M_\pi^2}
,\quad
X \equiv \frac{s_2 - s_1}{M_\pi^2}  ,
\lbl{X_Y_defined}
\ee
of the moduli squared of the amplitudes \rf{decay_amps},
\be
\vert {\cal A} (s_1,s_2.s_3) \vert^2 = \vert {\cal A} (s_0,s_0,s_0) \vert^2 [ 1 + g Y + h Y^2 + k X^2 ] .
\lbl{eq:h000param}
\ee
Bose symmetry forbids a term linear in $X$, and in the case of $K_L\to\pi^0\pi^0\pi^0$ requires in addition $g=0$ and $k=h/3$.
The normalization $\vert {\cal A} (s_0,s_0,s_0) \vert^2$ can only be fixed from the 
corresponding $K\to\pi\pi\pi$ partial widths. The latter
have also been measured, except for $K_S\to\pi^+\pi^-\pi^0$, whose partial width is reconstructed
from the measurement of the interference parameter $\lambda$ with the amplitude of the 
$K_L\to\pi^+\pi^-\pi^0$ decay mode,
\be\lbl{eq:lambda}
\lambda=\frac{\int_{Y_{min}}^{Y_{max}} dY \int_0^{X_{max}(Y)}dX 
{\cal A}^{K_L\to\pi^+\pi^-\pi^0} (X,Y) { }^* {\cal A}^{K_S\to\pi^+\pi^-\pi^0}(X,Y)}
{\int_{Y_{min}}^{Y_{max}} dY \int_0^{X_{max}(Y)}dX |{\cal A}^{K_L\to\pi^+\pi^-\pi^0}(X,Y)|^2} ,
\ee
where the integration runs over the whole decay region.
The list of observables and their experimental values are displayed in the first two 
columns of Table \ref{tab:fitsresults}. They are taken from ref. \cite{ParticleDataGroup:2022pth}
and correspond, as far as the partial widths are concerned, to the values called PDG average 
in Table 6 of ref. \cite{DAmbrosio:2022jmd}. For the Dalitz-plot observables, we differ from 
ref. \cite{DAmbrosio:2022jmd} in two respects:
\begin{itemize}
\item 
since we account for the presence of imaginary parts in the decay amplitudes,
the experimental value of Im$[\lambda]$ is included in the list of relevant observables. Both Re$[\lambda]$ and Im$[\lambda]$
have been measured with comparable precisions. We use the experimental values from refs. \cite{CPLEAR:1998nkj} and \cite{NA48:2005uiw},
which also provide the correlation between the real and imaginary parts of $\lambda$;
 \item as far as the quadratic slope $h$ of the $K_L\to\pi^0\pi^0\pi^0$ distribution 
 is concerned (recall that in this case we have $g=0$ and $k=h/3$), the value quoted by the PDG \cite{ParticleDataGroup:2022pth} 
 for $h$ is taken from ref. \cite{KTeV:2008gel}, and
is derived from a fit to the KTeV data using the Cabibbo-Isidori
rescattering model~\cite{Cabibbo:2005ez}. In this model, the parameter $h$
refers to only part of the full decay distribution. The
parametrisation~\rf{eq:h000param} obviously displays no cusp at the
$\pi^+\pi^-$ threshold in the case of $K_L\to\pi^0\pi^0\pi^0$, but it can still 
be used provided one avoids the vicinity of
the cusp region.  The amplitudes generated through the Omn\`es-Khuri-Treiman
equations do implement $\pi\pi$
rescattering but isospin-breaking corrections would need to be introduced before one can
properly describe the cusp region, which is not the case in the present analysis (this was done in
ref.~\cite{Colangelo:2018jxw} in the case of the $\eta\to 3\pi$ amplitudes).
In order to circumvent this issue, we will instead use an earlier determination
of the quadratic slope $h$ by the NA48 experiment~\cite{NA48:2001jrj}, 
\be
h=-6.1(1.0)\cdot 10^{-3} .
\ee
\end{itemize}

\begin{table}[th]
\centering
\begin{tabular}{|l|c||c|c||c|c|}
\hline
\\[-0.375cm]
~~~~~observables & exp$^{\rm al}$ values  & fit values & $\chi^2$    &   $\alpha$-fit values & $\chi^2(\alpha$-fit)
\\ \hline\hline
$\Gamma[K_L\to\pi^0\pi^0\pi^0]$ &$2.5417(352)$ &2.5532&0.1 & 2.6113  & 3.9   
\\
$\Gamma[K_L\to\pi^+\pi^-\pi^0]$ &$1.6200(102)$ &1.6185&0.02 &1.6105  & 0.9    
\\
$\Gamma[K^+\to\pi^0\pi^0\pi^+]$ &$0.9438(150)$ &0.8984&9.2 &0.9133& 4.1   
\\
$\Gamma[K^+\to\pi^+\pi^+\pi^-]$ &$2.9590(218)$ &2.9865&1.7 & 2.9779& 0.8  
\\
$\Gamma[K_S\to\pi^+\pi^-\pi^0]$ &$0.0026(7)$ &0.0031 &$-$ &0.0031&$-$  
\\ \hline
$h[[K_L\to\pi^0\pi^0\pi^0]$ &$-0.0061(10)$     &$-0.00644$ &0.1 &$-0.00645$ &0.1 
\\ \hline
$g[K_L\to\pi^+\pi^-\pi^0]$      &$0.678(8)$  & 0.675& 0.1 & 0.677 &0.02
\\  
$h[K_L\to\pi^+\pi^-\pi^0]$      &$0.076(6)$  & 0.082 & 0.9 & 0.082 &0.9  
\\  
$k[K_L\to\pi^+\pi^-\pi^0]$      &$0.0099(15)$ & 0.011& 0.3 & 0.011 &0.5  
\\  \hline
$g[K^+\to\pi^0\pi^0\pi^+]$      &$0.626(7)$  & 0.625 & 0.04 & 0.624 & 0.1 
\\ 
$h[K^+\to\pi^0\pi^0\pi^+]$      &$0.052(8)$  & 0.058 & 0.6 & 0.063 & 1.9 
\\ 
$k[K^+\to\pi^0\pi^0\pi^+]$      &$0.0054(35)$ & 0.011 & 2.7 & 0.012 & 3.5  
\\ \hline
$g[K^+\to\pi^+\pi^+\pi^-]$      &$-0.21134(17)$ &$-0.21134$ & 0 &$-0.21134$ & 0
\\
$h[K^+\to\pi^+\pi^+\pi^-]$      &$0.0185(4)$   &~0.0185 & 0 & 0.0185     & 0
\\
$k[K^+\to\pi^+\pi^+\pi^-]$      &$-0.00463(14)$ &$-0.00464$ & 0 &$-0.00464$ & 0
\\ \hline
Re$[\lambda]$                   &$0.0334(52)$   &0.0360 & 0.4   & 0.0356   & 0.2 
\\
Im$[\lambda]$                  &$-0.0108(48)~\,~ $   &$-0.0092~\,~$ & 0.11  &$-0.0091$  & 0.13 
\\ \hline\hline
\end{tabular}
\caption{The two first columns show the experimental observables and their input values used for the fits.
The third and fourth columns give the results from the fits described in subsection \ref{subsec:fit},
and the two last columns for the $\alpha$-fit described in subsection \ref{subsec:alpha-fit}. The widths are expressed in
units of $10^{-18}$ GeV. The width $\Gamma[K_S\to\pi^+\pi^-\pi^0]$ is not an independent
quantity, being reconstructed from $\Gamma[K_L\to\pi^+\pi^-\pi^0]$ and $\lambda$. It is thus 
not an input for the fits, and is only given for completeness.}
\label{tab:fitsresults}
\end{table}

\noindent
The list of experimental observables and the values we use are collected in the two first columns of Table \ref{tab:fitsresults}.
In order to proceed with the determination of the real parts of the subtraction constants, we take the 
modulus squared of the various amplitudes generated by our solutions \rf{gen_sol}, express the imaginary 
parts of the ten subtraction constants $\mu_n$, $\nu_n$ and ${\tilde\nu}_n$, as described in the preceding subsection, 
in terms of their real parts and fit the latter, in the decay regions, to the experimental descriptions \rf{eq:h000param} using the values 
given in Table \ref{tab:fitsresults}. For this fit, ${\tilde\mu}_1$ is set to zero.
The outcome of this procedure as far as the subtraction constants are
concerned is displayed in Table \ref{tab:subtractions}. The third column in Table \ref{tab:fitsresults} 
gives the values of the observables that result from this fit and the fourth column provides the 
contribution of each observable to the total $\chi^2$ whose value is $\chi^2_{\rm tot}=16.1$. 
Given that there are six degrees of freedom, the quality of the fit as measured by the value of $\chi^2_{\rm tot}$ is 
certainly not optimal. As discussed in detail in ref. \cite{DAmbrosio:2022jmd} this can most likely be ascribed to 
tensions among some of the experimental partial widths. 
Notwithstanding this issue with the high value of the total $\chi^2$, and given the 16 independent experimental data listed 
in Table \ref{tab:fitsresults}, one might wonder why 
not also include the complex subtraction constant ${\tilde\mu}_1$ into the set of parameters to be fitted, 
since this would still leave us with four real degrees of freedom. The reason why this does 
actually not work can be understood upon looking more closely at the expression for the observable $\lambda$,
which turns out, according to the eqs. \rf{gen_sol} and \rf{decay_amps}, to represent the only observable sensitive to ${\tilde\mu}_1$.
We defer the discussion of this issue to the subsection \ref{subsec:mu1} below.

\begin{table}[ht]
\begin{center}
\begin{tabular}{c|cccc}
\hline\hline
& 
\\[-0.35cm]
$n$   &   $\mu_n$   &   $\nu_n$   &   ${\tilde\nu}_n$   &   ${\tilde\mu}_n$
\\[0.05cm]
\hline
& 
\\[-0.35cm]
$0$   &   $~~~~~~\,\,~ +3.8(4) - i 0.570(2)$         &   $\qquad\qquad\ \,\quad -4.7(3) - i 2.37(10) \cdot 10^{-2}$   &   $\, -2.04(1.2) + i 4.8(5) \cdot 10^{-4}$   &   $-$
\\
$1$   &   $\,~ -676.1(4.0) + i 7.27(2)$     &   $~\, +26.7(2.8) + i 0.30(1)$               &   $+433.2(12.2) - i 6.0(1.3) \cdot 10^{-4}$&   n.d.
\\
$2$   &   $~\, +559.7(18.5) - i 16.80(5)$   & $~\! -46.0(14.2) - i 0.74(3)$                &   $-~\,~$              &   $-$
\\
$3$   &   $-1072.6(13.6)  + i 7.57 (2)$ & $+123.9(12.6) - i 0.28(1)$               &   $-~\,~$              &   $-$
\\
\hline\hline
\end{tabular}
\end{center}
\caption{Values of the real and imaginary parts of the subtraction constants. $\mu_0$ and $\nu_0$ are dimensionless numbers, 
$\mu_1 , \nu_1 , {\tilde\nu}_0$ are expressed in units of GeV$^{-2}$, 
and the remaining subtraction constants in units of GeV$^{-4}$.}\label{tab:subtractions}
\end{table}

\subsection{Description of the decay amplitudes as complex polynomials}\label{subsec:alpha-fit}

As already mentioned, in the absence of isospin violation the decay amplitudes are smooth functions
over the whole decay region. It is thus possible to describe them in terms 
of polynomials in the variables $s_i-s_0$, equivalently in the variables $X$ and $Y$ defined
in eq. \rf{X_Y_defined}, corresponding to their Taylor expansions 
around the centre of the decay region $s_1=s_2=s_3=s_0$. From a phenomenological point 
of view, the decay amplitudes are well described by such polynomials truncated at
second order, at least away from the narrow region of the cusp in the two cases mentioned 
at the beginning of the preceding subsection.
In order to construct such a description we are thus led to consider the Taylor expansions
of the single-variable functions at $s=s_0$. As before, the first task consists in dealing with the 
ambiguities the latter are beset with. The construction of the appropriate Taylor invariants proceeds 
as previously, but starting now from the Taylor expansions of the fundamental functions at $s=s_0$.
We denote the corresponding Taylor coefficients with lower-case latin letters, i.e. in equations
\rf{Taylor_sol_1} and \rf{Taylor_sol_2} we replace $s$ by $s-s_0$ and at the same time $A_I^{(n)}$ 
by $a_I^{(n)}$, and so on. The values of these coefficients are again provided by 
our numerical determination of the fundamental functions in eq. \rf{gen_sol}.
For our purposes, the relevant Taylor invariants we can construct read\footnote{Any 
combination of Taylor invariants is also a Taylor invariant. The reason for making these specific choices 
and for denoting the Taylor invariants with these particular set of barred Greek letters will become clear soon.}
\bea\lbl{Taylor_coeffs_s0}
&& 
\sum_{n=0}^3 \left(\!\!
\begin{tabular}{c}
 $\mu_n$ \\
 $\nu_n$
\end{tabular}
\!\!\right)
\left[ - a_0^{(n)} - \frac{4}{3} a_2^{(n)}\right]
\equiv
\left(\!\!\!
\begin{tabular}{c}
${\bar\alpha}_1$ \\[0.15cm]
$-{\bar\alpha}_3/2$
\end{tabular}
\!\!\right)      ,
\nonumber\\ 
&&
\sum_{n=0}^3 \left(\!\!
\begin{tabular}{c}
 $\mu_n$ \\
 $\nu_n$
\end{tabular}
\!\!\right) 
\left[ 3 a_1^{(n)} + b_0^{(n)} - \frac{5}{3} b_2^{(n)} \right] M_\pi^2
\equiv
\left(\!\!\!
\begin{tabular}{c}
${\bar\beta}_1$ \\[0.15cm]
$-{\bar\beta}_3/2$
\end{tabular}
\!\!\right)      ,
\nonumber\\ 
&&
\sum_{n=0}^3 \left(\!\!
\begin{tabular}{c}
 $\mu_n$ \\
 $\nu_n$
\end{tabular}
\!\!\right) 
\left[ - c_0^{(n)} - \frac{4}{3} c_2^{(n)} \right] \frac{M_\pi^4}{2}
\equiv
\left(\!\!\!
\begin{tabular}{c}
${\bar\zeta}_1$ \\[0.15cm]
${\bar\zeta}_3$
\end{tabular}
\!\!\right)    ,
\nonumber\\ 
&&
\sum_{n=0}^3 \left(\!\!
\begin{tabular}{c}
 $\mu_n$ \\
 $\nu_n$
\end{tabular}
\!\!\right)  
\left[ - c_0^{(n)} + \frac{5}{3} c_2^{(n)} + 3 b_1^{(n)} \right] \frac{M_\pi^4}{2} 
\equiv
\left(\!\!\!
\begin{tabular}{c}
${\bar\xi}_1$ \\[0.15cm]
${\bar\xi}_3$
\end{tabular}
\!\!\right)     ,
\nonumber\\
\\
&&
\sum_{n=0,1} {\tilde\nu}_n [ 3 {\tilde a}^{(n)}_1 - {\tilde b}^{(n)}_2 ]  \frac{\sqrt{3}}{2} M_\pi^2 \equiv {\bar\gamma}_3    ,
\nonumber\\
&&
\sum_{n=0,1} {\tilde\nu}_n [ - 3 {\tilde b}^{(n)}_1 - {\tilde c}^{(n)}_2 ] \frac{3}{4} M_\pi^4  \equiv {\bar\xi}_3'       .
\nonumber
\eea
Let us observe in passing that again ${\tilde\mu}_1$ is not involved in these Taylor invariants.

The next step consists in expressing the physical amplitudes as polynomials in the variables $s_i-s_0$ 
truncated at second order, with coefficients given by the Taylor invariants. This calculation is straightforward, 
starting from eq. \rf{decay_amps}, then inserting the Taylor expansions at $s=s_0$ of the single-particle functions \rf{gen_sol}, 
and finally checking that the various Taylor coefficients indeed combine to yield the Taylor invariants of eq. \rf{Taylor_coeffs_s0}. 
The result of these manipulations reads
\bea\lbl{Dalitz}
{\cal A}^{K_L\to\pi^0\pi^0\pi^0}(s_1,s_2,s_3) &=& 
- 3 ( {\bar\alpha}_1 + {\bar\alpha}_3 ) - 3 ( {\bar\zeta}_1 - 2 {\bar\zeta}_3 )  \!\left( Y^2 + \frac{X^2}{3} \right)
,
\nonumber\\
{\cal A}^{K_L\to\pi^+\pi^-\pi^0}(s_1,s_2,s_3) &=& 
{\bar\alpha}_1 + {\bar\alpha}_3 - ( {\bar\beta}_1 + {\bar\beta}_3 ) Y
+ ( {\bar\zeta}_1 - 2 {\bar\zeta}_3 ) \!\left( Y^2 + \frac{X^2}{3} \right)\!
+ ( {\bar\xi}_1 - 2 {\bar\xi}_3 ) \!\left( Y^2 - \frac{X^2}{3} \right)\!      ,
\nonumber\\
{\cal A}^{K_S\to\pi^+\pi^-\pi^0}(s_1,s_2,s_3) &=& 
\frac{2}{3} \sqrt{3} {\bar\gamma}_3 X - \frac{4}{3} {\bar\xi}_3 { }\!\!' \,X Y   ,
\\
{\cal A}^{K^+\to\pi^+\pi^+\pi^-}(s_1,s_2,s_3) &=& 
2 {\bar\alpha}_1 - {\bar\alpha}_3 + ( {\bar\beta}_1 - \frac{{\bar\beta}_3}{2} + \sqrt{3} {\bar\gamma}_3 ) Y
+ 2 ( {\bar\zeta}_1 + {\bar\zeta}_3 ) \!\left( \! Y^2 + \frac{X^2}{3} \right)\!
- ( {\bar\xi}_1 + {\bar\xi}_3 - {\bar\xi}_3 ' ) \!\left( \! Y^2 - \frac{X^2}{3} \right)\!
,
\nonumber\\
{\cal A}^{K^+\to\pi^0\pi^0\pi^+}(s_1,s_2,s_3) &=&
- ( {\bar\alpha}_1 - \frac{{\bar\alpha}_3}{2} ) + ( {\bar\beta}_1 - \frac{{\bar\beta}_3}{2} - \sqrt{3} {\bar\gamma}_3 ) Y
- ( {\bar\zeta}_1 + {\bar\zeta}_3 ) \!\left( \! Y^2 + \frac{X^2}{3} \right)\!
- ( {\bar\xi}_1 + {\bar\xi}_3 + {\bar\xi}_3 ' )\!\left( \! Y^2 - \frac{X^2}{3} \right)\!   .
\nonumber
\eea
These expressions reproduce the well known Dalitz expansions of the $K\to3\pi$
amplitudes, implementing the constraints coming from isospin conservation
\cite{Dalitz:1956da,Weinberg:1960zza,Barton:1963mg,Zemach:1963bc,Li:1979wa}
(we follow here the notation of~\cite{Li:1979wa,Kambor:1991ah}). The expansion
parameters which appear in eq. \rf{Dalitz} are complex numbers (even when CP is conserved).
Several approximations to their imaginary parts have been considered in the literature,
for instance one-loop results~\cite{DAmbrosio:1994vba} obtained in the low-energy expansion. 
But most of the time these imaginary parts have simply been discarded in fits to the experimental
data relying on the expansions \rf{Dalitz}~\cite{Devlin:1978ye,Kambor:1991ah,Bijnens:2002vr,DAmbrosio:2022jmd}. 
Our solutions of the Omn\`es-Khuri-Treiman equations instead allow one to use the expressions \rf{Dalitz}
together with realistic estimates of the imaginary parts. These are simply obtained from
eqs. \rf{Taylor_coeffs_s0}. Indeed, the linear relations which we have obtained between the
real and imaginary parts of the subtraction parameters $\mu_n$, $\nu_n$, 
$\tilde{\nu}_n$ (via a matching to the low-energy expansion around $s=0$)
translate, via eqs. \rf{Taylor_coeffs_s0}, into a set of linear relations 
expressing the imaginary parts of the parameters $\bar{\alpha}_i$,
$\bar{\beta}_i$, $\bar{\zeta}_i$, $\bar{\xi}_i$, $\bar{\gamma}_3$,
$\bar{\xi}'_3$ in terms of their real parts.  
We can then determine the latter by performing 
a fit to the 16 observables listed in Table \ref{tab:fitsresults} starting directly from the expressions \rf{Dalitz} 
of the amplitudes, in a similar manner as done for the fit carried out in the case where the amplitude coefficients are assumed 
to be strictly real, as described for instance in ref \cite{DAmbrosio:2022jmd}. 

The outcome of this fit (called `$\alpha$-fit' to differentiate it from the fit described in subsection \ref{subsec:fit}) is shown in 
Table  \ref{tab:alpha_values}.  
For the sake of comparison, we also show the values of the real amplitude parameters displayed in Table 5
of ref. \cite{DAmbrosio:2022jmd} together with their imaginary parts obtained from a one-loop 
calculation  \cite{DAmbrosio:1994vba,Bijnens:2002vr}.
One observes that the imaginary parts of the 
subtraction constants are generically much smaller (in absolute values) than their real parts,
but that there are substantial differences between their one-loop values and the values 
generated by the solutions of the Omn\`es-Khuri-Treiman equations, showing the effect of 
the resummation of final-state interactions at work in the dispersive approach.
As a side remark, let us mention that determinations of $\bar{\alpha}_1$ and $\bar{\beta}_1$, based on an approximate
solution of the Khuri-Treiman equations (cf. footnote \ref{footnote}), were also obtained by the 
authors of ref. \cite{Kambor:1993tv}.
The value they give for $\bar{\alpha}_1$ is close to the one shown in the central column of Table \ref{tab:alpha_values},
whereas their imaginary part for $\bar{\beta}_1$ remains closer to the one-loop value.

The penultimate column of Table \ref{tab:fitsresults} shows the values obtained for the various observables 
from the $\alpha$-fit. They are close to the values obtained from the direct fit described in subsection \ref{subsec:fit},
with a comparable value of the total $\chi^2$ value, $\chi^2_{\rm tot}=17.6$ (for the same number of degrees of freedom), but the 
contributions from the individual observables to the total $\chi^2$ value are somewhat different.
Likewise, evaluating the complex amplitude parameters 
${\bar\alpha}_{1,3}, {\bar\beta}_{1,3}, {\bar\gamma}_3, {\bar\zeta}_{1,3}, {\bar\xi}_{1,3}, {\bar\xi}_3'$
directly from their definitions \rf{Taylor_coeffs_s0} and from the fitted values of the subtraction 
constants $\mu_n$, $\nu_n$, ${\tilde\nu}_n$ given in Table \ref{tab:subtractions} would lead to values close to the ones obtained from the $\alpha$-fit.

\begin{table}[hb]
\centering  
\begin{tabular}{|c|| r || r|} \hline
$\bar{\alpha}_1$ &$ 91.02(22)  +i18.47(4)$ & $92.87(23)+i12.07(3)$\\
$\bar{\beta}_1 $ & $-23.73(16)  -i16.32(4)$&$-26.50(16)-i12.39(3)$\\
$\bar{\zeta}_1 $ & $-0.33(2)   +i0.27(1)$& $-0.11(2)+i0.367(4)$ \\
$\bar{\xi}_1   $ &$-1.42(12)   -i0.50(1)$& $-1.35(12) -i0.191(1)$\\ \hline 
$\bar{\alpha}_3$ &$-6.88(27)   -i1.40(5)$& $-7.13(27) -i0.93(4)$\\
$\bar{\beta}_3 $ & $-2.79(24)   +i0.92(4)$& $-2.53(24) +i0.74(3)$\\
$\bar{\zeta}_3 $ & $0.014(15)  -i0.06(1)$&$-0.05(2) -i0.041(3)$ \\
$\bar{\xi}_3   $ &$0.11(7)   -i0.049(4)$ &$ 0.03(7) -i0.018(1)$ \\ \hline 
$\bar{\gamma}_3$ &$2.88(9)   -i0.102(3)$ &$ 2.80(9) -i0.132(4)$ \\
$\bar{\xi}'_3  $ &$0.59(16)   -i0.088(4)$ &$-0.30(16) -i0.096(3)$\\ \hline 
\end{tabular}
\caption{The values, in units of $10^{-8}$, of the amplitudes parameters of eq. \rf{Dalitz} as obtained from 
the $\alpha$-fit (see main text).
For comparison the last column shows the values obtained in ref. \cite{DAmbrosio:2022jmd} from a fit to the experimental data
shown in Table \ref{tab:fitsresults} assuming the amplitude coefficients to be real, 
and to which the 
imaginary parts computed in the one-loop approximation \cite{DAmbrosio:1994vba,Bijnens:2002vr} have been added
afterwards.}
\label{tab:alpha_values}
\end{table}

\subsection{The case of ${\tilde\mu}_1$}\label{subsec:mu1}

The sole subtraction constant that remains undetermined so far is ${\tilde\mu}_1$. It appears only 
in the single-variable function ${\tilde M}_1 (s)$, for which it provides the normalization.
This function by itself accounts for the whole $\Delta I = 1/2$ contribution to the 
$K_S\to\pi^+\pi^-\pi^0$ amplitude (if higher partial waves are neglected), but this contribution starts only at the level of cubic 
terms in the expansion of the amplitudes around the point $s_i=s_0$: $K_S\to\pi^+\pi^-\pi^0$ is 
the only decay mode whose amplitude, up to quadratic order in $X$ and $Y$, receives contributions from the $\Delta I = 3/2$
transition alone. While cubic terms are suppressed in the low-energy expansion,
contributions with $\Delta I = 3/2$ are expected to be suppressed with respect to the
ones with $\Delta I = 1/2$. It seems thus reasonable to assume that the $\Delta I = 1/2$
cubic term could contribute to the decay rate on par with the $\Delta I = 3/2$ linear and quadratic terms.
With this term added, the amplitude reads (the Taylor coefficient ${\tilde c}_1$ is a Taylor invariant,
of dimension GeV$^{-2}$)
\be
{\cal A}^{K_S\to\pi^+\pi^-\pi^0}(s_1,s_2,s_3) =
\frac{2}{3} \sqrt{3} {\bar\gamma}_3 X - \frac{4}{3} {\bar\xi}_3 { }\!\!' \,X Y
+ \frac{1}{4} {\tilde\mu}_1 {\tilde c}_1 M_\pi^6 X (X^2 - 9 Y^2) + \cdots   .
\lbl{K_S_extended}
\ee
The ellipsis indicates cubic terms from the $\Delta I = 3/2$ transition Hamiltonian,
which should be suppressed as compared to the term proportional to ${\tilde\mu}_1$
that has been retained. The whole available experimental information on the $K_S\to\pi^+\pi^-\pi^0$ 
decay mode is contained in the overlap parameter $\lambda$, which also plays a role in the determination 
of the two subtraction constants ${\tilde\nu}_{0}$ and ${\tilde\nu}_{1}$.  Keeping also the third term in \rf{K_S_extended},
we obtain ($\lambda$ is without dimension and the numerical coefficients are expressed in
appropriate powers of GeV according to Table \ref{dimensions})
\be\lbl{eq:lambexpansion}
10^{4}\lambda=\tilde{\nu}_0(6.64-i1.66)+\tilde{\nu}_1(0.86-i0.22)
+\tilde{\mu}_1 (-0.84-i0.22) \cdot 10^{-3} \ .     
\ee
We see that the contribution proportional to
$\tilde{\mu}_1$ is extremely suppressed. This suppression can actually be traced back to the presence 
of zeros in the integrand that defines $\lambda$ in eq.~\rf{eq:lambda}. This is the reason why the fit 
in subsection \ref{subsec:fit} remains, for all practical purposes, silent about $\tilde{\mu}_1$.
In order for the contribution proportional to $\tilde{\mu}_1$ to be comparable with the experimental
error on $\lambda$ we would need to have $\tilde{\mu}_1\sim 200 \tilde{\nu}_1$ (i.e. ${\tilde\mu}_1\sim 8\cdot 10^4$ GeV$^{-4}$). 
Assuming a usual size for the $\Delta{I}=1/2$ enhancement, we would rather expect $\tilde{\mu}_1$ to be larger in magnitude 
than $\tilde{\nu}_1$ by a factor $10-20$. An increase in precision by a factor of ten, 
at least, in $\lambda$ would then be necessary in order for this observable to become eventually sensitive 
to the value of $\tilde{\mu}_1$.

\subsection{Strong phases in the $K\to\pi\pi\pi$ decay amplitudes}

\begin{figure}[ht]
\centering
\includegraphics[width=0.45\linewidth]{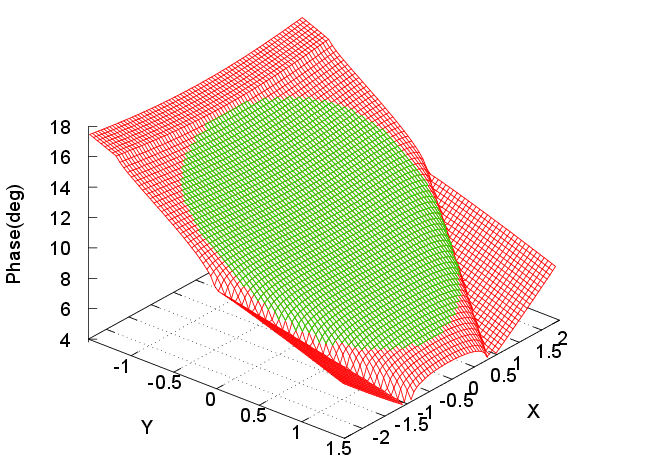}\includegraphics[width=0.45\linewidth]{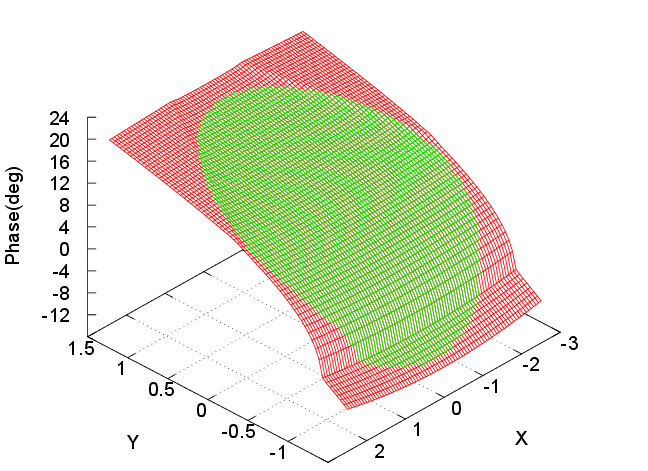}
\caption{Strong phases (in degrees) of the $K^+\to\pi^+\pi^+\pi^-$ amplitude (left)
  and the $K^+\to\pi^0\pi^0\pi^+$ amplitude (right) derived from the solutions of the Omn\`es-Khuri-Treiman
equations. The areas in green correspond to the physical decay regions.}
\label{fig:K+phases}
\end{figure}

Finally, let us close this section by recalling that the knowledge of rescattering phases generated by the strong-interaction
dynamics is an important input in order to predict the size of direct violations of CP invariance in
$K\to\pi\pi\pi$ decays, which are expected to be quite small 
\cite{Grinstein:1985ut,DAmbrosio:1991oli,Isidori:1991xx,DAmbrosio:1994vba}.
In these studies, the input for the strong phases was limited to values generated at one loop in the 
low-energy expansion.
Our solutions of the Omn\`es-Khuri-Treiman equations provide a more precise
determination of the strong phases of the $K\to\pi\pi\pi$ amplitudes covering the
whole Dalitz plot, which could be useful for estimating the size of local CP-violation
effects. We do not wish to enter a full discussion of this interesting topic here,
but simply illustrate, in fig.~\ref{fig:K+phases}, our results for the local phases generated 
by the elastic $\pi\pi$ rescattering in the case of the two charged kaon decay
amplitudes for $K^+\to\pi^+\pi^+\pi^-$ and $K^+\to\pi^0\pi^0\pi^+$.

\section{Summary and conclusions}\label{sec:conclusions}
\setcounter{equation}{0}

In the present work, we have implemented elastic pion-pion final-state rescattering for the amplitudes
$K\pi\to\pi\pi$ within a dispersive framework of the Omn\`es-Khuri-Treiman-Sawyer-Wali type under the assumptions 
that invariance under both CP and isospin symmetry is preserved. Starting from the isospin decomposition
\textit{\`a la} Wigner-Eckart of the various amplitudes, we have identified, upon imposing the additional
restrictions provided by crossing, a set of four invariant amplitudes that allows us to describe all the 
thirteen initial scattering amplitudes. Assuming furthermore that the imaginary parts of the partial waves higher 
than the P wave can be neglected, we have provided a decomposition of the invariant amplitudes in terms
of a set of nine single-variable functions, see eq. \rf{decay_amps}. 

The single-variable functions are analytic in the complex plane except for a right-hand cut starting at $s=4M_\pi^2$ 
and whose discontinuity is given by unitarity, here restricted to two-pion states. 
As a consequence of crossing, this ensures that the $K\pi\to\pi\pi$ amplitudes
have the correct analyticity properties in the region of the Mandelstam plane below the inelastic thresholds.
The single-variable functions thus obey a set of subtracted dispersion relations where the only input are the elastic $\pi\pi$ 
phase shifts for the S and P waves, which are well known. Using these phase shifts, we have solved these dispersion 
relations numerically, and obtained the solutions in terms of linear combinations of a set of fundamental functions
modulated by the subtraction constants. The latter have been determined, apart from one, by matching the amplitudes constructed 
this way with available experimental data on the $K\to\pi\pi\pi$ decay modes, assuming that the imaginary parts of the lowest
Taylor invariants of the single-variable functions vanish at the origin. The 17 independent fundamental functions, which do not 
depend on the subtraction constants, have been collected, in numerical form, in a series of ancillary files.

We have discussed a few applications of our results to the phenomenology of $K\to\pi\pi\pi$ decays.
We have reconsidered the description of the decay amplitudes as second-order polynomials in the
Dalitz-plot variables from the point of view of the dispersive approach and have shown that the 
corresponding amplitude coefficients, usually assumed to be real, have generically larger 
imaginary parts than those generated at the one-loop level. We have also determined the strong 
phases of the decay amplitudes of the charged kaon locally in the Dalitz plot, which can be useful for the study 
of possible direct violations of CP in the $K\to\pi\pi\pi$ decays.

Other approaches 
to determine the values of the subtraction constants could be considered \cite{Colangelo:2018jxw}. 
For instance one could make use of the behaviour of the single-variable functions in the vicinity
of the cusp at $s=4M_\pi^2$, where a non-relativistic description in terms of a power expansion 
with respect to the variable $q=\sqrt{s-4M_\pi^2}$ can be used. This can then be compared to the
expressions obtained within an effective-theory framework in refs. \cite{Colangelo:2006va,Bissegger:2007yq,Gasser:2011ju}.
So far, only the amplitudes for the charged kaon and the long-lived neutral kaon have 
been worked out within this framework, but a similar treatment of
the amplitude for the decay mode $K_S \to \pi^+ \pi^- \pi^0$ is missing. Therefore, a 
determination of all the subtraction constants (and in particular of ${\tilde\mu}_1$)
is probably not possible by following this approach at the moment. In any case, the single-variable functions
corresponding to alternative determinations of the subtraction constants can easily be reconstructed
from the expressions given in eq. \rf{gen_sol} and the fundamental functions
tabulated in the ancillary files. 

The results obtained in this work fill a gap in the recent literature to the extent that a full 
dispersive treatment of final-state interactions, albeit restricted to the elastic regime and to the absence
of isospin-breaking effects, for all $K\to\pi\pi\pi$ amplitudes was not available before our work. The main objective 
of this study, however, is to prepare the field for applications to a class of radiative kaon-decay modes, 
for which more precise experimental results are expected in the future. This issue is left for forthcoming work.

\vfill\newpage

\appendix

\def\theequation{\Alph{section}.\arabic{equation}}

\section{Restrictions on the isospin amplitudes from the crossing properties}\label{app:crossing}
\setcounter{equation}{0}

In this appendix, we show how the crossing properties \rf{crossing_isospin-amplitudes} of the 
nine isospin amplitudes in eqs. \rf{isospin_amplitudes_1} and \rf{isospin_amplitudes_3}
lead to the general representations \rf{inv_amps_1} and \rf{inv_amps_3} in 
terms of only four independent functions. In the next appendix we provide an 
alternative derivation using the so-called spurion formalism.

The situation in the case of amplitudes for $K\pi\to\pi\pi$ scattering
is slightly more complicated, from an algebraic point of view, than 
in the case of elastic pion-pion scattering. This is mainly due to
the increased number of isospin amplitudes, although the amplitudes
${\vec{\cal M}}_1(s,t,u)$ and ${\vec{\cal M}}_3(s,t,u)$ can be discussed 
separately, which is what we will do.

\indent

Starting with the $\Delta I = 1/2$ amplitudes ${\vec{\cal M}}_1(s,t,u)$
we define the function
\be
{\cal M}_1 (s \vert t,u) = - {\cal A}_4 (s,t,u) 
= - \frac{1}{3} {\cal M}_1^{2;\frac{3}{2}} (s,t,u) + \frac{1}{3} {\cal M}_1^{0;\frac{1}{2}} (s,t,u)  ,
\ee
where it is understood that in this paragraph ${\cal A}_i (s,t,u)$ refers 
only to the $\Delta I = \frac{1}{2}$ component of the corresponding amplitude,
related to the isospin amplitudes through the matrix ${\mathbf{R}}_1$ in eq. \rf{amplitudes_iso}.
Since ${\cal A}_4 (s,t,u)$ corresponds to an amplitude with two neutral pions in the final state,
Bose symmetry requires
\be
{\cal M}_1 (s \vert t,u) = {\cal M}_1 (s \vert u,t)
\ee
while crossing gives
\be
{\cal A}_5 (s,t,u) = - {\cal A}_4 (u,t,s) = {\cal M}_1 (u \vert s,t)  ,
\ee
In addition, from the relations
\be
{\cal M}_1^{2;\frac{3}{2}} (s,t,u) = {\cal A}_1 (s,t,u) = {\cal A}_2 (s,t,u) + {\cal A}_4 (s,t,u) + {\cal A}_5 (s,t,u)
\ee
and ${\cal A}_2 (s,t,u)= {\cal A}_1 (u,t,s)$, one deduces that
\be
{\cal A}_2 (s,t,u) - {\cal A}_2 (u,t,s) = {\cal M}_1 (s \vert u,t) - {\cal M}_1 (u \vert s,t)
\lbl{constraint_A2}
\ee
Next, from the ${\mathbf{R}}_1$ part of eq. \rf{amplitudes_iso} we establish the relations
\be
\frac{2}{3} {\cal M}_1^{2;\frac{3}{2}} (s,t,u) + \frac{1}{3} {\cal M}_1^{0;\frac{1}{2}} (s,t,u) =
\sqrt{2} {\cal A}_{13} (s,t,u) = {\cal A}_2 (s,t,u) + {\cal A}_5 (s,t,u) = {\cal A}_1 (s,t,u) - {\cal A}_4 (s,t,u)  .
\ee
But ${\cal A}_{13} (s,t,u)$ has to be fully symmetric under permutations of the variables $s$, $t$, $u$. Therefore one obtains
\be
{\cal A}_{2} (s,t,u) = {\cal M}_1 (s \vert u,t) + {\cal M}_1 (t \vert s,u) + {\widetilde{\cal M}} (s,t,u)
\ee
where ${\widetilde{\cal M}} (s,t,u)$ has the same symmetry properties as ${\cal A}_{13} (s,t,u)$. This expression  
of ${\cal A}_{2} (s,t,u)$ also satisfies the previous condition \rf{constraint_A2}. Thus, from the analysis involving 
so far only the amplitudes with a charged kaon and the fully symmetric amplitude ${\cal A}_{13} (s,t,u)$, we have 
obtained
\be
\sqrt{2} {\cal A}_{13} (s,t,u) = {\cal M}_1 (s \vert u,t) + {\cal M}_1 (t \vert s,u) + {\cal M}_1 (u \vert s,t) + {\widetilde{\cal M}} (s,t,u)
\ee
and 
\bea 
{\cal M}_1^{2;\frac{3}{2}} (s,t,u) &=&  {\cal M}_1 (t \vert s,u) + {\cal M}_1 (u \vert s,t) + {\widetilde{\cal M}} (s,t,u) 
\nonumber\\
{\cal M}_1^{0;\frac{1}{2}} (s,t,u) &=&  {\cal M}_1 (t \vert s,u) + {\cal M}_1 (u \vert s,t) + 3 {\cal M}_1 (s \vert u,t) + {\widetilde{\cal M}} (s,t,u) 
\nonumber\\
\frac{1}{3} {\cal M}_1^{1;\frac{3}{2}} (s,t,u) + \frac{2}{3} {\cal M}_1^{1;\frac{1}{2}} (s,t,u) &=&
{\cal M}_1 (t \vert s,u) - {\cal M}_1 (u \vert s,t) + {\widetilde{\cal M}} (s,t,u)   .
\eea
The left-hand side of the last relation, which follows from the expression of
${\cal A}_2 (s,t,u)$ in terms of the isospin amplitudes ${\vec{\cal M}}_1 (s,t,u)$ and the two preceding relations,
involves a combination of two amplitudes where the final pions
are in an isospin state $I=1$. Bose symmetry then requires this combination, and hence the combination
occuring on the right-hand side of this relation, to be antisymmetric under exchange of $t$ and $u$.
This immediately implies ${\widetilde{\cal M}} (s,t,u)=0$. Looking now at the amplitudes involving 
the neutral kaon, one finds the relation
\be
\frac{1}{\sqrt{2}} {\cal M}_1^{2;\frac{3}{2}} (s,t,u) = {\cal A}_{13} (s,t,u) + {\cal A}_{7} (s,t,u)
- \frac{1}{2} {\cal A}_{9} (s,t,u) + \frac{1}{2} {\cal A}_{11} (s,t,u)
\ee
This means
\be 
{\cal A}_{7} (s,t,u) + \frac{1}{2} {\cal A}_{7} (u,t,s) - \frac{1}{2} {\cal A}_{7} (u,s,t) =
- \frac{1}{\sqrt{2}} {\cal M}_1 (s \vert t,u)  ,
\ee
from which one deduces that 
\be
{\cal A}_{7} (s,t,u) = - \frac{1}{\sqrt{2}} [ {\cal M}_1 (s \vert t,u) + {\cal N}_1 (s,t,u) ]
\ee
with $2 {\cal N}_1 (s,t,u) + {\cal N}_1 (u,t,s) - {\cal N}_1 (u,s,t) = 0$, which in turn implies
that ${\cal N}_1 (s,t,u) = - {\cal N}_1 (t,s,u)$. Another relation,
\be
\frac{1}{\sqrt{2}} {\cal M}_1^{1;\frac{3}{2}} (s,t,u) = 
{\cal A}_{13} (s,t,u) + {\cal A}_{7} (s,t,u) - \frac{1}{2} {\cal A}_{9} (s,t,u) - \frac{3}{2} {\cal A}_{11} (s,t,u)  ,
\ee
leads to
\be
{\cal M}_1^{1;\frac{3}{2}} (s,t,u) = {\cal M}_1 (t \vert s,u) - {\cal M}_1 (u \vert s,t) - 2 {\cal N}_1 (u,s,t)
\ee
and implies ${\cal N}_1 (s,t,u) = - {\cal N}_1 (u,t,s)$. Finally, from the expression of ${\cal A}_7 (s,t,u)$
in terms of the isospin amplitudes and the relations obtained so far, one deduces that
\be
{\cal M}_1^{1;\frac{3}{2}} (s,t,u) - {\cal M}_1^{1;\frac{1}{2}} (s,t,u) = - 3 {\cal N}_1 (s,t,u)
\ee
and therefore that ${\cal N}_1 (s,t,u)=-{\cal N}_1 (s,u,t)$. Thus, eventually ${\cal N}_1 (s,t,u)$
turns out to be antisymmetric under a permutation of any two of its arguments. Putting everything 
discussed in this paragraph together then allows to reproduce the relations displayed in eq. \rf{inv_amps_1}.

We now turn to the relations \rf{inv_amps_3}, where it is understood that in this paragraph 
${\cal A}_i (s,t,u)$ now refers only to the $\Delta I = \frac{3}{2}$ component of the corresponding 
amplitude. We start again with the definition
\be
{\cal M}_3 (s \vert t,u) = - {\cal A}_4 (s,t,u) 
= \frac{1}{3} {\cal M}_3^{2;\frac{3}{2}} (s,t,u) - \frac{2}{3} {\cal M}_3^{2;\frac{1}{2}} (s,t,u)
+ \frac{1}{3} {\cal M}_3^{0;\frac{1}{2}} (s,t,u)   
\ee
with ${\cal M}_3 (s \vert t,u) = {\cal M}_3 (s \vert u,t)$. From the decomposition
of the amplitudes ${\cal A}_i(s,t,u)$ in terms of the isospin amplitudes one obtains
then the following relation  
\be
{\cal A}_7 (s,t,u) = \sqrt{2} {\cal M}_3 (s \vert t,u) + {\cal A}_7 (u,t,s) - {\cal A}_7 (u,s,t) .
\ee
This relation in turn implies that 
\be
{\cal A}_7 (s,t,u) = \sqrt{2} {\cal M}_3 (s \vert t,u) + \sqrt{2} {\cal N}_3 (u,t,s)
\ee
with
\be
{\cal N}_3 (s,t,u) = - {\cal N}_3 (s,u,t) ,~~~~~
{\cal N}_3 (s,t,u) + {\cal N}_3 (t,u,s) + {\cal N}_3 (u,s,t) = 0.
\ee
Likewise, one has
\be
{\cal A}_1(s,t,u) + {\cal A}_2(s,t,u) - {\cal A}_4(s,t,u) - \sqrt{2} {\cal A}_9(s,t,u)
+ \frac{1}{\sqrt{2}} {\cal A}_{11}(s,t,u) + \sqrt{2} {\cal A}_{13}(s,t,u) = 0  ,
\ee
which rewrites as
\be
\sqrt{2} {\cal A}_{13}(s,t,u) = - {\cal A}_2(s,t,u) - {\cal A}_2(u,t,s)
- {\cal M}_3 (s \vert t,u) - {\cal N}_3 (s,t,u)
- {\cal M}_3 (u \vert t,s) - {\cal N}_3 (u,t,s)   .
\ee
Adding to this relation the two other ones that are obtained by cyclic
permutations of the variables $s$, $t$, $u$, one finds
\be
3 \sqrt{2} {\cal A}_{13}(s,t,u) = - 2 [ {\cal A}_2(s,t,u) + {\cal A}_2(u,s,t) + {\cal A}_2(t,u,s)
+ {\cal M}_3 (s \vert t,u) + {\cal M}_3 (t \vert s,u) + {\cal M}_3 (u \vert s,t) ]  .
\ee
As a last relation, we take
\be
{\cal A}_1 (s,t,u) + \sqrt{2} {\cal A}_{11} (s,t,u) = {\cal M}_3^{1;\frac{3}{2}} (s,t,u) .
\ee
Upon symmetrizing with respect to $t$ and $u$, taking into account that
${\cal M}_3^{1;\frac{3}{2}} (s,t,u) + {\cal M}_3^{1;\frac{3}{2}} (s,u,t)=0$, it gives
\bea
{\cal M}_3^{2;\frac{3}{2}} (s,t,u) &=&
{\cal A}_1 (s,t,u) = - \frac{1}{\sqrt{2}} [ {\cal A}_{11} (s,t,u) + {\cal A}_{11} (s,u,t) ] =
\frac{1}{\sqrt{2}} [ {\cal A}_{7} (u,s,t) + {\cal A}_{7} (t,s,u) ] 
\nonumber\\
&=& {\cal M}_3 (u \vert s,t) + {\cal N}_3 (u,s,t) + {\cal M}_3 (t \vert s,u) + {\cal N}_3 (t,s,u)  .
\eea
The expressions obtained for ${\cal A}_1 (s,t,u)$, ${\cal A}_4 (s,t,u)$, ${\cal A}_7 (s,t,u)$
and of the amplitudes related to them by crossing allow to generate five independent equations 
involving the isospin amplitudes ${\vec{\cal M}}_3 (s,t,u)$ and the functions ${\cal M}_3$, 
${\cal N}_3$ that can then be solved. The result of this straightforward exercice is given by 
the relations \rf{inv_amps_3}.

\section{Invariant amplitudes from the spurion formalism}\label{app:spurions}
\setcounter{equation}{0}

The decomposition of the $K\pi\to\pi\pi$ amplitudes into invariant amplitudes can also be obtained
from the spurions formalism \cite{Bell:1966,Bouchiat:1967wjs,Kambor:1992he}. This amounts to describing the isospin-changing nature
of the weak interactions by introducing fictituous particles having the same
quantum numbers as the weak Hamiltonian. In particular, these spurions have zero
momentum. They correspond to an iso-spinor ${\rm s}_{1/2}$ in the case of the $\Delta I =1/2$ 
transitions, and to an isospin vector-spinor ${\rm s}_{3/2}^a$ in the case of the $\Delta I = 3/2$ 
transitions. In the latter case, the unwanted isospin $1/2$ component in the
decomposition $\frac{1}{2} \otimes 1 = \frac{1}{2} \oplus \frac{3}{2}$ is projected out
through the condition $\tau^a {\rm s}^a_{3/2}=0$. One then considers processes like
$K\pi\to{\rm s}_{1/2}\pi\pi$ and $K\pi\to{\rm s}_{3/2}^a\pi\pi$, and their respective
isospin decompositions. Explicitly, one writes [$\alpha$, $\beta$... are indices 
in the fundamental representation of the $SU(2)$ isospin group]
\be
({\rm s}_{1/2})_\alpha = \Big( 
\!\!
\begin{tabular}{c}
 $\zeta_1$ \\
 $\zeta_2$
\end{tabular}
\!\!\Big)
,~
({\rm s}_{1/2})^\alpha = (\zeta_2 , - \zeta_1 ) 
,\quad
({\rm s}_{3/2}^a)_\alpha = \Big( 
\!\!
\begin{tabular}{c}
 $\xi^a_1$ \\
 $\xi^a_2$
\end{tabular}
\!\!\Big)   ,
\ee
for these spinors and for their $SU(2)$-conjugates. The constraint that enforces that
${\rm s}^a_{3/2}$ is a pure isospin $3/2$ object translates into the relations
\be
\xi^1_2 - i \xi^2_2 + \xi^3_1 = 0 , \quad \xi^1_1 + i \xi^2_1 - \xi^3_2 = 0
.
\lbl{one-half_proj}
\ee
The isospin structure of amplitudes for the processes $K \to \pi^a\pi^b$ 
or $K \to \pi^b \pi^c \pi^d$ are then expressed in terms of $SU(2)$-invariant
structures constructed with the $SU(2)$-invariant tensors $\delta^{ab}$, $\epsilon^{abc}$,
$(\tau^a)_\alpha^{\ \beta}$, and the spinors ${\rm s}_{1/2}$ and ${\rm s}_{3/2}^a$, multiplied
by Lorentz-invariant amplitudes.

For instance, in the case of the $K\to\pi\pi$ amplitudes, this procedure gives [generalized Bose symmetry 
has also to be taken into account]
\be
{\cal A} (K \to \pi^a\pi^b) = - \frac{1}{\sqrt{3}} \, A_0 \delta^{ab} ({\rm s}_{1/2})^\alpha K_\alpha 
- \frac{1}{\sqrt{6}} \, A_2 \left[ ({\rm s}_{3/2}^a)^\beta (\tau^b)_\beta^{\ \alpha} K_\alpha +
({\rm s}_{3/2}^b)^\beta (\tau^a)_\beta^{\ \alpha} K_\alpha \right]
,
\ee
where the spinor $K_\alpha$ describes the kaon states. For the physical processes one needs,
on the one hand, to identify the physical states in terms of isospin components and, on the other
hand, to project the spurions on their physical $I_3 = + 1/2$ components.
The first step is straightforward, and one has [these definitions correspond to the Condon and 
Shortley phase convention]
\be
\pi^0 = \pi^3 ,\quad \pi^\pm = \mp \frac{1}{\sqrt{2}} \big( \pi^1 \mp i \pi^2 \big)
,\quad
K^+ \to K_\alpha = \Big( 
\!\!
\begin{tabular}{c}
 1 \\
 0
\end{tabular}
\!\!\Big) 
, ~
K^0 \to K_\alpha= \Big( 
\!\!
\begin{tabular}{c}
 0 \\
 1
\end{tabular}
\!\!\Big)
.
\ee
For the second step, as far as the spurions ${\rm s}_{1/2}$ is concerned, one simply
takes [up to an overall normalization that is absorbed by the Lorentz-invariant amplitudes]
$\zeta_1=1$ and $\zeta_2=0$, i.e. $({\rm s}_{1/2})^\alpha = (0 , - 1 )$, so that
$\tau^3 {\rm s}_{1/2} = + {\rm s}_{1/2}$. In the case of  ${\rm s}_{3/2}^a$, the 
corresponding condition reads
\be
\left[ 
(T^3)^{ab} \delta_{\alpha}^{\ \beta} + \delta^{ab} \frac{(\tau^3)_\alpha^{\ \beta}}{2}
\right] ({\rm s}_{3/2}^b)_\beta = \frac{1}{2} ({\rm s}_{3/2}^a)_\alpha
,
\ee
where $T^a$ denote the generators of $SU(2)$ in the adjoint representation, i.e.
$(T^a)^{bc} = -i \epsilon^{abc}$. The above condition then gives $\xi^1_1 = \xi^2_1 = \xi^3_2 =0$
and $\xi^2_2 = i \xi^1_2$. Combined with the condition $\tau^a {\rm s}^a_{3/2}=0$, this
then gives [again, up to an overall normalization]
\be
({\rm s}_{1/2})^\alpha = (0,-1) , \quad ({\rm s}^a_{3/2})^\alpha = \Big( (\frac{1}{2} , 0) , (\frac{i}{2} , 0) , (0 , 1) \Big)
, \quad {\rm s}_{3/2}^a \tau^a = (0,0)       .
\ee
With these ingredients, one indeed recovers the usual isospin decomposition of the
$K\to\pi\pi$ amplitudes:
\be
{\cal A} (K^0 \to \pi^0\pi^0) = + \frac{1}{\sqrt{3}} \, A_0 + \sqrt{\frac{2}{3}} \, A_2
,\quad
{\cal A} (K^0 \to \pi^+\pi^-) = - \frac{1}{\sqrt{3}} \, A_0 + \frac{1}{\sqrt{6}} \, A_2
,\quad
{\cal A} (K^+ \to \pi^+\pi^0) = + \frac{\sqrt{3}}{2} \, A_2     .
\ee

The corresponding representation of the $K\to\pi\pi\pi$ amplitudes reads
[contracted spinor indices are not shown]
\bea
{\cal A}^{K\to\pi^a\pi^b\pi^c}(s_a,s_b,s_c) &=& 
- \, \delta^{ab} ({\rm s}_{1/2} \tau^c K) {\cal F}_1 (s_a,s_b,s_c)
- \delta^{bc} ({\rm s}_{1/2} \tau^a K) {\cal F}_1 (s_b,s_c,s_a)
- \delta^{ca} ({\rm s}_{1/2} \tau^b K) {\cal F}_1 (s_c,s_a,s_b)
\nonumber\\
&&
+ \, i \epsilon^{abc} ({\rm s}_{1/2} K) {\cal G}_1 (s_a,s_b,s_c)
\nonumber\\
&&
+ \, \delta^{ab} ({\rm s}_{3/2}^c K) {\cal F}_3 (s_a,s_b,s_c)
+ \delta^{bc} ({\rm s}_{3/2}^a K) {\cal F}_3 (s_b,s_c,s_a)
+ \delta^{ca} ({\rm s}_{1/2}^b K) {\cal F}_3 (s_c,s_a,s_b)
\nonumber\\
&&
+ \, i \epsilon^{abd} \left( {\rm s}_{3/2}^c \tau^d K + {\rm s}_{3/2}^d \tau^c K \right) {\cal G}_3 (s_a,s_b,s_c)
\nonumber\\
&&
+ \, i \epsilon^{bcd} \left( {\rm s}_{3/2}^a \tau^d K + {\rm s}_{3/2}^d \tau^a K \right) {\cal G}_3 (s_b,s_c,s_a)
\nonumber\\
&&
+ \, i \epsilon^{cad} \left( {\rm s}_{3/2}^b \tau^d K + {\rm s}_{3/2}^d \tau^b K \right) {\cal G}_3 (s_c,s_a,s_b)
\eea
with
\be
{\cal F}_{1,3} (s_b,s_a,s_c) = + {\cal F}_{1,3} (s_a,s_b,s_c)
,\quad
{\cal G}_3 (s_b,s_a,s_c) = - {\cal G}_3 (s_a,s_b,s_c)
,
\ee
and the function ${\cal G}_1 (s_a,s_b,s_c)$ is antisymmetric under 
permutation of any two of its arguments.
The terms involving the amplitudes ${\cal F}_1 (s_a,s_b,s_c)$ and
${\cal G}_1 (s_a,s_b,s_c)$, proportional to
${\rm s}_{1/2}$, account for the $\Delta I = 1/2$ part of the transitions, whereas the $\Delta I = 3/2$
part is described by the two amplitudes ${\cal F}_3 (s_a,s_b,s_c)$ and
${\cal G}_3 (s_a,s_b,s_c)$, proportional to ${\rm s}_{3/2}$.

Two remarks need to be
made at this stage:
\begin{itemize}
 \item One would also expect a contribution of the type
\be
 i \epsilon^{abd} \left( {\rm s}_{3/2}^c \tau^d K - {\rm s}_{3/2}^d \tau^c K \right) {\,^{\mbox{\tiny[3]}}{\hat{\cal G}}} (s_a,s_b,s_c)
 + {\rm cyclic}
\ee
to ${\cal A}^{K^n\to\pi^a\pi^b\pi^c}(s_a,s_b,s_c)$. However, this contribution is redundant. 
Indeed, writing
\be
 \epsilon^{abd} \left( {\rm s}_{3/2}^c \tau^d  - {\rm s}_{3/2}^d \tau^c  \right)^\alpha
 = 
 - \delta^{ac} \epsilon^{bfg} ({\rm s}_{3/2}^f \tau^g )^\alpha
 + \delta^{bc} \epsilon^{afg} ({\rm s}_{3/2}^f \tau^g )^\alpha
 ,
\ee
and computing
\be
 \epsilon^{afg} ({\rm s}_{3/2}^f \tau^g )^\alpha
 = - i ({\rm s}_{3/2}^a)^\alpha
\ee
using the condition \rf{one-half_proj}, one recovers the structure
corresponding to the amplitude ${\cal F}_3 (s_a,s_b,s_c)$.
 \item A Schouten-type identity in isospin space combined with the condition \rf{one-half_proj} provides the identity
\be
\epsilon^{abd} \left( {\rm s}_{3/2}^c \tau^d  + {\rm s}_{3/2}^d \tau^c  \right)
+
\epsilon^{bcd} \left( {\rm s}_{3/2}^a \tau^d  + {\rm s}_{3/2}^d \tau^a  \right)
+
\epsilon^{cad} \left( {\rm s}_{3/2}^b \tau^d  + {\rm s}_{3/2}^d \tau^b  \right)
= 0
.
\ee
This means that one can shift the function ${\cal G}_3 (s_a,s_b,s_c)$ by a function 
$\Delta {\cal G}_3 (s_a,s_b,s_c)$, 
which is necessarily completely antisymmetric under any permutation of $s_a$, $s_b$ and $s_c$. Therefore,
\be
{\cal G}_3 (s_a,s_b,s_c) + {\cal G}_3 (s_b,s_c,s_a) + {\cal G}_3 (s_c,s_a,s_b)
\to
{\cal G}_3 (s_a,s_b,s_c) + {\cal G}_3 (s_b,s_c,s_a) + {\cal G}_3 (s_c,s_a,s_b)
+ 3 \Delta {\cal G}_3 (s_a,s_b,s_c)
,
\ee
and one can choose the function $\Delta {\cal G}_3 (s_a,s_b,s_c)$ such as to
enforce the condition
\be
{\cal G}_3 (s_a,s_b,s_c) + {\cal G}_3 (s_b,s_c,s_a) + {\cal G}_3 (s_c,s_a,s_b) = 0
.
\ee
\end{itemize}
Working out the different amplitudes explicitly allows to make the link with the amplitudes
${\cal M}_{1,3} (s \vert t,u)$ and ${\cal N}_{1,3} (s,t,u)$. This correspondence reads 
\bea
&&\hspace{2.0cm}
 {\cal M}_1 (s_c \vert s_a , s_b) = - \sqrt{2} \, {\cal F}_1 (s_a,s_b,s_c) ,  ~~~~~ 
{\cal N}_1 (s_c , s_a , s_b) = - \sqrt{2} \, {\cal G}_1 (s_a,s_b,s_c)  ,
\nonumber\\
\\
&&\hspace{-0.5cm}
{\cal M}_3 (s_c \vert s_a , s_b) = - \frac{1}{\sqrt{2}} {\cal F}_3 (s_a,s_b,s_c)   
 - \frac{3}{\sqrt{2}} \left[ {\cal G}_3 (s_c,s_a,s_b) + {\cal G}_3 (s_c,s_b,s_a) \right]   , ~~~~~
{\cal N}_3 (s_c,s_a,s_b) = - 3 \sqrt{2} \,  {\cal G}_3 (s_a,s_b,s_c)   .
\nonumber 
\eea

\section{Polynomial redefinitions of the single-variable functions}\label{app:polynomial}
\setcounter{equation}{0}

In this appendix, we provide some detail on the procedure that allows for the identification
of the set of arbitrary parameters that describe the ambiguity of the decomposition of the 
$K\pi\to\pi\pi$ amplitudes into single-variable functions. This procedure is analogous to
the one followed in ref. \cite{Colangelo:2018jxw} in the case of the $\eta\to\pi\pi\pi$ 
amplitudes, and we refer the reader to it for further details. The first step consists in
eliminating the variable $s$ in favour of the variables $t$ and $u$, now considered as independent.
We will work with the four independent functions ${\cal M}_{1,3} (s \vert t , u)$ 
and ${\cal N}_{1,3} (s,t,u)$, using their expressions in terms of single-variable functions, given 
in eq. \rf{iso_single}.

Starting with the $\Delta I = 1/2$
amplitudes, one establishes the following identities:
\bea
( \partial_t - \partial_u )^2 {\cal N}_1 (3s_0 - t - u , t,u)
\!\!&=&\!\!
\frac{4}{3} \left[ \partial_t M_1^{1 ; \frac{3}{2}} (t) - \partial_t M_1^{1 ; \frac{1}{2}} (t) \right]
-
\frac{4}{3} \left[ \partial_u M_1^{1 ; \frac{3}{2}} (u) - \partial_u M_1^{1 ; \frac{1}{2}} (u) \right]
\\
&&\!\!\!\!\!\!\!
- \, \frac{2}{3} (2u + t - 3 s_0) \left[ \partial_t^2 M_1^{1 ; \frac{3}{2}} (t) - \partial_t^2 M_1^{1 ; \frac{1}{2}} (t) \right] 
+
\frac{2}{3} (2t + u - 3 s_0) \left[ \partial_u^2 M_1^{1 ; \frac{3}{2}} (u) - \partial_u^2 M_1^{1 ; \frac{1}{2}} (u) \right]
,
\nonumber
\eea
and
\be
\partial_u \partial_t^2 ( \partial_t - \partial_u ) {\cal M}_1 (3s_0 - t - u \vert t,u)
= - \frac{2}{3} \left[ \partial_t^3 M_1^{1 ; \frac{3}{2}} (t) + 2 \partial_t^3 M_1^{1 ; \frac{1}{2}} (t) \right]
.
\ee
We perform now a polynomial shift on the single-variable functions and require that the amplitudes
${\cal N}_1 (3s_0 - t - u , t,u)$ and ${\cal M}_1 (3s_0 - t - u , t,u)$ remain the same under this shift.
The first identity tells us that, in order to cancel the terms on the right-hand side
that involve both $t$ and $u$, the combination
$\partial_t^2 \delta M_1^{1 ; \frac{3}{2}} (t) - \partial_t^2 \delta M_1^{1 ; \frac{1}{2}} (t)$
must be at most linear in $t$. The second identity requires that
$\delta M_1^{1 ; \frac{3}{2}} (t) + 2 \delta M_1^{1 ; \frac{1}{2}} (t)$ is
at most quadratic in $t$.  
The structure of ${\cal M}_1 (s \vert t,u)$ then requires that $\delta M_1^{2 ; \frac{3}{2}} (s)$
is at most cubic in $s$. One finally finds that the transformations allowed for the functions
$M_1^{I ; {\cal I}} (s)$ are as given in eq. \rf{transf_1}. 
These transformations depend on a total of eight arbitrary
parameters $a_1$, $b_1,\ldots$ $h_1$. 

In the case of the $\Delta I = 3/2$
amplitudes, the starting point is provided by similar identities, i.e.
\bea
( \partial_t - \partial_u )^2 {\cal N}_3 (3s_0 - t - u , t,u)
\!\!&=&\!\!
\partial_t^2 M_3^{2 ; \frac{3}{2}} (t) - \partial_t^2 M_3^{2 ; \frac{1}{2}} (t)
- 
\partial_u^2 M_3^{2 ; \frac{3}{2}} (u) + \partial_u^2 M_3^{2 ; \frac{1}{2}} (u)
\nonumber\\
&&\!\!\!\!\!\!\!
+ \, \frac{2}{3} \left[ \partial_t M_3^{1 ; \frac{3}{2}} (t) - \partial_t M_3^{1 ; \frac{1}{2}} (t) \right]
- \frac{2}{3} \left[ \partial_u M_3^{1 ; \frac{3}{2}} (u) - \partial_u M_3^{1 ; \frac{1}{2}} (u) \right]
\nonumber\\
&&\!\!\!\!\!\!\!
- \, \frac{1}{3} (2u + t - 3 s_0) \left[ \partial_t^2 M_3^{1 ; \frac{3}{2}} (t) - \partial_t^2 M_3^{1 ; \frac{1}{2}} (t) \right] 
+
\frac{1}{3} (2t + u - 3 s_0) \left[ \partial_u^2 M_3^{1 ; \frac{3}{2}} (u) - \partial_u^2 M_3^{1 ; \frac{1}{2}} (u) \right]
,
\nonumber
\eea
and
\be
\partial_u \partial_t^2 ( \partial_t - \partial_u ) {\cal M}_3 (3s_0 - t - u \vert t,u)
= - \frac{2}{3} \left[ 4 \partial_t^3 M_3^{1 ; \frac{3}{2}} (t) - \partial_t^3 M_3^{1 ; \frac{1}{2}} (t) \right]
.
\ee
For the same reason as before, the first identity again tells us that the combination
$\partial_t^2 \delta M_3^{1 ; \frac{3}{2}} (t) - \partial_t^2 \delta M_3^{1 ; \frac{1}{2}} (t)$
has to be at most linear in $t$. Explicit substitution into the expression of $\delta {\cal N}_3$
shows that $\delta M_3^{1 ; \frac{3}{2}} (s) - \delta M_3^{1 ; \frac{1}{2}} (s)$ actually consists
of a polynomial of at most second degree in $s$. The same conclusion applies to
$4 \delta M_3^{1 ; \frac{3}{2}} (s) - \delta M_3^{1 ; \frac{1}{2}} (s)$, owing to the second
identity.  
It then follows that $\delta M_3^{2 ; \frac{3}{2}} (s)$ and $\delta M_3^{2 ; \frac{1}{2}} (s)$
are polynomials of at most third degree in $s$, and an explicit calculation leads to the expressions
given in eq. \rf{transf_3}. 
Altogether, these transformations now depend on a total of nine arbitrary
parameters $a_3$, $b_3,\ldots$ $h_3$, $k_3$.

\section{Quadratic asymptotic conditions}\label{app:quadratic}
\setcounter{equation}{0}

In this appendix we briefly discuss the situation where the amplitudes are required
to follow an asymptotic behaviour that is at most quadratic in the independent variables
$s$ and $\tau$,
\be
\lim_{\lambda\to + \infty} {\cal M}_{1,3} (\lambda s , \lambda\tau) = {\cal O} (\lambda^2)
, \ \ \lim_{\lambda\to + \infty} {\cal N}_{1,3} (\lambda s , \lambda\tau) = {\cal O} (\lambda^2)
.
\lbl{asympt_2}
\ee
For the individual single-variable amplitudes, this means
\bea
M_1^{2 ; \frac{3}{2}} (s) \!&\asymp &\! \frac{B_1}{3} s^3 + \cdots 
\nonumber\\
M_1^{1 ; \frac{3}{2}} (s) \!&\asymp &\! 2 A_1 s^3 + (B_1 - 6 A_1 s_0) s^2 + \cdots 
\nonumber\\
M_1^{1 ; \frac{1}{2}} (s) \!&\asymp &\! - A_1 s^3 + (B_1 + 3 A_1 s_0) s^2 +\cdots 
\nonumber\\
M_1^{0 ; \frac{1}{2}} (s) \!&\asymp &\! - \frac{4B_1}{9} s^3 + \cdots 
.
\lbl{asympt_M1_quad}
\eea
and
\bea
M_3^{2 ; \frac{3}{2}} (s) \!&\asymp &\! \frac{1}{9} (A_3 + 2 C_3) s^3 + \cdots 
\nonumber\\
M_3^{2 ; \frac{1}{2}} (s) \!&\asymp &\! \frac{1}{9} (4A_3 - C_3) s^3 + \cdots
\nonumber\\
M_3^{1 ; \frac{3}{2}} (s) \!&\asymp &\! A_3 s^2 + B_3 s + \cdots 
\nonumber\\
M_3^{1 ; \frac{1}{2}} (s) \!&\asymp &\! C_3 s^2 + D_3 s + \cdots 
\nonumber\\
M_3^{0 ; \frac{3}{2}} (s) \!&\asymp &\! - \frac{2}{27} (5 A_3 + C_3) s^3  + \cdots 
.
\lbl{asympt_M3_quad}
\eea
By an appropriate choice of $a_1$ and $b_1$ in Eq. \rf{transf_1}, namely $a_1=-A_1$ and $b_1=-B_1$,
one can achieve that both $M_1^{1 ; \frac{3}{2}} (s)$ and $ M_1^{1 ; \frac{1}{2}} (s)$ grow
only linearly with $s$, while both $M_1^{2 ; \frac{3}{2}} (s)$ and $ M_1^{0 ; \frac{1}{2}} (s)$ grow
only quadratically, cf. Eq. \rf{transf_1}. The remaining parameters $c_1$ to $h_1$
can then be used to constrain the behaviour of the functions $M_1^{1 ; \frac{3}{2}} (s)$,
$ M_1^{1 ; \frac{1}{2}} (s)$ and $M_1^{2 ; \frac{3}{2}} (s)$ at $s=0$. Explicitly, one ends up with
\be
M_1^{0 ; \frac{1}{2}} (s) \asymp s^2 , \ \ M_1^{1 ; \frac{3}{2}} (s) \asymp s ,
\ \ M_1^{1 ; \frac{1}{2}} (s) \asymp s , \ \ M_1^{2 ; \frac{3}{2}} (s) \asymp s^2 ,
\ee
and 
\be
M_1^{1 ; \frac{3}{2} \prime} (0) = M_1^{1 ; \frac{3}{2}} (0) = 0 ,
\ \ M_1^{1 ; \frac{1}{2} \prime} (0) = M_1^{1 ; \frac{1}{2}} (0) = 0,
\ \ M_1^{2 ; \frac{3}{2} \prime} (0) = M_1^{2 ; \frac{3}{2}} (0) = 0 .
\ee
Likewise, using the transformations \rf{transf_3} with $a_3 = - A_3$, $b_3 = - B_3$, $c_3 = - C_3$, $d_3 = - D_3$, 
one achieves
\be
M_3^{1 ; \frac{3}{2}} (s) \asymp {\rm constant},
\ \ M_3^{1 ; \frac{1}{2}} (s) \asymp {\rm constant} ,
\ \ M_3^{2 ; \frac{3}{2}} (s) \asymp {\cal O}(s^2) ,
\ \ M_3^{2 ; \frac{1}{2}} (s) \asymp {\cal O}(s) ,
\ \ M_3^{0 ; \frac{3}{2}} (s) \asymp {\cal O}(s^2)
.
\ee
Finally, with the remaining five parameters of the transformations 
\rf{transf_3} one can, for instance,  enforce the additional conditions
\be
M_3^{1 ; \frac{3}{2}} (0) = M_3^{1 ; \frac{1}{2}} (0) = 0 ,
\ \ M_3^{2 ; \frac{3}{2}} (0) = M_3^{2 ; \frac{3}{2}\prime} (0) = 0 ,
\ \ M_3^{2 ; \frac{1}{2}} (0) = 0
.
\ee
With these choices, the dispersion relations now take the following form:
\bea
M_1^{2 ; \frac{3}{2}} (s) &=&  
{\tilde\gamma}_{1,2} s^2 + \frac{s^{3}}{\pi}
\int_{4 M_\pi^2}^\infty \frac{dx}{x^3} \, \frac{{\rm Abs}\,M_1^{2 ; \frac{3}{2}} (x)}{x - s}
, 
\nonumber\\
M_1^{1 ; \frac{3}{2}} (s) &=& \frac{s^{2}}{\pi}
\int_{4 M_\pi^2}^\infty \frac{dx}{x^2} \, \frac{{\rm Abs}\,M_1^{1 ; \frac{3}{2}} (x)}{x - s}
, 
\nonumber\\
M_1^{1 ; \frac{1}{2}} (s) &=& \frac{s^{2}}{\pi}
\int_{4 M_\pi^2}^\infty \frac{dx}{x^2} \, \frac{{\rm Abs}\,M_1^{1 ; \frac{1}{2}} (x)}{x - s}
, 
\nonumber\\
M_1^{0 ; \frac{1}{2}} (s) &=& {\tilde\alpha}_{1,0} + {\tilde\beta}_{1,0} s + {\tilde\gamma}_{1,0} s^2 + \frac{s^{3}}{\pi}
\int_{4 M_\pi^2}^\infty \frac{dx}{x^3} \, \frac{{\rm Abs}\,M_1^{0 ; \frac{1}{2}} (x)}{x - s}
,
\eea
and
\bea
M_3^{2 ; \frac{3}{2}} (s) &=&  
{\tilde\gamma}_{3,2} s^2 + \frac{s^{3}}{\pi}
\int_{4 M_\pi^2}^\infty \frac{dx}{x^3} \, \frac{{\rm Abs}\,M_3^{2 ; \frac{3}{2}} (x)}{x - s}
, 
\nonumber\\
M_3^{2 ; \frac{1}{2}} (s) &=&  
{\tilde\beta}_{3,2} s + \frac{s^{2}}{\pi}
\int_{4 M_\pi^2}^\infty \frac{dx}{x^2} \, \frac{{\rm Abs}\,M_3^{2 ; \frac{3}{2}} (x)}{x - s}
, 
\nonumber\\
M_3^{1 ; \frac{3}{2}} (s) &=& \frac{s}{\pi}
\int_{4 M_\pi^2}^\infty \frac{dx}{x} \, \frac{{\rm Abs}\,M_3^{1 ; \frac{3}{2}} (x)}{x - s}
, 
\nonumber\\
M_3^{1 ; \frac{1}{2}} (s) &=& \frac{s}{\pi}
\int_{4 M_\pi^2}^\infty \frac{dx}{x} \, \frac{{\rm Abs}\,M_3^{1 ; \frac{1}{2}} (x)}{x - s}
, 
\nonumber\\
M_3^{0 ; \frac{3}{2}} (s) &=& {\tilde\alpha}_{3,0} + {\tilde\beta}_{3,0} s + {\tilde\gamma}_{3,0} s^2 + \frac{s^{3}}{\pi}
\int_{4 M_\pi^2}^\infty \frac{dx}{x^3} \, \frac{{\rm Abs}\,M_3^{0 ; \frac{3}{2}} (x)}{x - s}
.
\eea

\section{Comparison with ref. \cite{Bijnens:2002vr}}\label{app:comp_BDP}
\setcounter{equation}{0}

The authors of ref. \cite{Bijnens:2002vr} compute the $K\to\pi\pi\pi$
at one-loop order in the low-energy expansion and present their result
in terms of a set of nine independent single-variable functions
$M_i (s)$ that we call here $M_i^{\rm BDP} (s)$ in order to avoid
confusion with some of the functions of the set we are using in the 
present work. We provide here the relations between the two sets of functions,
since they may be useful if one wishes, for instance, to apply the results of the one-loop 
calculation of ref. \cite{Bijnens:2002vr} to our set of functions. These relations read
\bea\lbl{comp_BDP}
M_0^{\rm BDP} (s) \!\!\!&=&\!\!\! - M_0 (s) - \frac{4}{3} M_2 (s) + 2 N_0 (s) + \frac{8}{3} N_2 (s)   ,
\nonumber\\
M_2^{\rm BDP} (s) \!\!\!&=&\!\!\! - M_2 (s) + 2 N_2 (s)   ,
\nonumber\\
M_3^{\rm BDP} (s) \!\!\!&=&\!\!\!  M_1 (s) - 2 N_1 (s)   ,
\nonumber\\
M_4^{\rm BDP} (s) \!\!\!&=&\!\!\! {\tilde N}_2 (s)   ,
\nonumber\\
M_5^{\rm BDP} (s) \!\!\!&=&\!\!\! {\tilde M}_1 (s) + {\tilde N}_1 (s)   ,
\\
M_9^{\rm BDP} (s) \!\!\!&=&\!\!\! - M_1 (s) - N_1 (s) + \frac{3}{2} {\tilde N}_1 (s)    ,
\nonumber\\
M_{10}^{\rm BDP} (s) \!\!\!&=&\!\!\! 2 M_2 (s) + 2 N_2 (s) + {\tilde N}_2 (s)   ,
\nonumber\\
M_{11}^{\rm BDP} (s) \!\!\!&=&\!\!\! M_0 (s) + \frac{1}{3} M_2 (s) + N_0 (s) + \frac{1}{3} N_2 (s) - \frac{1}{2} {\tilde N}_2 (s)   ,
\nonumber\\
M_{12}^{\rm BDP} (s) \!\!\!&=&\!\!\! M_1 (s) + N_1 (s) + \frac{3}{2} {\tilde N}_1 (s)    ,
\nonumber
\eea
and, conversely,
\bea\lbl{comp_BDP_rev}
M_0 (s) \!\!\!&=&\!\!\! \frac{1}{9} \left[ - 3 M_0^{\rm BDP} (s) + 4 M_2^{\rm BDP} (s) + 4 M_4^{\rm BDP} (s) - M_{10}^{\rm BDP} (s) + 6 M_{11}^{\rm BDP} (s) \right]   ,
\nonumber\\
M_1 (s) \!\!\!&=&\!\!\! \frac{1}{3} \left[ M_3^{\rm BDP} (s) - M_9^{\rm BDP} (s) + M_{12}^{\rm BDP} (s) \right]   ,
\nonumber\\
{\tilde M}_1 (s) \!\!\!&=&\!\!\!  \frac{1}{3} \left[ 3 M_5^{\rm BDP} (s) - M_9^{\rm BDP} (s) - M_{12}^{\rm BDP} (s) \right]   ,
\nonumber\\
M_2 (s) \!\!\!&=&\!\!\! \frac{1}{3} \left[ - M_2^{\rm BDP} (s) - M_4^{\rm BDP} (s) + M_{10}^{\rm BDP} (s) \right]   ,
\nonumber\\
N_0 (s) \!\!\!&=&\!\!\! \frac{1}{18} \left[ 6 M_0^{\rm BDP} (s)- 8 M_2^{\rm BDP} (s) + 4 M_4^{\rm BDP} (s) - M_{10}^{\rm BDP} (s) + 6 M_{11}^{\rm BDP} (s) \right]   ,
\\
N_1 (s) \!\!\!&=&\!\!\! \frac{1}{6} \left[ - 2 M_3^{\rm BDP} (s) - M_9^{\rm BDP} (s) + M_{12}^{\rm BDP} (s) \right]    ,
\nonumber\\
{\tilde N}_1 (s) \!\!\!&=&\!\!\! \frac{1}{3} \left[ M_9^{\rm BDP} (s) + M_{12}^{\rm BDP} (s) \right]   ,
\nonumber\\
N_2 (s) \!\!\!&=&\!\!\! \frac{1}{6} \left[ 2 M_2^{\rm BDP} (s) - M_4^{\rm BDP} (s) + M_{10}^{\rm BDP} (s) \right]  ,
\nonumber\\
{\tilde N}_2 (s) \!\!\!&=&\!\!\! M_4^{\rm BDP} (s)   .
\nonumber
\eea
The missing entries in the list \rf{comp_BDP}, i.e. the redundant functions $M_{1,6,7,8}^{\rm BDP} (s)$, can be obtained
from the relations (B.3) in ref. \cite{Bijnens:2002vr}.


\begin{thebibliography}{99}



  
  

\bibitem{Khuri:1960zz} 
  N.~N.~Khuri and S.~B.~Treiman,
  Phys.\ Rev.\  {\bf 119}, 1115 (1960).
  
  
  
\bibitem{Sawyer:1960hrc}
R.~F.~Sawyer and K.~C.~Wali,
Phys. Rev. \textbf{119}, 1429 (1960).
  
  
  
\bibitem{Bronzan:1963mby}
J.~B.~Bronzan and C.~Kacser,
Phys. Rev. \textbf{132}, 2703 (1963).
  
  

\bibitem{Kacser:1963zz}   
  C.~Kacser,
  Phys.\ Rev.\  {\bf 132}, 2712 (1963).
  
  

\bibitem{Bronzan:1964zz}
J.~B.~Bronzan,
Phys. Rev. \textbf{134}, B687 (1964).
  
  
\bibitem{Aitchison:1664}
I. J. R. Aitchison, Nuovo Cim. 35, 434 (1964).


\bibitem{Aitchison:1966lpz}
I.~J.~R.~Aitchison and R.~Pasquier,
Phys. Rev. \textbf{152}, 1274 (1966).


\bibitem{Pasquier:1968zz}
R.~Pasquier and J.~Y.~Pasquier,
Phys. Rev. \textbf{170}, 1294 (1968).


\bibitem{Pasquier:1969dt}
R.~Pasquier and J.~Y.~Pasquier,
Phys. Rev. \textbf{177}, 2482 (1969).
  
  

\bibitem{Neveu:1970tn} 
  A.~Neveu and J.~Scherk,
  Annals Phys.\  {\bf 57}, 39 (1970).
  
  
\bibitem{Aitchison:2015jxa}
I.~J.~R.~Aitchison,
\textit{`Unitarity, Analyticity and Crossing Symmetry in Two- and Three-hadron Final State Interactions},
[arXiv:1507.02697 [hep-ph]].
  
  

\bibitem{Anisovich:1993kn} 
  A.~V.~Anisovich,
  Phys.\ Atom.\ Nucl.\  {\bf 58}, 1383 (1995)
  [Yad.\ Fiz.\  {\bf 58N8}, 1467 (1995)].
  
  

\bibitem{Kambor:1995yc} 
  J.~Kambor, C.~Wiesendanger and D.~Wyler,
  Nucl.\ Phys.\ B {\bf 465}, 215 (1996).
  
  

\bibitem{Anisovich:1996tx} 
  A.~V.~Anisovich and H.~Leutwyler,
  Phys.\ Lett.\ B {\bf 375}, 335 (1996)
  [hep-ph/9601237].
  
  

\bibitem{Guo:2015zqa}
P.~Guo, I.~V.~Danilkin, D.~Schott, C.~Fern\'andez-Ram\'\i{}rez, V.~Mathieu and A.~P.~Szczepaniak,
Phys. Rev. D \textbf{92}, 054016 (2015)
[arXiv:1505.01715 [hep-ph]].
  
  
  
\bibitem{Colangelo:2018jxw}
G.~Colangelo, S.~Lanz, H.~Leutwyler and E.~Passemar,
Eur. Phys. J. C \textbf{78}, 947 (2018)
[arXiv:1807.11937 [hep-ph]].



\bibitem{Albaladejo:2017hhj}
M.~Albaladejo and B.~Moussallam,
Eur. Phys. J. C \textbf{77}, 508 (2017)
[arXiv:1702.04931 [hep-ph]].
  
  

\bibitem{Gasser:2018qtg}
J.~Gasser and A.~Rusetsky,
Eur. Phys. J. C \textbf{78}, 906 (2018)
[arXiv:1809.06399 [hep-ph]].


  
  
\bibitem{Descotes-Genon:2014tla} 
  S.~Descotes-Genon and B.~Moussallam,
  Eur.\ Phys.\ J.\ C {\bf 74}, 2946 (2014)
  [arXiv:1404.0251 [hep-ph]].



\bibitem{Niecknig:2012sj}
F.~Niecknig, B.~Kubis and S.~P.~Schneider,
Eur. Phys. J. C \textbf{72}, 2014 (2012)
[arXiv:1203.2501 [hep-ph]].



\bibitem{Kubis:2014gka}
B.~Kubis and F.~Niecknig,
Phys. Rev. D \textbf{91}, 036004 (2015)
[arXiv:1412.5385 [hep-ph]].



\bibitem{Isken:2017dkw}
T.~Isken, B.~Kubis, S.~P.~Schneider and P.~Stoffer,
Eur. Phys. J. C \textbf{77}, 489 (2017)
[arXiv:1705.04339 [hep-ph]].
  
  
  
\bibitem{Albaladejo:2018gif} 
  M.~Albaladejo {\it et al.} [JPAC Collaboration],
  Eur.\ Phys.\ J.\ C {\bf 78}, 574 (2018)
  [arXiv:1803.06027 [hep-ph]].



\bibitem{JPAC:2020umo}
M.~Albaladejo et al. [JPAC Collaboration],
Eur. Phys. J. C \textbf{80}, 1107 (2020)
[arXiv:2006.01058 [hep-ph]].



\bibitem{Dax:2020dzg}
M.~Dax, D.~Stamen and B.~Kubis,
Eur. Phys. J. C \textbf{81}, 221 (2021)
[arXiv:2012.04655 [hep-ph]].



\bibitem{Stamen:2022eda}
D.~Stamen, T.~Isken, B.~Kubis, M.~Mikhasenko and M.~Niehus,
\textit{Analysis of rescattering effects in $3\pi$ final states}
[arXiv:2212.11767 [hep-ph]].











\bibitem{Kambor:1991ah}
J.~Kambor, J.~H.~Missimer and D.~Wyler,
Phys. Lett. B \textbf{261}, 496-503 (1991).


\bibitem{Bijnens:2002vr}
J.~Bijnens, P.~Dhonte and F.~Borg,
Nucl. Phys. B \textbf{648}, 317-344 (2003)
[arXiv:hep-ph/0205341 [hep-ph]].


\bibitem{Devlin:1978ye}
T.~J.~Devlin and J.~O.~Dickey,
Rev. Mod. Phys. \textbf{51}, 237 (1979).


\bibitem{DAmbrosio:2022jmd}
G.~D'Ambrosio, M.~Knecht and S.~Neshatpour,
Phys. Lett. B \textbf{835}, 137594 (2022)
[arXiv:2209.02143 [hep-ph]].









\bibitem{Cappiello:1992kk}
L.~Cappiello, G.~D'Ambrosio and M.~Miragliuolo,
Phys. Lett. B \textbf{298}, 423 (1993).


\bibitem{Kambor:1993tv}
J.~Kambor and B.~R.~Holstein,
Phys. Rev. D \textbf{49}, 2346-2355 (1994)
[arXiv:hep-ph/9310324 [hep-ph]].


\bibitem{Cohen:1993ta}
A.~G.~Cohen, G.~Ecker and A.~Pich,
Phys. Lett. B \textbf{304}, 347 (1993).


\bibitem{Donoghue:1997rr}
J.~F.~Donoghue and F.~Gabbiani,
Phys. Rev. D \textbf{56}, 1605 (1997)
[arXiv:hep-ph/9702278 [hep-ph]].


\bibitem{Donoghue:1998ur}
J.~F.~Donoghue and F.~Gabbiani,
Phys. Rev. D \textbf{58}, 037504 (1998)
[arXiv:hep-ph/9803331 [hep-ph]].


\bibitem{DAmbrosio:1996cak}
G.~D'Ambrosio and J.~Portol\'es,
Phys. Lett. B \textbf{386}, 403 (1996)
[erratum: Phys. Lett. B \textbf{389}, 770 (1996); erratum: Phys. Lett. B \textbf{395}, 389 (1997)]
[arXiv:hep-ph/9606213 [hep-ph]].



\bibitem{DAmbrosio:1998gur}
G.~D'Ambrosio, G.~Ecker, G.~Isidori and J.~Portol\'es,
JHEP \textbf{08}, 004 (1998)
[arXiv:hep-ph/9808289 [hep-ph]].


\bibitem{Gabbiani:1998tj}
F.~Gabbiani,
Phys. Rev. D \textbf{59}, 094022 (1999)
[arXiv:hep-ph/9812419 [hep-ph]].






\bibitem{Goudzovski:2022scl}
E.~Goudzovski, E.~Passemar, J.~Aebischer, S.~Banerjee, D.~Bryman, A.~Buras, V.~Cirigliano, N.~Christ, A.~Dery and F.~Dettori, \textit{et al.}
\textit{Weak Decays of Strange and Light Quarks},
[arXiv:2209.07156 [hep-ex]].

  
\bibitem{Aebischer:2022vky}
J.~Aebischer, A.~J.~Buras and J.~Kumar,
[arXiv:2203.09524 [hep-ph]].


\bibitem{NA62KLEVER:2022nea}
 [NA62/KLEVER, US Kaon Interest Group, KOTO and LHCb],
\textit{Searches for new physics with high-intensity kaon beams},
[arXiv:2204.13394 [hep-ex]].


\bibitem{HIKE:2022qra}
E.~Cortina Gil \textit{et al.} [HIKE],
\textit{HIKE, High Intensity Kaon Experiments at the CERN SPS: Letter of Intent}
[arXiv:2211.16586 [hep-ex]].








\bibitem{NA482:2005wht}
J.~R.~Batley \textit{et al.} [NA48/2],
Phys. Lett. B \textbf{633}, 173-182 (2006)
[arXiv:hep-ex/0511056 [hep-ex]].



\bibitem{KTeV:2008gel}
E.~Abouzaid \textit{et al.} [KTeV],
Phys. Rev. D \textbf{78}, 032009 (2008)
[arXiv:0806.3535 [hep-ex]].





\bibitem{ParticleDataGroup:2022pth}
R.~L.~Workman \textit{et al.} [Particle Data Group],
PTEP \textbf{2022}, 083C01 (2022)
doi:10.1093/ptep/ptac097

  

\bibitem{CondonShortley1953}
E. U. Condon and G. H. Shortley, \textit{Theory of Atomic Spectra}, Cambridge Univ. Press, 1953.



\bibitem{deSwart:1963pdg}
J.~J.~de Swart,
Rev. Mod. Phys. \textbf{35}, 916 (1963)
[erratum: Rev. Mod. Phys. \textbf{37}, 326-326 (1965)].

  
  
  
\bibitem{Gaillard:1974nj}                          
  M.~K.~Gaillard and B.~W.~Lee,
  Phys.\ Rev.\ Lett.\  {\bf 33}, 108 (1974).


\bibitem{Altarelli:1974exa} 
  G.~Altarelli and L.~Maiani,
  Phys.\ Lett.\  {\bf 52B}, 351 (1974).


\bibitem{Shifman:1975tn} 
  M.~A.~Shifman, A.~I.~Vainshtein and V.~I.~Zakharov,
  Nucl.\ Phys.\ B {\bf 120}, 316 (1977).


\bibitem{Witten:1976kx} 
  E.~Witten,
  Nucl.\ Phys.\ B {\bf 122}, 109 (1977).


\bibitem{Wise:1979at}   
  M.~B.~Wise and E.~Witten,
  Phys.\ Rev.\ D {\bf 20}, 1216 (1979).
  


\bibitem{Gilman:1979bc} 
  F.~J.~Gilman and M.~B.~Wise,
  Phys.\ Rev.\ D {\bf 20}, 2392 (1979).






\bibitem{Chew:1960iv}
G.~F.~Chew and S.~Mandelstam,
Phys. Rev. \textbf{119}, 467 (1960).


\bibitem{Petersen:1977cs}
J.~L.~Petersen,
\textit{The pi pi Interaction}, 
CERN Yellow Report CERN-1977-004.













\bibitem{Stern:1993rg}
J.~Stern, H.~Sazdjian and N.~H.~Fuchs,
Phys. Rev. D \textbf{47}, 3814 (1993)
[arXiv:hep-ph/9301244 [hep-ph]].


\bibitem{Zdrahal:2008bd}
M.~Zdr\'ahal and J.~Novotn\'y,
Phys. Rev. D \textbf{78}, 116016 (2008)
[arXiv:0806.4529 [hep-ph]].


\bibitem{Kampf:2008ts}
K.~Kampf, M.~Knecht, J.~Novotn\'y and M.~Zdr\'ahal,
Nucl. Phys. B Proc. Suppl. \textbf{186}, 334 (2009)
[arXiv:0810.1906 [hep-ph]].


\bibitem{Zdrahal:2009ns}
M.~Zdr\'ahal, K.~Kampf, M.~Knecht and J.~Novotn\'y,
PoS \textbf{EFT09}, 063 (2009)
[arXiv:0905.4868 [hep-ph]].


\bibitem{Kampf:2019bkf}
K.~Kampf, M.~Knecht, J.~Novotn\'y and M.~Zdr\'ahal,
Phys. Rev. D \textbf{101}, 074043 (2020)
[arXiv:1911.11762 [hep-ph]].









\bibitem{Dalitz:1956da}
R.H. Dalitz, Proc. Phys. Soc. A \textbf{69}, 527 (1956).


\bibitem{Weinberg:1960zza}         
S.~Weinberg,
Phys. Rev. Lett. \textbf{4}, 87 (1960)
[erratum: Phys. Rev. Lett. \textbf{4}, 585 (1960)].


\bibitem{Barton:1963mg}
G.~Barton, C.~Kacser, S.P. Rosen, Phys. Rev. \textbf{130}, 783 (1963).


\bibitem{Zemach:1963bc}
C.~Zemach,
Phys. Rev. \textbf{133}, B1201 (1964).


\bibitem{DAmbrosio:1994vba}
G.~D'Ambrosio, G.~Isidori, A.~Pugliese and N.~Paver,
Phys. Rev. D \textbf{50}, 5767 (1994)
[erratum: Phys. Rev. D \textbf{51}, 3975 (1995)]
[arXiv:hep-ph/9403235 [hep-ph]].








\bibitem{Martin:1976mb}
B.~R.~Martin, D.~Morgan and G.~Shaw,
\textit{Pion Pion Interactions in Particle Physics},
Academic Press Inc., London, 1976.


\bibitem{NA482:2010dug}
J.~R.~Batley \textit{et al.} [NA48/2],
Eur. Phys. J. C \textbf{70}, 635 (2010).


\bibitem{Cirigliano:2009rr}
V.~Cirigliano, G.~Ecker and A.~Pich,
Phys. Lett. B \textbf{679}, 445 (2009)
[arXiv:0907.1451 [hep-ph]].


\bibitem{Colangelo:2001df}
G.~Colangelo, J.~Gasser and H.~Leutwyler,
Nucl. Phys. B \textbf{603}, 125 (2001)
[arXiv:hep-ph/0103088 [hep-ph]].


\bibitem{Kaminski:2006qe}
R.~Kaminski, J.~R.~Pelaez and F.~J.~Yndurain,
Phys. Rev. D \textbf{77}, 054015 (2008)
[arXiv:0710.1150 [hep-ph]].


\bibitem{Ananthanarayan:2000ht}
B.~Ananthanarayan, G.~Colangelo, J.~Gasser and H.~Leutwyler,
Phys. Rept. \textbf{353}, 207 (2001)
[arXiv:hep-ph/0005297 [hep-ph]].


\bibitem{Hyams:1973zf}
B.~Hyams, C.~Jones, P.~Weilhammer, W.~Blum, H.~Dietl, G.~Grayer, W.~Koch, E.~Lorenz, G.~Lutjens and W.~Manner, \textit{et al.}
Nucl. Phys. B \textbf{64}, 134 (1973).


\bibitem{Moussallam:1999aq}
B.~Moussallam,
Eur. Phys. J. C \textbf{14}, 111 (2000)
[arXiv:hep-ph/9909292 [hep-ph]].

\bibitem{Losty:1973et}
M.~J.~Losty, V.~Chaloupka, A.~Ferrando, L.~Montanet, E.~Paul, D.~Yaffe, A.~Zieminski, J.~Alitti, B.~Gandois and J.~Louie,
Nucl. Phys. B \textbf{69}, 185 (1974).


\bibitem{Hoogland:1977kt}
W.~Hoogland, S.~Peters, G.~Grayer, B.~Hyams, P.~Weilhammer, W.~Blum, H.~Dietl, G.~Hentschel, W.~Koch and E.~Lorenz, \textit{et al.}
Nucl. Phys. B \textbf{126}, 109 (1977).


\bibitem{Moussallam:2011zg}
B.~Moussallam,
Eur. Phys. J. C \textbf{71}, 1814 (2011)
[arXiv:1110.6074 [hep-ph]].








\bibitem{CPLEAR:1998nkj}
A.~Angelopoulos \textit{et al.} [CPLEAR],
Eur. Phys. J. C \textbf{5}, 389 (1998)   .


\bibitem{NA48:2005uiw}
J.~R.~Batley \textit{et al.} [NA48],
Phys. Lett. B \textbf{630}, 31 (2005)
[arXiv:hep-ex/0510008 [hep-ex]].


\bibitem{Cabibbo:2005ez}
N.~Cabibbo and G.~Isidori,
JHEP \textbf{03}, 021 (2005)
[arXiv:hep-ph/0502130 [hep-ph]].


\bibitem{NA48:2001jrj}
A.~Lai \textit{et al.} [NA48],
Phys. Lett. B \textbf{515}, 261 (2001)
[arXiv:hep-ex/0106075 [hep-ex]].







\bibitem{Li:1979wa}
L.F. Li, L.~Wolfenstein, Phys. Rev. D \textbf{21}, 178 (1980)







\bibitem{Grinstein:1985ut}
B.~Grinstein, S.~J.~Rey and M.~B.~Wise,
Phys. Rev. D \textbf{33}, 1495 (1986).


\bibitem{DAmbrosio:1991oli}
G.~D'Ambrosio, G.~Isidori and N.~Paver,
Phys. Lett. B \textbf{273}, 497 (1991).


\bibitem{Isidori:1991xx}
G.~Isidori, L.~Maiani and A.~Pugliese,
Nucl. Phys. B \textbf{381}, 522 (1992).









\bibitem{Colangelo:2006va}
G.~Colangelo, J.~Gasser, B.~Kubis and A.~Rusetsky,
Phys. Lett. B \textbf{638}, 187-194 (2006)
[arXiv:hep-ph/0604084 [hep-ph]].


\bibitem{Bissegger:2007yq}
M.~Bissegger, A.~Fuhrer, J.~Gasser, B.~Kubis and A.~Rusetsky,
Phys. Lett. B \textbf{659}, 576-584 (2008)
[arXiv:0710.4456 [hep-ph]].


\bibitem{Gasser:2011ju}
J.~Gasser, B.~Kubis and A.~Rusetsky,
Nucl. Phys. B \textbf{850}, 96-147 (2011)
[arXiv:1103.4273 [hep-ph]].









  

\bibitem{Bell:1966}
  J. S. Bell, \textit{Weak interactions and current algebras},
  Proc. of the 1966 CERN School of Physics, CERN Report 66-29 vol. 1 (1966).
  
  
  
\bibitem{Bouchiat:1967wjs}   
C.~Bouchiat and P.~Meyer,
Phys. Lett. B \textbf{25}, 282 (1967).


\bibitem{Kambor:1992he}
J.~Kambor, J.~F.~Donoghue, B.~R.~Holstein, J.~H.~Missimer and D.~Wyler,
Phys. Rev. Lett. \textbf{68}, 1818 (1992).












 
\end{thebibliography}
\end{document}